\documentclass{aa}  

\usepackage{subfig}
\usepackage{xspace}
\usepackage{amsmath}
\usepackage{graphicx}
\usepackage{natbib}
\usepackage{lscape}

\usepackage{multirow}

\begin{document}
\newcommand{\um}{\ensuremath{\mu\mathrm{m}}}
\newcommand{\lnir}{\ensuremath{L_{\rm AGN}^{\rm NIR}}}
\newcommand{\lmir}{\ensuremath{L_{\rm AGN}^{\rm MIR}}}
\newcommand{\lx}{\ensuremath{L_{\rm AGN}^{\rm X}}}
\newcommand{\oiv}{\ensuremath{{\rm [O IV]} \lambda 25.89 \um{}}}
\newcommand{\loiv}{\ensuremath{L_{\rm AGN}^{\rm [OIV]}}}
\newcommand{\fnir}{\ensuremath{f_{\rm AGN}^{\rm NIR}}}
\newcommand{\brg}{\ensuremath{{\rm Br}\gamma}}

\title{Obscuration in AGNs: near-infrared luminosity relations and dust colors}

\author{L. Burtscher \inst{1}
	\and
	G. Orban de Xivry \inst{1}
	\and
	R. I. Davies \inst{1}
	\and
	A. Janssen \inst{1}
	\and
	D. Lutz \inst{1}
	\and
	D. Rosario \inst{1}
	\and
	A. Contursi \inst{1}
	\and
	R. Genzel \inst{1}
	\and
	J. Graci\'{a}-Carpio \inst{1}
	\and
	M.-Y. Lin \inst{1}
	\and
	A. Schnorr-M\"uller \inst{1}
	\and
	A. Sternberg \inst{2}
	\and
	E. Sturm \inst{1}
	\and
	L. Tacconi \inst{1}
}

\institute{Max-Planck-Institut f\"ur extraterrestrische Physik,
		Postfach 1312, Gie\ss enbachstr., 85741 Garching, Germany\\
		\email{burtscher@mpe.mpg.de}
	\and
		Raymond and Beverly Sackler School of Physics \& Astronomy, Tel Aviv University, Ramat Aviv 69978, Israel
}

\date{Draft version, \today}                                           % Activate to display a given date or no date
%   \date{Received ; accepted }

% limit to 300 words!

\abstract{We combine two approaches to isolate the AGN luminosity at near-infrared wavelengths and relate the near-IR pure AGN luminosity to other tracers of the AGN. Using integral-field spectroscopic data of an archival sample of 51 local AGNs, we estimate the fraction of non-stellar light by comparing the nuclear equivalent width of the stellar 2.3 \um{} CO absorption feature with the intrinsic value for each galaxy. We compare this fraction to that derived from a spectral decomposition of the integrated light in the central arc second and find them to be consistent with each other.
Using our estimates of the near-IR AGN light, we find a strong correlation with presumably isotropic AGN tracers. We show that a significant offset exists between type 1 and type 2 sources in the sense that type 1 sources are 7 (10) times brighter in the near-IR at $\log \lmir{}$ = 42.5 ($\log \lx{}$ = 42.5). These offsets only becomes clear when treating infrared type 1 sources as type 1 AGNs.

All AGNs have very red near-to-mid-IR dust colors. This, as well as the range of observed near-IR temperatures, can be explained with a simple model with only two free parameters: the obscuration to the hot dust and the ratio between the warm and hot dust areas. We find obscurations of $A_{\rm V}^{\rm hot} = 5 \ldots 15$ mag for infrared type 1 sources and $A_{\rm V}^{\rm hot} = 15 \ldots 35$ mag for type 2 sources. The ratio of hot dust to warm dust areas of about 1000 is nicely consistent with the ratio of radii of the respective regions as found by infrared interferometry.}


\maketitle

%%%%%%%%%%%%%%%%%%%%%%%%%%%%%%%%%%%%%%%%%%%%%%%%%%%%%%%%%%%%%%%%%%%%%%%%%%%%%%%%%%%%%%%%%%%
%%%%%%%%%%%%%%%%%%%%%%%%%%%%%%%%%%%%%%%%%%%%%%%%%%%%%%%%%%%%%%%%%%%%%%%%%%%%%%%%%%%%%%%%%%%
%%%%%%%%%%%%%%%%%%%%%%%%%%%%%%%%%%%%%%%%%%%%%%%%%%%%%%%%%%%%%%%%%%%%%%%%%%%%%%%%%%%%%%%%%%%
%%%%%%%%%%%%%%%%%%%%%%%%%%%% INTRODUCTION %%%%%%%%%%%%%%%%%%%%%%%%%%%%
%%%%%%%%%%%%%%%%%%%%%%%%%%%%%%%%%%%%%%%%%%%%%%%%%%%%%%%%%%%%%%%%%%%%%%%%%%%%%%%%%%%%%%%%%%%
%%%%%%%%%%%%%%%%%%%%%%%%%%%%%%%%%%%%%%%%%%%%%%%%%%%%%%%%%%%%%%%%%%%%%%%%%%%%%%%%%%%%%%%%%%%
%%%%%%%%%%%%%%%%%%%%%%%%%%%%%%%%%%%%%%%%%%%%%%%%%%%%%%%%%%%%%%%%%%%%%%%%%%%%%%%%%%%%%%%%%%%

\section{Introduction}

Most, if not all, Active Galactic Nuclei (AGNs) have nuclear dust structures on parsec scales that are made responsible for the line of sight obscuration in edge-on systems \citep{antonucci1993} and the collimation of the ionizing radiation from the AGN \citep[e.g.][]{fischer2013}. While these ``tori'' were initially invoked to explain the differences between edge-on and face-on systems, it is increasingly realized that they are not simple ``donuts''. Instead, they are part of a complex and dynamical accretion flow towards the central super-massive black hole \citep[e.g.][]{norman1988, wada2002, vollmer2008, wada2009, schartmann2009, schartmann2010, hopkins2010b, hopkins2012}. Multi-scale hydrodynamical simulations are successful in predicting observed properties such as the scale-size and inferred mass of the parsec-scale ``torus'' \citep{schartmann2009,schartmann2010} and explain the appearance of dusty disk structures as a dynamical consequence of accretion \citep{hopkins2012}.

This region (we will keep calling it torus for the sake of simplicity) is very prominently seen by its thermal emission. It re-radiates a substantial fraction of the bolometric luminosity of the AGN central engine in near- to mid-infrared wavelengths \citep[e.g.][]{prieto2010}. While most tori are unresolved even in the highest resolution single-dish observations \citep[e.g.][]{asmus2014}, more than two dozen have now been spatially resolved using near- and mid-infrared interferometry \citep[e.g.][]{kishimoto2011,burtscher2013} and their mid-infrared sizes have been measured to range between about 0.1 and 10 pc for Seyfert galaxies and quasars up to a bolometric luminosity of $L_{\rm bol} = 10^{47}$ erg/s.

In infrared observations, the most prominent difference between face-on (type 1) and edge-on (type 2) systems is seen in the Silicate feature at 9.7 \um{} which is often in emission in type 1 and mostly in absorption in type 2 objects \citep[e.g.][]{hoenig2010}. Models with continuous \citep[e.g.][]{schartmann2005} and clumpy \citep[e.g.][]{nenkova2002} dust distributions have been invoked to explain the observed differences between the various classes of AGNs and to relate them to torus properties such as inclination angle, number of clumps or opening angle \citep[see][for a recent review]{hoenig2013b}. It is unclear, however, how accurately torus parameters can be retrieved from SED fits as the assumptions about the torus structure (e.g. power-law distribution in clump density) may not be correct. Interferometric observations have found that essentially all infrared bright tori consist of two spatially distinct components which are very different in size \citep{burtscher2013} which seems to contradict the simple cloud distributions assumed in models. Moreover, in a comparison between torus models, \citet{feltre2012} found that most of the differences in the SEDs between models arise not from the dust morphology but from the model assumptions.

On the other hand, all models predict significantly different intensities from dust emission at infrared wavelengths between edge-on and face-on inclinations. In many studies, however, no such differences were observed \citep{lutz2004,gandhi2009,asmus2011}, thereby challenging the classical, inclination-dependent AGN unification scheme.

In this study we compile a large archival sample of high-resolution near-IR integral field spectroscopic data of local AGNs and determine the fraction of non-stellar light in the near-IR with two complementary methods. We describe the sample and the data analysis in Section \ref{sec:sample} and the decomposition of the nuclear light into AGN and stellar light in Section \ref{sec:decompose}. We then compare the near-IR AGN luminosity with other tracers of the AGN luminosity and discuss the luminosity relations in the context of obscuration and isotropy (Section \ref{sec:discussion}). We summarize our findings in Section \ref{sec:conclusions}.

%%%%%%%%%%%%%%%%%%%%%%%%%%%%%%%%%%%%%%%%%%%%%%%%%%%%%%%%%%%%%%%%%%%%%%%%%%%%%%%%%%%%%%%%%%%
%%%%%%%%%%%%%%%%%%%%%%%%%%%%%%%%%%%%%%%%%%%%%%%%%%%%%%%%%%%%%%%%%%%%%%%%%%%%%%%%%%%%%%%%%%%
%%%%%%%%%%%%%%%%%%%%%%%%%%%%%%%%%%%%%%%%%%%%%%%%%%%%%%%%%%%%%%%%%%%%%%%%%%%%%%%%%%%%%%%%%%%
%%%%%%%%%%%%%%%%%%%%%%%%%%%% SAMPLE, DATA REDUCTION, OTHER DATA %%%%%%%%%%%%%%%%%%%%%%%%%%%%
%%%%%%%%%%%%%%%%%%%%%%%%%%%%%%%%%%%%%%%%%%%%%%%%%%%%%%%%%%%%%%%%%%%%%%%%%%%%%%%%%%%%%%%%%%%
%%%%%%%%%%%%%%%%%%%%%%%%%%%%%%%%%%%%%%%%%%%%%%%%%%%%%%%%%%%%%%%%%%%%%%%%%%%%%%%%%%%%%%%%%%%
%%%%%%%%%%%%%%%%%%%%%%%%%%%%%%%%%%%%%%%%%%%%%%%%%%%%%%%%%%%%%%%%%%%%%%%%%%%%%%%%%%%%%%%%%%%

\section{Sample selection and data reduction}
\label{sec:sample}

We select observations of all objects from the ESO SINFONI \citep{eisenhauer2003c,bonnet2004} archive that have an AGN classification in SIMBAD. Additionally we impose a distance limit ($\lesssim$ 60 Mpc) in order to well resolve the nuclear region of the AGN. At the maximum distance, our resolution is about 30 pc (300 pc) in the medium resolution cubes with (without) AO.
We searched the ESO archive for SINFONI observations in the H+K or K gratings with any pixel scale and required only 10 minutes or more of time on target with seeing better than about 1 arcsec.\footnote{To find these observations, we downloaded the headers of all SINFONI observations from the ESO archive and parsed the relevant data into an SQLite database. From that we constructed a list of unique coordinates which we parsed through SIMBAD to retrieve the object identifications and classifications based on these coordinates. From this list we filtered out local AGNs with sufficient data taken with appropriate settings, sufficient signal to noise and under good atmospheric conditions.} In addition we add one source observed with Gemini NIFS (NGC~4051), the reduced cube of which was kindly provided to us by R.~A.~Riffel. This resulted in a sample of 51 active galaxies. The list of targets and observations can be found in Tab.~\ref{tab:sources}.

Out of the 51 objects, 40 have been observed with AO (6 with laser guide star) and 11 without. We detect non-stellar continuum associated with the AGN in the near-IR in 31 sources (see Section~\ref{sec:decompose}). Only five of these sources have been observed without AO. The median seeing of these cubes is 0.9 arcsec.

For the distance we adopt the median value of redshift-independent distance measurements from NED, where available, else the luminosity distance computed by NED using a $\Lambda$CDM cosmology with $H_0$ = 73 km/s/Mpc, $\Omega_{\rm matter} = 0.27, \Omega_{\rm vacuum} = 0.73$. For those galaxies where our X-ray reference (see below) quotes luminosities we adopt the distance used in the particular reference.

\begin{landscape}
%%
%% This table is generated by IDL, but the following columns have been modified by hand: source name, observing dates, reference or ESO prog. ID, distance ref
%% referee process: added column "physical resolution" by hand
%%
\begin{table}
\caption{\label{tab:sources}Table of sources and relevant SINFONI data. In the column seeing, $^n$ marks data taken without AO. Distance references: $^a$: redshift-independent estimate from NED, $^b$: redshift-based distance from NED, $^c$: \citet{levenson2006}, $^d$: \citet{asmus2011}, $^e$: \citet{castangia2013}, $^f$: \citet{osullivan2001}, $^g$: \citet{blandhawthorn1997}, $^h$: \citet{panessa2006}, $^i$: \citet{liu2011}, $^j$: \citet{harris2010}}
\vspace{-0.5cm}
\begin{center}
\tiny
\begin{tabular}{|l|l|l|l|l|l|l|l|l|l|}
Source name & redshift & distance [Mpc] &resolution [pc] & type [NED] & grat1 & opti1 & avg. seeing & observing dates & reference or ESO prog. ID \\
\hline
Circinus galaxy& 0.0014&  4.2$^a$&0.5&           Sy1h&  K& 25&0.7&                                       2004-07-14,2004-07-20&                      \citet{muellersanchez2006}\\
    ESO 428-G14& 0.0056& 24.3$^c$&11.8&            Sy2&  K&100&1.0&                                       2010-12-21,2011-01-10&           86.B-0484(PI:Mueller-Sanchez)\\
    ESO 548-G81& 0.0145& 58.5$^b$&28.4&            Sy1&  K&100&1.2&                                                  2008-10-21&                   82.B-0709(PI:Beckert)\\
        IC 1459& 0.0060& 28.4$^a$&13.7&          LINER&  K&100&0.9&                                                  2004-10-01&                             SV\_SINFONI\\
        IC 5063& 0.0113& 41.7$^a$&50.5&           Sy1h&H+K&250&1.2$^n$&                                       2012-05-21,2012-06-05&                  89.B-0971(PI:Nesvadba)\\
           M 87& 0.0042& 17.1$^d$&8.3&          LINER&  K&100&0.9&                                                  2010-02-26&                    84.B-0568(PI:Prieto)\\
  MCG-05-23-016& 0.0085& 39.5$^b$&19.1&           Sy1i&  K&100&1.0&                                       2010-12-29,2011-03-04&           86.B-0484(PI:Mueller-Sanchez)\\
        NGC 289& 0.0054& 21.5$^a$&10.4&              -&H+K&100&0.8&                                                  2009-08-29&                   83.B-0620(PI:Fischer)\\
        NGC 613& 0.0050& 20.0$^e$&24.2&            Sy?&H+K&250&0.9$^n$&                            2005-10-22,2005-11-12,2005-11-16&                              \citet{boeker2008}\\
        NGC 676& 0.0050& 19.5$^d$&9.5&            Sy2&H+K&100&1.1&                                                  2009-08-29&                   83.B-0620(PI:Fischer)\\
       NGC 1052& 0.0050& 20.0$^f$&9.7&           Sy3h&  K&100&1.0&      2010-01-03/04/05/06/07&                      \citet{muellersanchez2013}\\
       NGC 1068& 0.0038& 14.4$^g$&7.0&           Sy1h&H+K&100&0.7&                                       2005-10-05,2005-10-06&                      \citet{muellersanchez2009}\\
       NGC 1097& 0.0042& 17.0$^a$&8.2&           Sy3b&H+K&100&0.6&                                                  2005-10-10&                              \citet{davies2007}\\
       NGC 1365& 0.0056& 17.9$^a$&8.7&          Sy1.8&H+K&100&0.9&                 2010-11-17/18,2010-12-02,2011-02-01&                86.B-0635(PI:Schartmann)\\
       NGC 1386& 0.0029& 16.2$^a$&8.9&           Sy1i&  K&100&1.0&                                                  2010-10-04&           86.B-0484(PI:Mueller-Sanchez)\\
       NGC 1566& 0.0050& 12.2$^a$&5.9&          Sy1.5&  K&100&1.2&                                                  2008-10-31&                      \citet{muellersanchez2013}\\
       NGC 2110& 0.0076& 35.6$^a$&17.3&           Sy1i&  K&100&1.2&      2010-10-08,11-07,12-21/28,2011-01-14&           86.B-0484(PI:Mueller-Sanchez)\\
       NGC 2911& 0.0102& 52.0$^a$&25.2&            Sy3&  K&100&0.9&                                                  2009-01-28&                      \citet{muellersanchez2013}\\
       NGC 2974& 0.0066& 24.7$^a$&12.0&            Sy2&  K&100&1.0&                                                  2009-01-20&                   82.B-0709(PI:Beckert)\\
       NGC 2992& 0.0077& 31.6$^a$&15.3&           Sy1i&  K&100&0.7&                                                  2005-03-15&                              \citet{davies2007}\\
       NGC 3081& 0.0080& 26.5$^a$&12.8&           Sy1h&  K&100&0.9&                                                  2011-03-18&                      \citet{muellersanchez2013}\\
       NGC 3169& 0.0041& 18.8$^a$&9.1&            Sy?&  K&100&0.9&                                                  2010-02-24&                      \citet{muellersanchez2013}\\
       NGC 3227& 0.0039& 20.9$^a$&25.3&          Sy1.5&  K&250&0.7&                                                  2009-04-07&                              \citet{davies2007}\\
       NGC 3281& 0.0115& 48.7$^b$&23.6&            Sy2&H+K&100&1.2&                            2011-01-10,2011-02-10,2011-02-23&                86.B-0635(PI:Schartmann)\\
       NGC 3312& 0.0095& 44.7$^d$&21.7&            Sy3&  K&100&0.3&                                                  2009-02-08&                   82.B-0709(PI:Beckert)\\
       NGC 3627 (M 66)& 0.0023&  6.6$^h$&3.2&            Sy3&  K&100&0.7$^n$&                                       2007-03-20,2007-03-24&                            \citet{mazzalay2013}\\
       NGC 3783& 0.0096& 47.8$^a$&23.1&          Sy1.5&  K&100&0.9&                                                  2005-03-18&                              \citet{davies2007}\\
       NGC 4051& 0.0022& 13.3$^a$&6.4&           Sy1n&  K&100&-&                                                     2006-01&                          \citet{riffel_ra2008}\\
       NGC 4261& 0.0075& 29.4$^d$&35.6&           Sy3h&  K&250&1.0&                                                  2010-04-12&                 84.B-0086(PI:Krajnovic)\\
       NGC 4303& 0.0052& 15.2$^d$&18.4&            Sy2&  K&250&0.8$^n$&                                       2009-02-16,2009-02-17&                    82.B-0611(PI:Colina)\\
       NGC 4388& 0.0086& 21.4$^a$&25.9&           Sy1h&  K&250&0.9$^n$&                                       2011-04-11,2011-05-01&                              \citet{greene2014}\\
       NGC 4438& 0.0002& 13.9$^a$&6.8&           Sy3b&  K&100&0.8&                            2011-03-06,2011-03-28,2011-04-24&                  86.B-0781(PI:Casassus)\\
       NGC 4472& 0.0033& 17.1$^d$&20.7&            Sy2&  K&250&0.4&                                                  2009-04-23&                               \citet{rusli2013}\\
       NGC 4501 (M 88)& 0.0076& 16.8$^h$&8.1&            Sy2&  K&100&0.8&                                                  2008-03-11&                            \citet{mazzalay2013}\\
       NGC 4569 (M 90)&-0.0008& 12.4$^a$&6.0&            Sy?&  K&100&0.7&                                                  2008-03-07&                            \citet{mazzalay2013}\\
       NGC 4579 (M 58)& 0.0050& 16.8$^h$&10.2&           Sy3b&  K&100&1.0&                                       2008-03-07,2008-03-08&                            \citet{mazzalay2013}\\
       NGC 4593& 0.0083& 26.6$^a$&12.9&          Sy1.0&  K&100&0.7&                 2008-04-09,2009-03-20/21/22&                    80.B-0239(PI:Davies)\\
       NGC 4762& 0.0033& 16.8$^i$&8.1&              -&  K&100&0.9&                                       2008-03-08,2008-03-11&                    80.B-0336(PI:Saglia)\\
       NGC 5128 (Cen A)& 0.0018&  3.8$^j$&1.9&              -&  K&100&0.6&                                       2005-03-23,2005-04-03&                            \citet{neumayer2007}\\
       NGC 5135& 0.0137& 60.9$^b$&72.7&            Sy2&  K&250&1.1$^n$&                                                  2006-04-06&                      \citet{piqueraslopez2012b}\\
       NGC 5506& 0.0062& 23.8$^a$&11.6&           Sy1i&  K&100&1.0&                                                  2011-03-03&           86.B-0484(PI:Mueller-Sanchez)\\
       NGC 5643& 0.0039& 16.9$^a$&20.5&            Sy2&  K&250&0.7$^n$&                                                  2009-04-07&                               \citet{hicks2013}\\
       NGC 6300& 0.0037& 13.6$^a$&16.4&            Sy2&  K&250&0.8$^n$&                                       2009-05-08,2009-05-16&                               \citet{hicks2013}\\
       NGC 6814& 0.0052& 22.8$^a$&27.6&          Sy1.5&  K&250&0.8$^n$&                            2007-07-21,2009-06-06,2009-06-18&                      \citet{muellersanchez2011}\\
       NGC 7130& 0.0162& 63.6$^b$&77.1&          Sy1.9&  K&250&0.8$^n$&                                       2006-07-02,2006-07-30&                      \citet{piqueraslopez2012b}\\
       NGC 7135& 0.0088& 34.7$^a$&16.8&            AGN&H+K&100&0.8&                                                  2009-08-29&                   83.B-0620(PI:Fischer)\\
       NGC 7172& 0.0087& 33.9$^a$&16.4&            Sy2&H+K&100&1.0&                                                  2009-08-29&                              \citet{smajic2012}\\
       NGC 7469& 0.0159& 56.7$^a$&6.9&          Sy1.5&  K& 25&0.9&                                                  2004-07-13&                              \citet{davies2007}\\
       NGC 7496& 0.0055& 15.0$^a$&7.3&            Sy2&H+K&100&0.8&                                                  2009-08-29&                   83.B-0620(PI:Fischer)\\
       NGC 7582& 0.0053& 20.9$^a$&10.2&           Sy1i&  K&100&0.9&                                       2010-10-07,2010-10-08&           86.B-0484(PI:Mueller-Sanchez)\\
       NGC 7743& 0.0056& 19.2$^d$&23.3&            Sy2&  K&250&0.7$^n$&                                       2009-07-29,2009-08-14&                               \citet{hicks2013}\
\end{tabular}
\end{center}
\end{table}
\end{landscape}
\normalsize

Data reduction was performed with {\tt spred}, a custom package developed at MPE for the analysis of SINFONI data \citep{abuter2006}. For the sky subtraction, we used the routines {\tt mxcor} and {\tt skysub} \citep{davies2007b} which correlate reconstructed object and sky cubes spectrally and shift them such that their OH line wavelengths match before subtracting the scaled sky frame. This reduces sky subtraction residuals that otherwise occur because of a change in the wavelength scale between target and sky observations. This was followed by bad pixel and cosmic ray treatment using {\tt lac3d}, a 3D Laplacian Edge Detection algorithm based on L.A.COSMIC \citep{van_dokkum2001}.

The typical calibration uncertainty is better than 0.1 mag (or about 10 \% in flux) as determined from the standard deviation of aperture photometries of the individual observations per source (before combining). The major source of uncertainty is variations in the conversion factor within and between nights. Only for NGC~7743 is the calibration uncertain to about 0.5 mag because the observations were performed under variable conditions. However, this source does not show any signs for AGN emission in the near-IR and is therefore not used in our quantitative analysis.

\subsection{Mid-IR, X-ray and [O IV] samples}
In order to compare our near-IR derived AGN luminosities with other AGN indicators we collect X-ray and mid-IR data for all sources. Ground-based, high spatial resolution observations in the mid-infrared $N$ band (8--13 \um{}) exist for all objects with detected near-IR AGN light except for NGC~7496 and also for about half of the near-IR non-detections. The mid-IR data are mostly from VLT/VISIR observations whose high spatial resolution ensures minimum contamination by the host galaxy. Most of the near-IR detected galaxies (24) also appear in the 70-month BAT all-sky survey \citep{baumgartner2013} in the very hard X-ray band 14-195 keV. This waveband is particularly helpful for AGN studies since it is essentially unaffected by obscuration except for the most Compton-thick objects. For six further objects, absorption-corrected 2-10 keV fluxes have been published which we use with the conversion factor $L_{\rm 14-195 keV} \approx 5 \times L_{\rm 2-10 keV}$ taken from Fig. 6 of \citet{winter2009}. NGC~7496 is again the odd-one out which does not have a published hard X-ray flux although we detect non-stellar light in the near-IR. Conversely, only two near-IR non-detected sources are BAT-detected, but about half of the near-IR non-detections have published mid-IR fluxes.

Additionally, we collected \oiv{} observations for those objects with near-IR detections from the literature. This mid-infrared line is only little affected by dust and due to its high ionization potential of 54.9 eV it is also essentially unaffected by starlight. It has been shown to be a good isotropic tracer of the AGN luminosity \citep{rigby2009,diamondstanic2009,lamassa2010}.

In summary, out of the full sample of 51 AGNs, we can determine the near-IR AGN fraction in 31 objects (for the others we give limits). Except for NGC~7496, all of these sources are also detected in the mid-IR and hard X-rays and all except four have been detected in [O IV]. This leads to a final sample size of 30 (27) for our multi-band AGN relations (including [O IV]). Throughout this article we denote the AGN luminosity in the near-IR as derived from the dilution of the 2.3 \um{} CO absorption feature with \lnir{}, the nuclear mid-IR luminosity with \lmir{}, the hard X-ray luminosity with \lx{} and the \oiv{} luminosity by \loiv{}. Our collection of multi-wavelength data is presented in Tab.~\ref{tab:starfit}.

Errors for these luminosities are taken from the respective references. The median errors on \lmir{}, \lx{}, \loiv{} are 0.06, 0.02, 0.03 dex respectively; error bars are shown in the plots, but omitted in the Table for compactness.

%%%%%%%%%%%%%%%%%%%%%%%%%%%%%%%%%%%%%%%%%%%%%%%%%%%%%%%%%%%%%%%%%%%%%%%%%%%%%%%%%%%%%%%%%%%
%%%%%%%%%%%%%%%%%%%%%%%%%%%%%%%%%%%%%%%%%%%%%%%%%%%%%%%%%%%%%%%%%%%%%%%%%%%%%%%%%%%%%%%%%%%
%%%%%%%%%%%%%%%%%%%%%%%%%%%%%%%%%%%%%%%%%%%%%%%%%%%%%%%%%%%%%%%%%%%%%%%%%%%%%%%%%%%%%%%%%%%
%%%%%%%%%%%%%%%%%%%%%%%%%%%% STARFIT %%%%%%%%%%%%%%%%%%%%%%%%%%%%
%%%%%%%%%%%%%%%%%%%%%%%%%%%%%%%%%%%%%%%%%%%%%%%%%%%%%%%%%%%%%%%%%%%%%%%%%%%%%%%%%%%%%%%%%%%
%%%%%%%%%%%%%%%%%%%%%%%%%%%%%%%%%%%%%%%%%%%%%%%%%%%%%%%%%%%%%%%%%%%%%%%%%%%%%%%%%%%%%%%%%%%
%%%%%%%%%%%%%%%%%%%%%%%%%%%%%%%%%%%%%%%%%%%%%%%%%%%%%%%%%%%%%%%%%%%%%%%%%%%%%%%%%%%%%%%%%%%

\section{Stellar and AGN light decomposition, photometry and auxiliary data}
\label{sec:decompose}

We decompose the stellar and AGN light in the near-IR by two complementary methods:
\begin{enumerate}
	\item Using the radial profile of the stellar CO absorption feature and integrating in a nuclear aperture (Section \ref{sec:starfit})
	\item By fitting a stellar template and a blackbody to the spectrum extracted in the same nuclear aperture (Section \ref{sec:gox_fit})
\end{enumerate}

\subsection{Radial decomposition: The CO equivalent width radial profile}
\label{sec:starfit}

First we smooth the cubes by a simple 5x5 pixel boxcar filter in order to improve the spectral signal to noise and increase the significance of the fit. This comes at the cost of spatial resolution, but since we are averaging over an aperture larger than this at the end, this does not impact our results. We then use the custom-built routine {\tt starfit}\footnote{The routine has already been used for \citet{davies2006} and \citet{hicks2013}.} to fit a stellar template to the CO(2,0) stellar absorption feature at 2.29 \um{} at every spatial pixel of the smoothed cubes to obtain the stellar kinematics. We used a SINFONI spectrum of the M1 star HD~176617. The specific star does not matter, however, since we not only normalize the continuum level, but subtract it, so that the continuum is effectively set to zero and the EW of the CO feature has much less impact on the fit than in other methods; see \citet{engel2010b} for a detailed discussion on this. With the known velocity field, we determine the equivalent width (EW) of this CO feature in the standard wavelength range \citep[e.g.][]{origlia1993} of [2.2924,2.2977] \um{}. In AGNs this feature is diluted in the nucleus due to the strong non-stellar continuum and we use this effect to measure the AGN contribution.

\begin{figure*}
\subfloat{\includegraphics[trim=2cm 0.5cm 1.5cm 0.5cm, width=0.2\hsize]{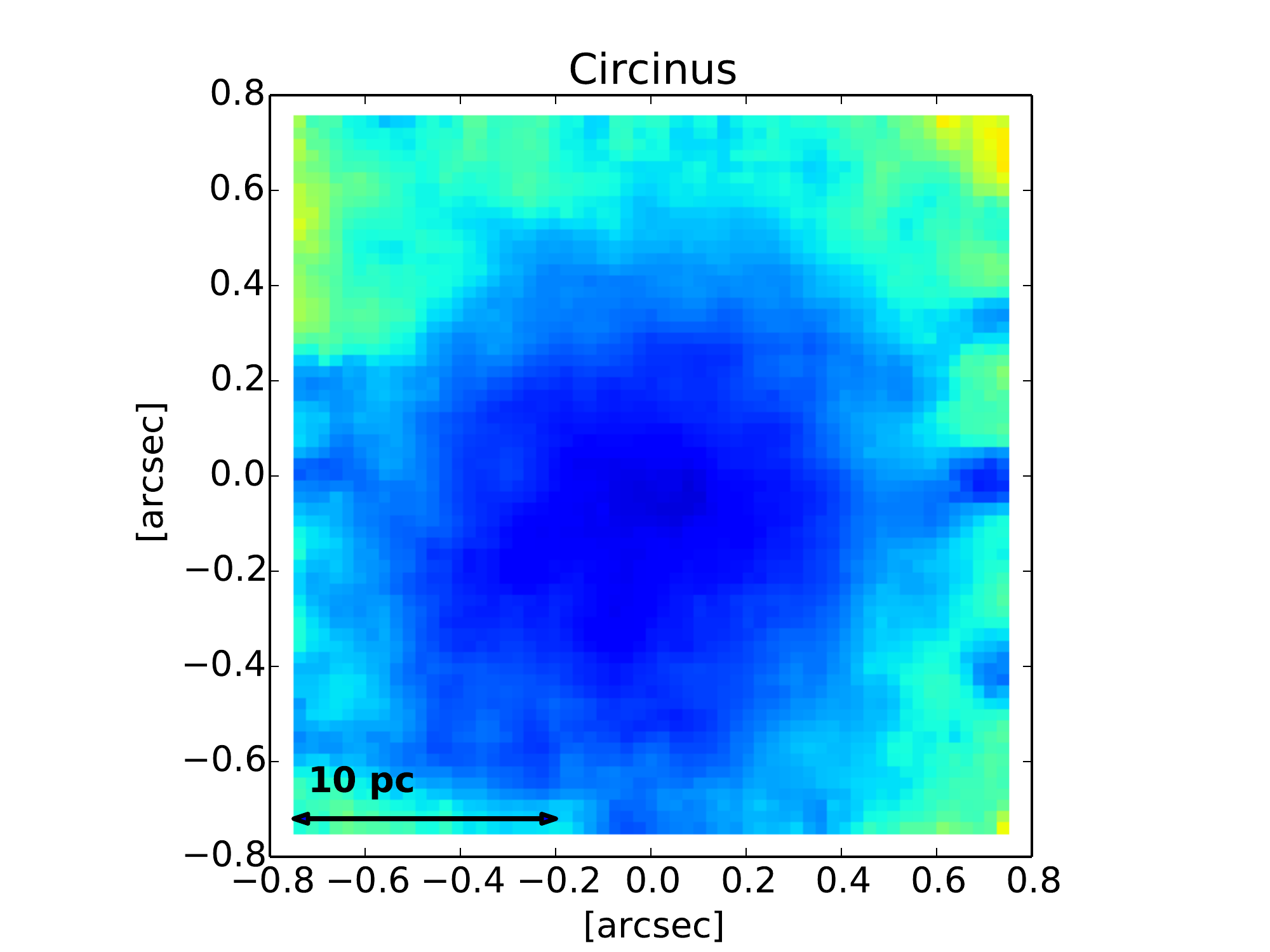}}
\subfloat{\includegraphics[trim=2cm 0.5cm 1.5cm 0.5cm, width=0.2\hsize]{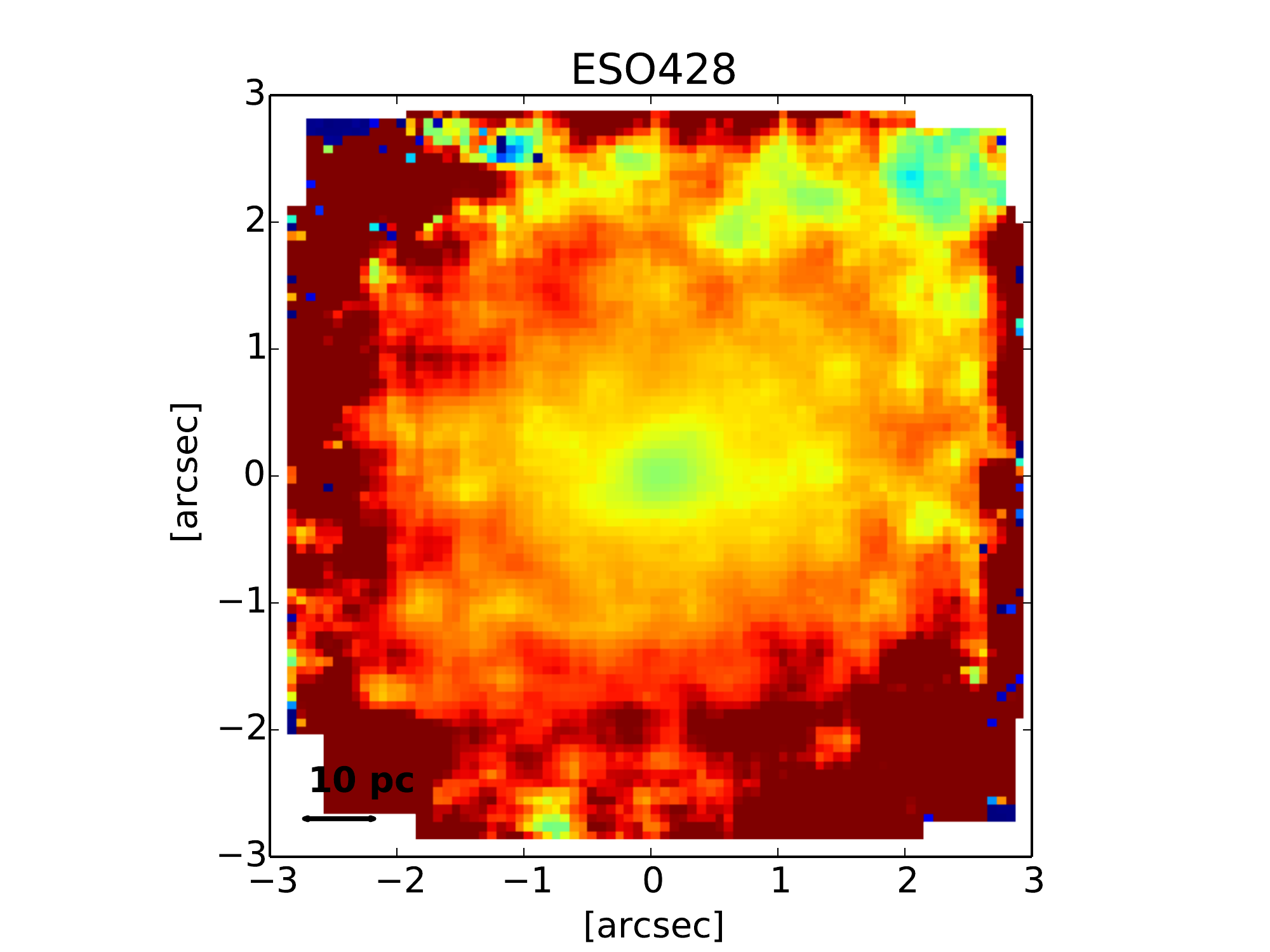}}
\subfloat{\includegraphics[trim=2cm 0.5cm 1.5cm 0.5cm, width=0.2\hsize]{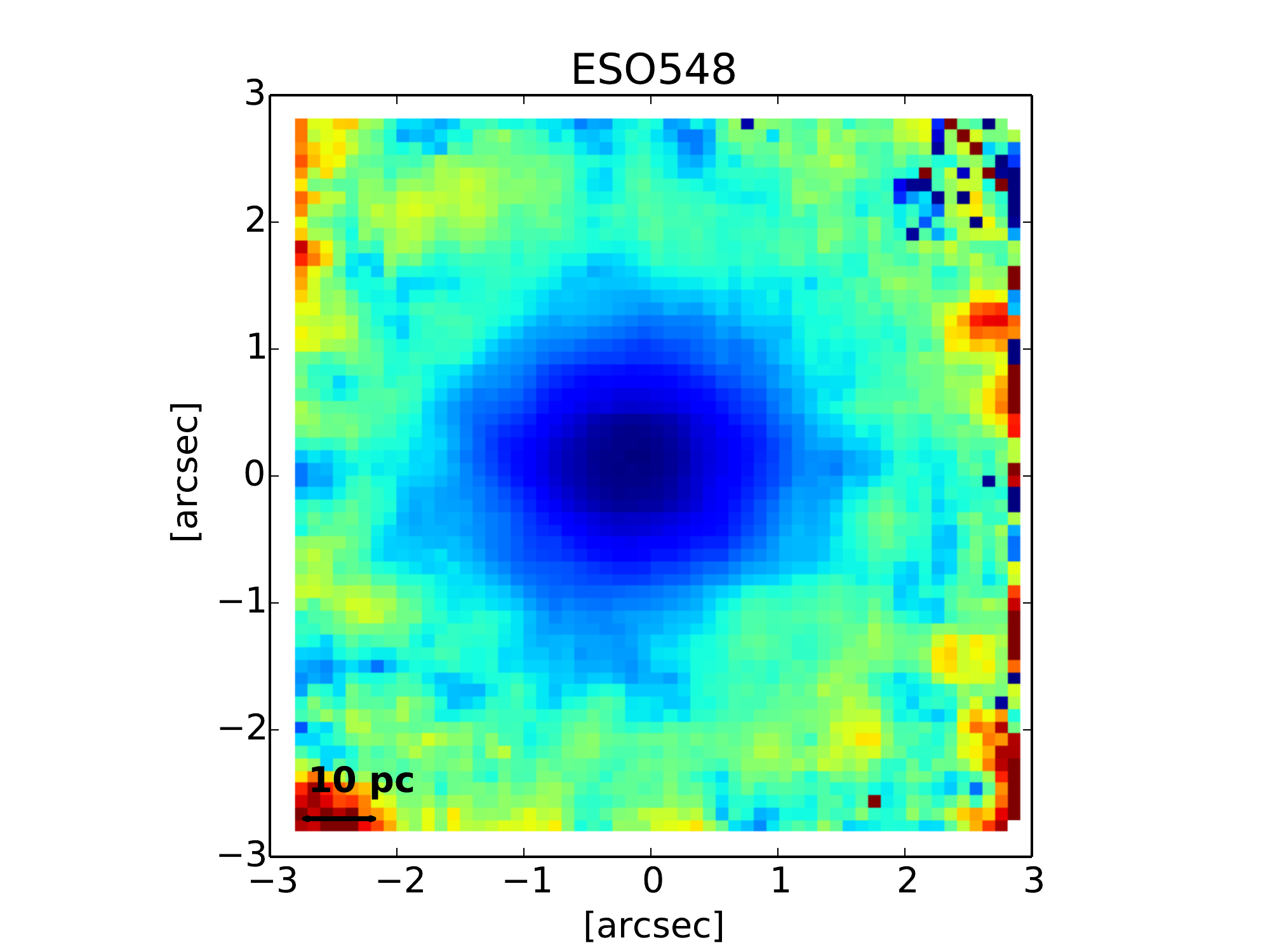}}
\subfloat{\includegraphics[trim=2cm 0.5cm 1.5cm 0.5cm, width=0.2\hsize]{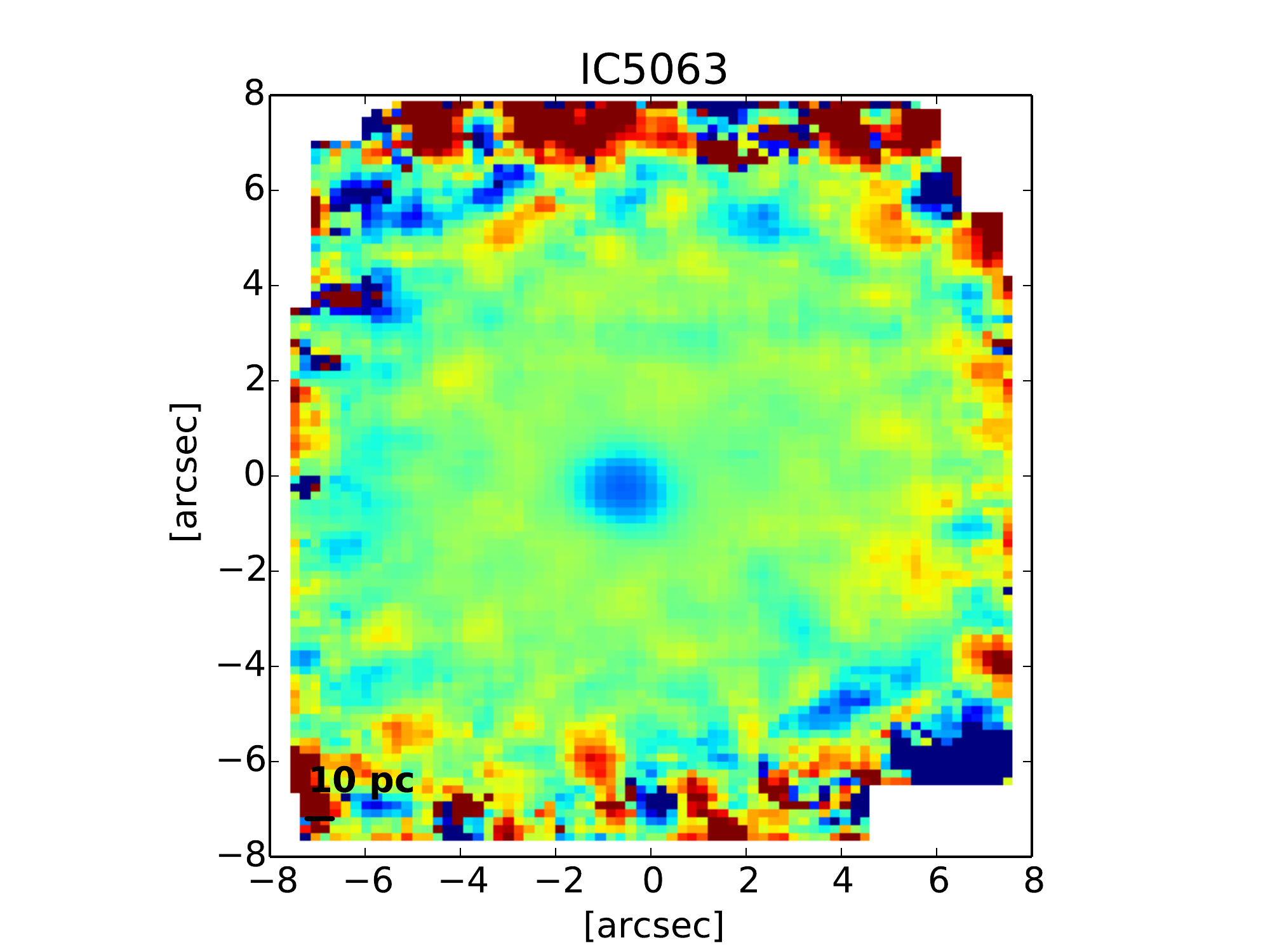}}
\subfloat{\includegraphics[trim=2cm 0.5cm 1.5cm 0.5cm, width=0.2\hsize]{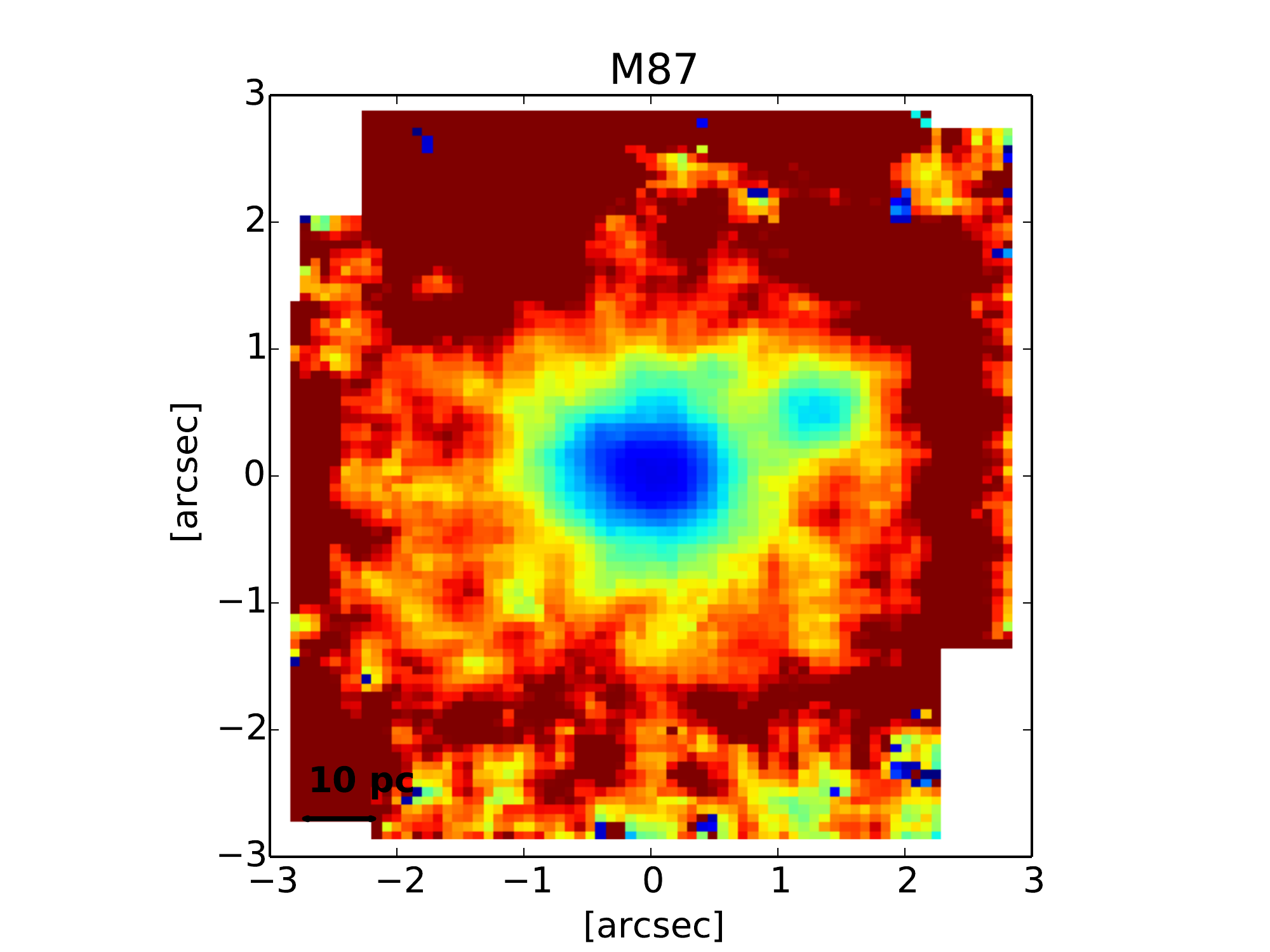}}\\
\subfloat{\includegraphics[trim=2cm 0.5cm 1.5cm 0.5cm, width=0.2\hsize]{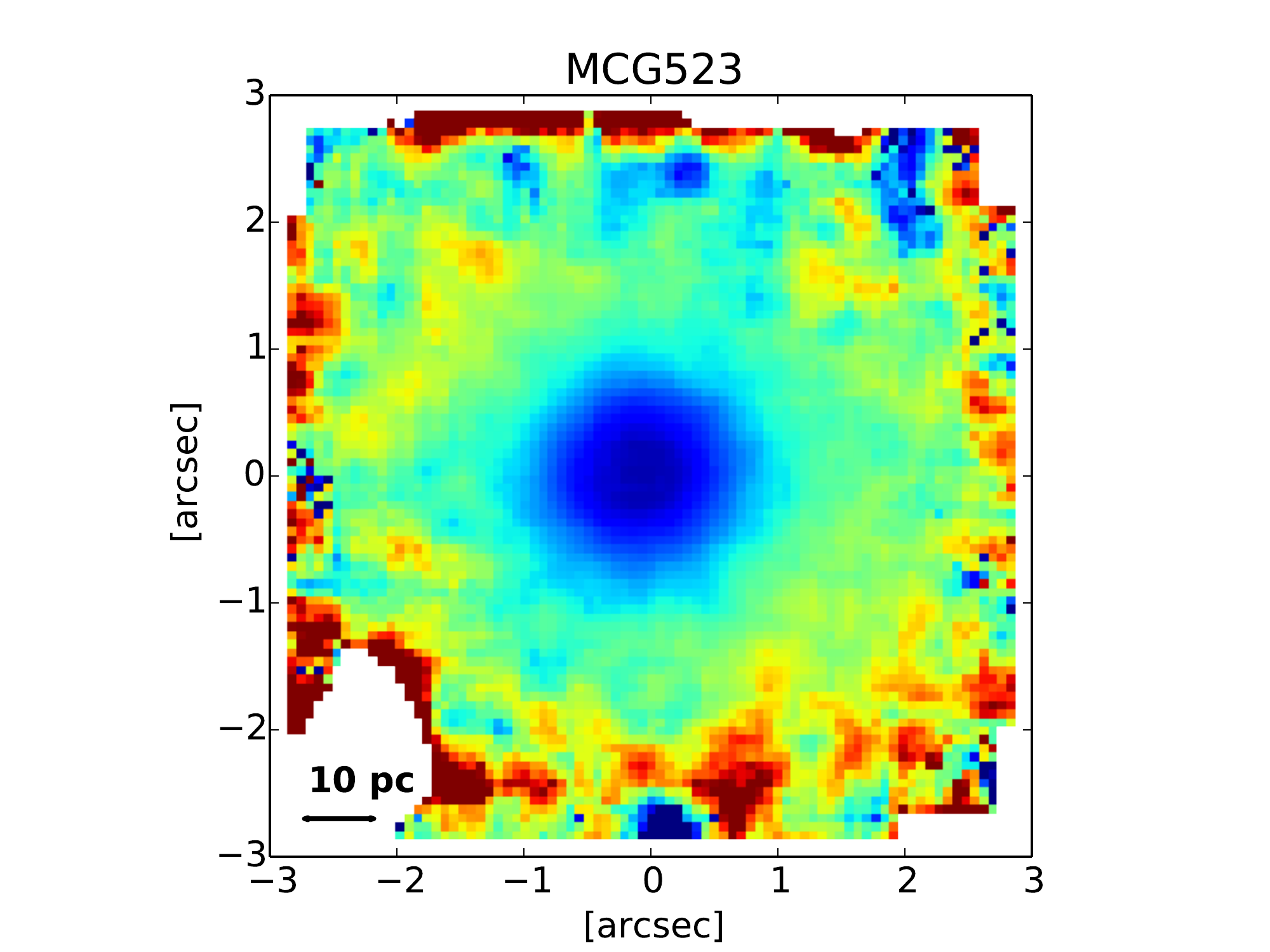}}
\subfloat{\includegraphics[trim=2cm 0.5cm 1.5cm 0.5cm, width=0.2\hsize]{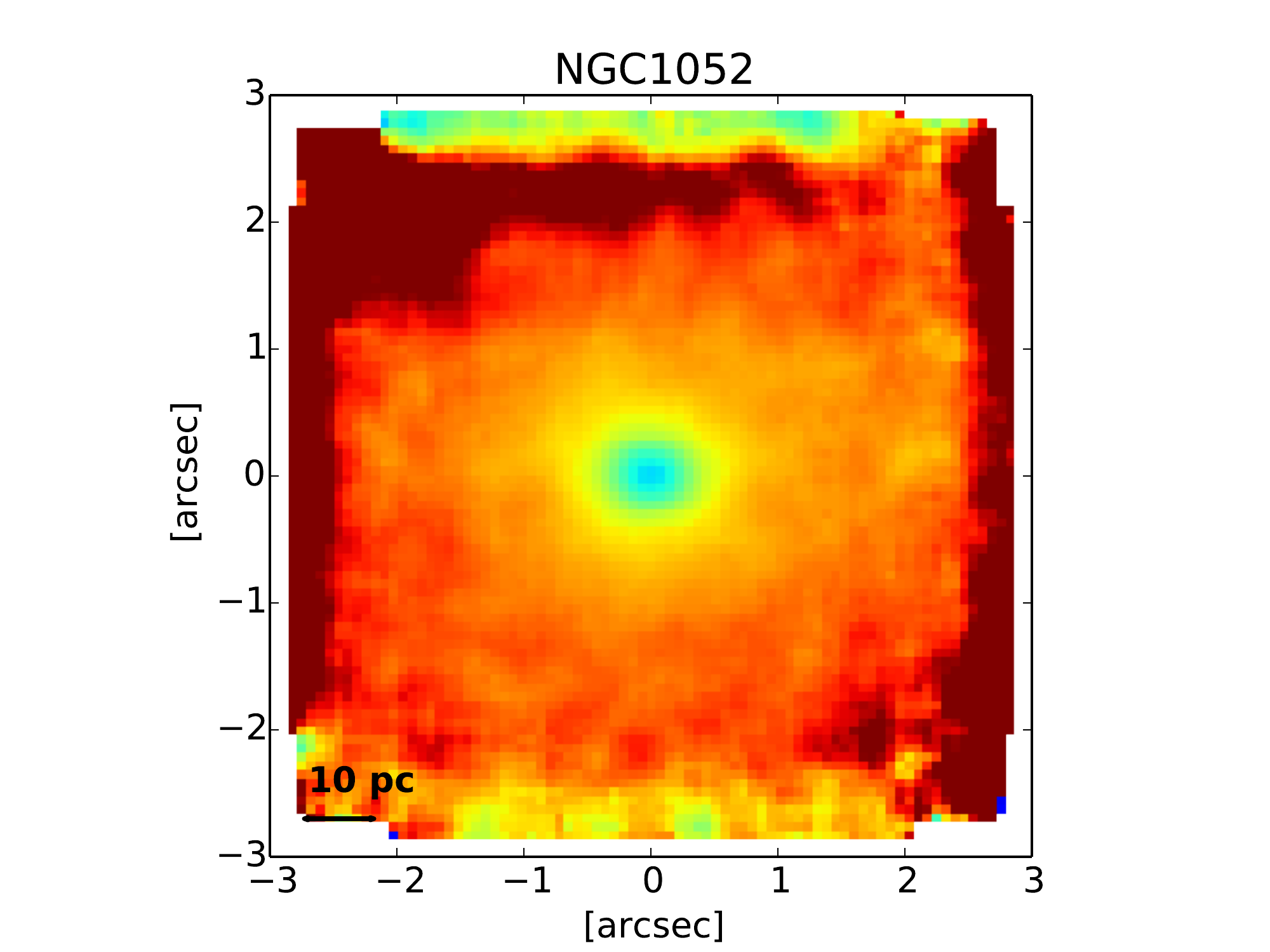}}
\subfloat{\includegraphics[trim=2cm 0.5cm 1.5cm 0.5cm, width=0.2\hsize]{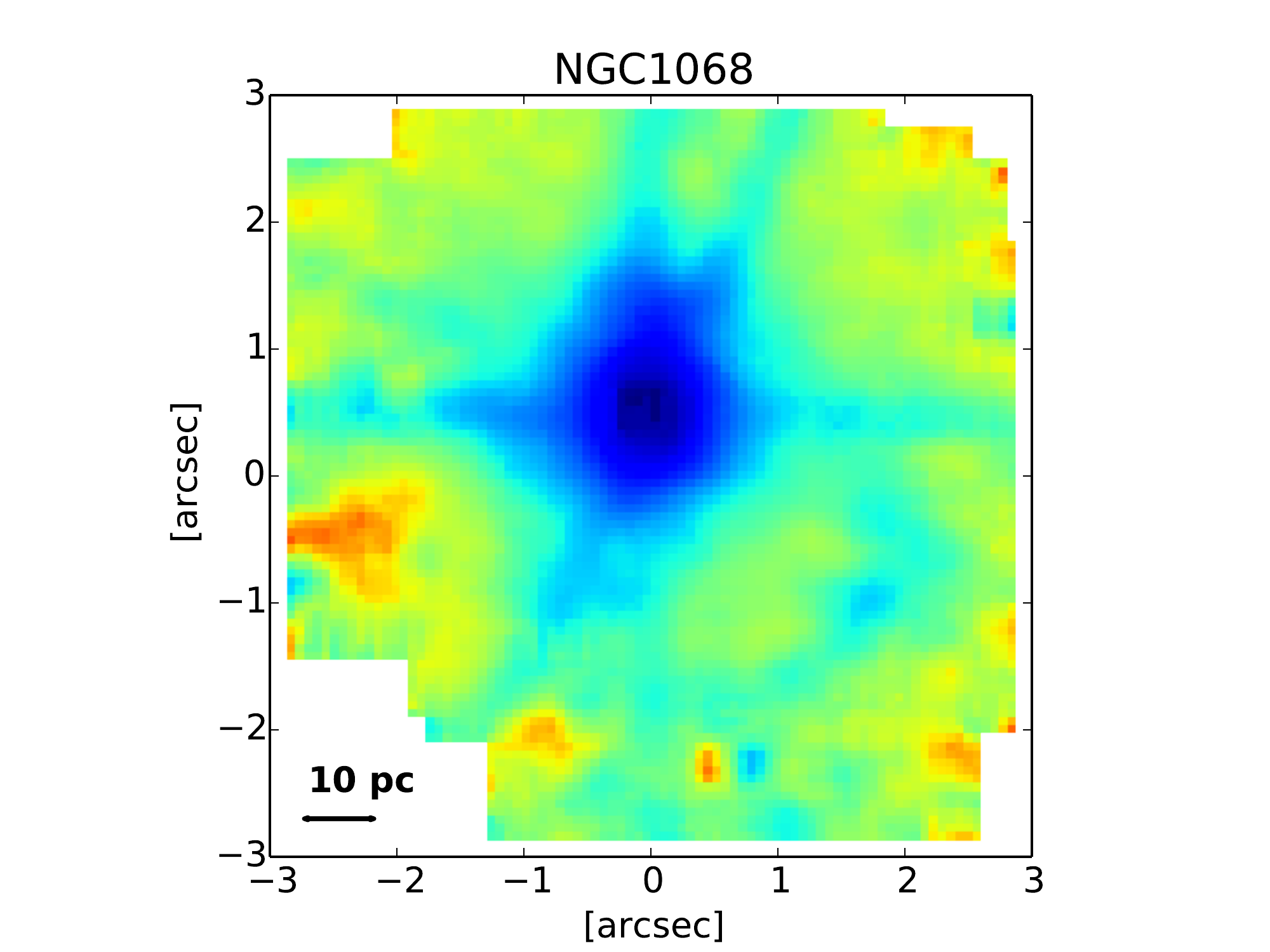}}
\subfloat{\includegraphics[trim=2cm 0.5cm 1.5cm 0.5cm, width=0.2\hsize]{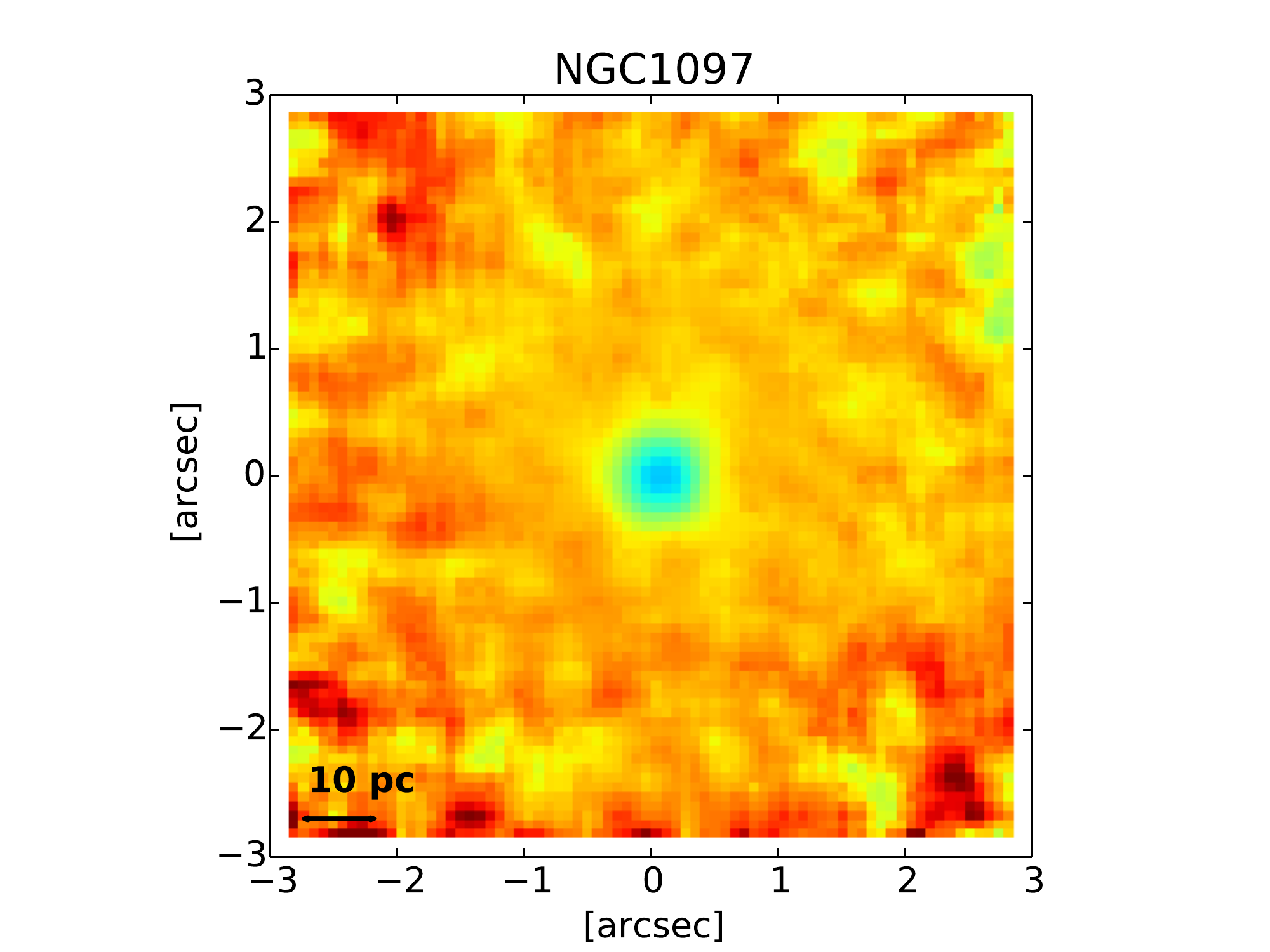}}
\subfloat{\includegraphics[trim=2cm 0.5cm 1.5cm 0.5cm, width=0.2\hsize]{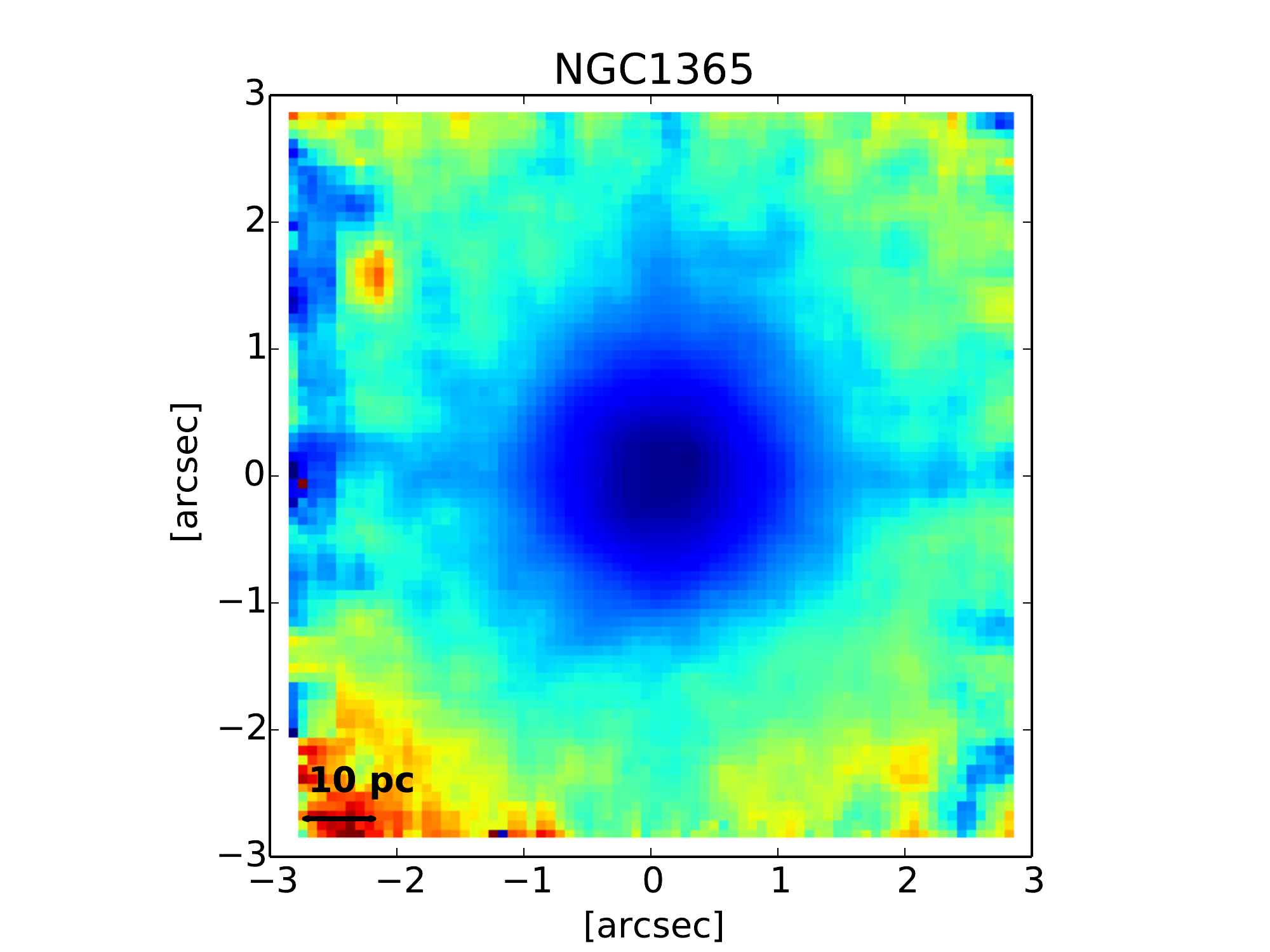}}\\
\subfloat{\includegraphics[trim=2cm 0.5cm 1.5cm 0.5cm, width=0.2\hsize]{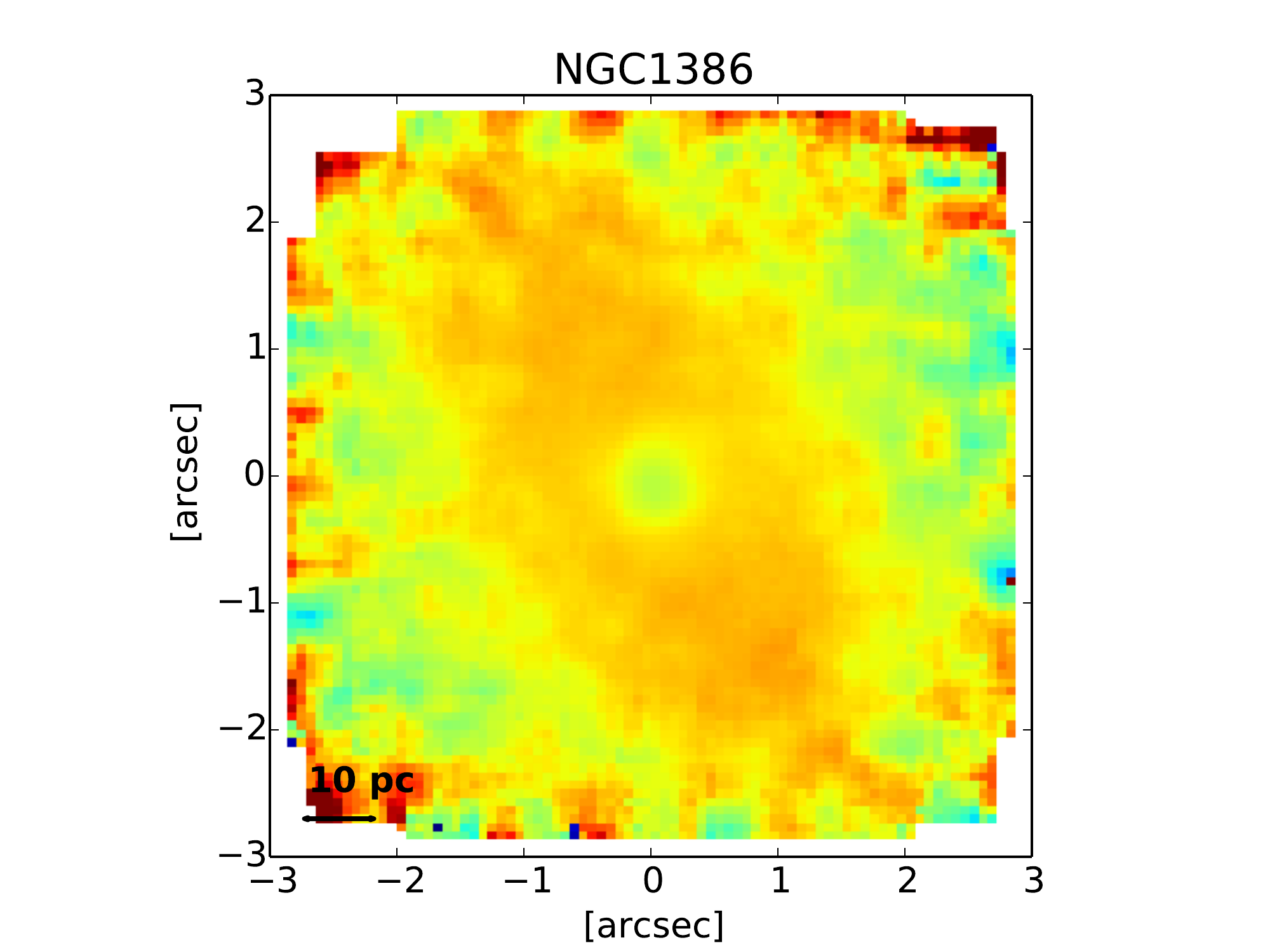}}
\subfloat{\includegraphics[trim=2cm 0.5cm 1.5cm 0.5cm, width=0.2\hsize]{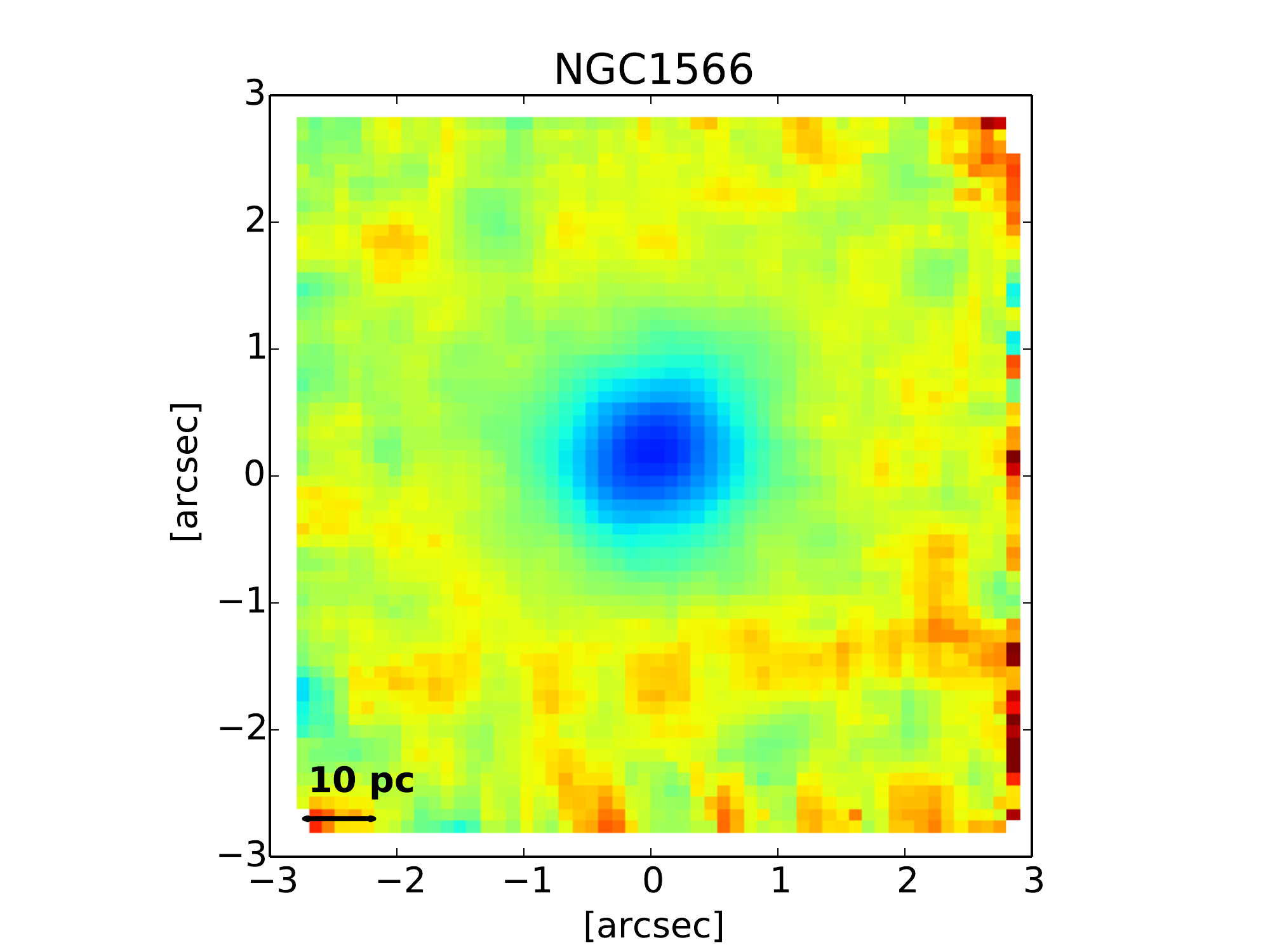}}
\subfloat{\includegraphics[trim=2cm 0.5cm 1.5cm 0.5cm, width=0.2\hsize]{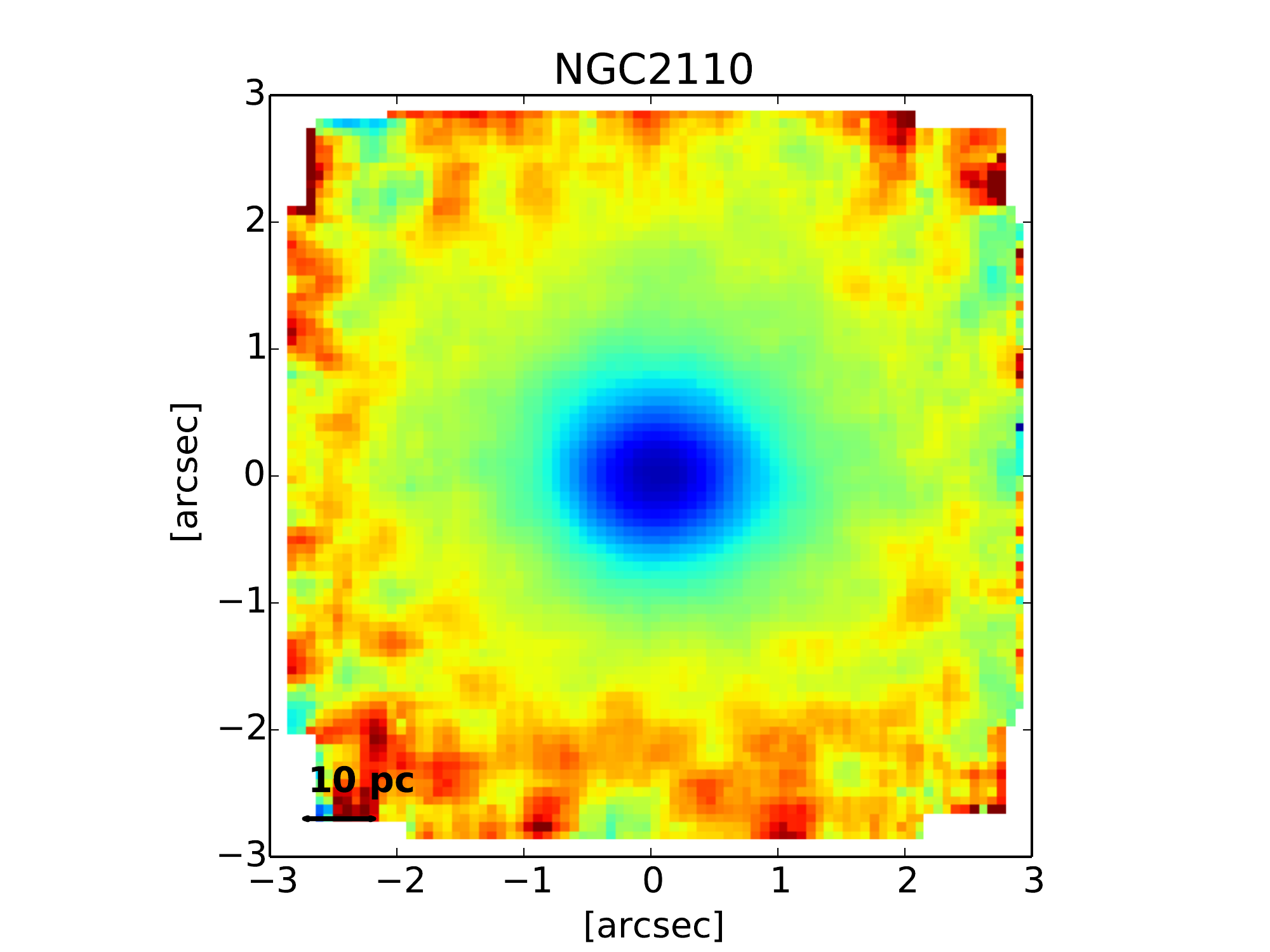}}
\subfloat{\includegraphics[trim=2cm 0.5cm 1.5cm 0.5cm, width=0.2\hsize]{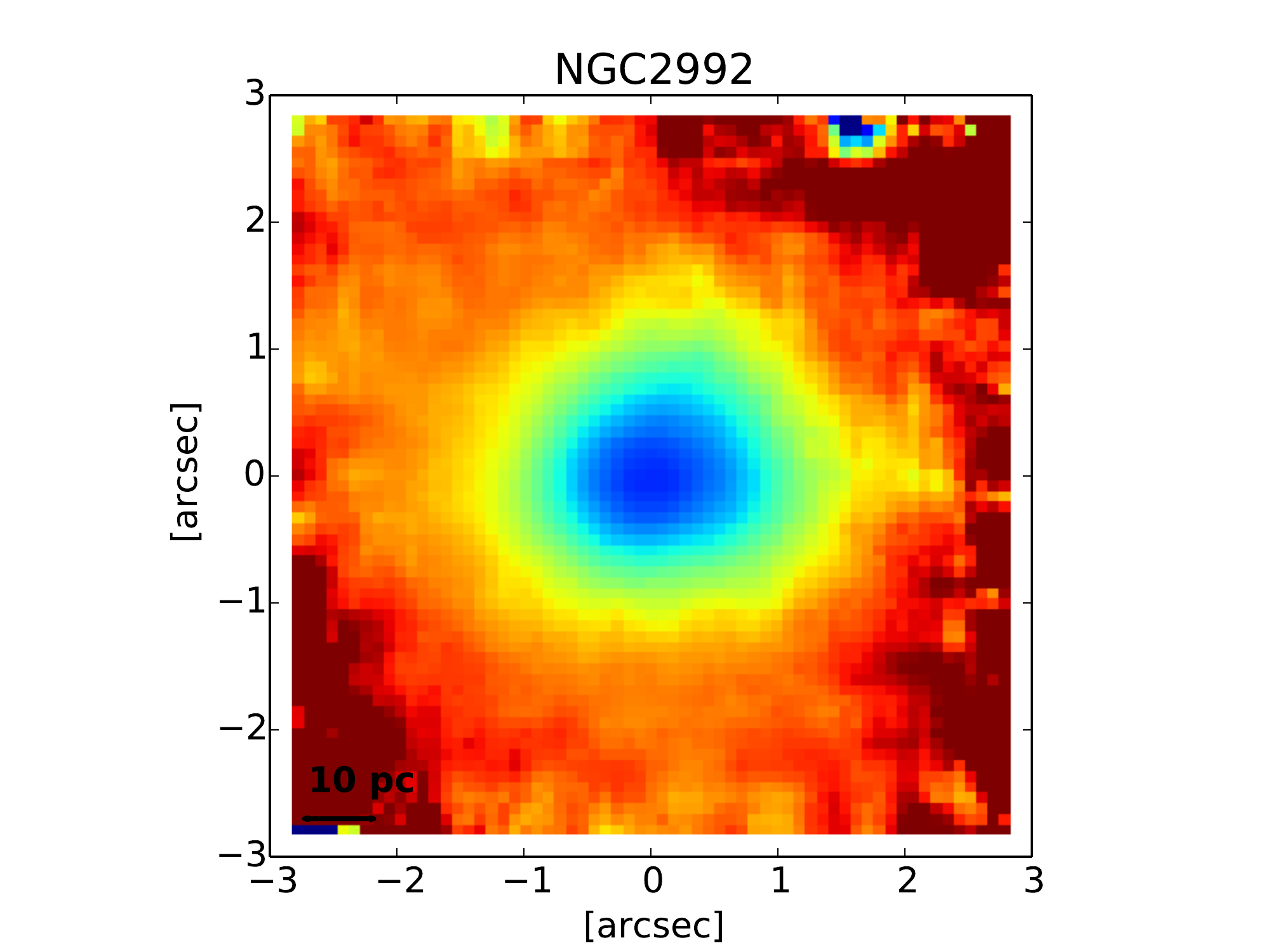}}
\subfloat{\includegraphics[trim=2cm 0.5cm 1.5cm 0.5cm, width=0.2\hsize]{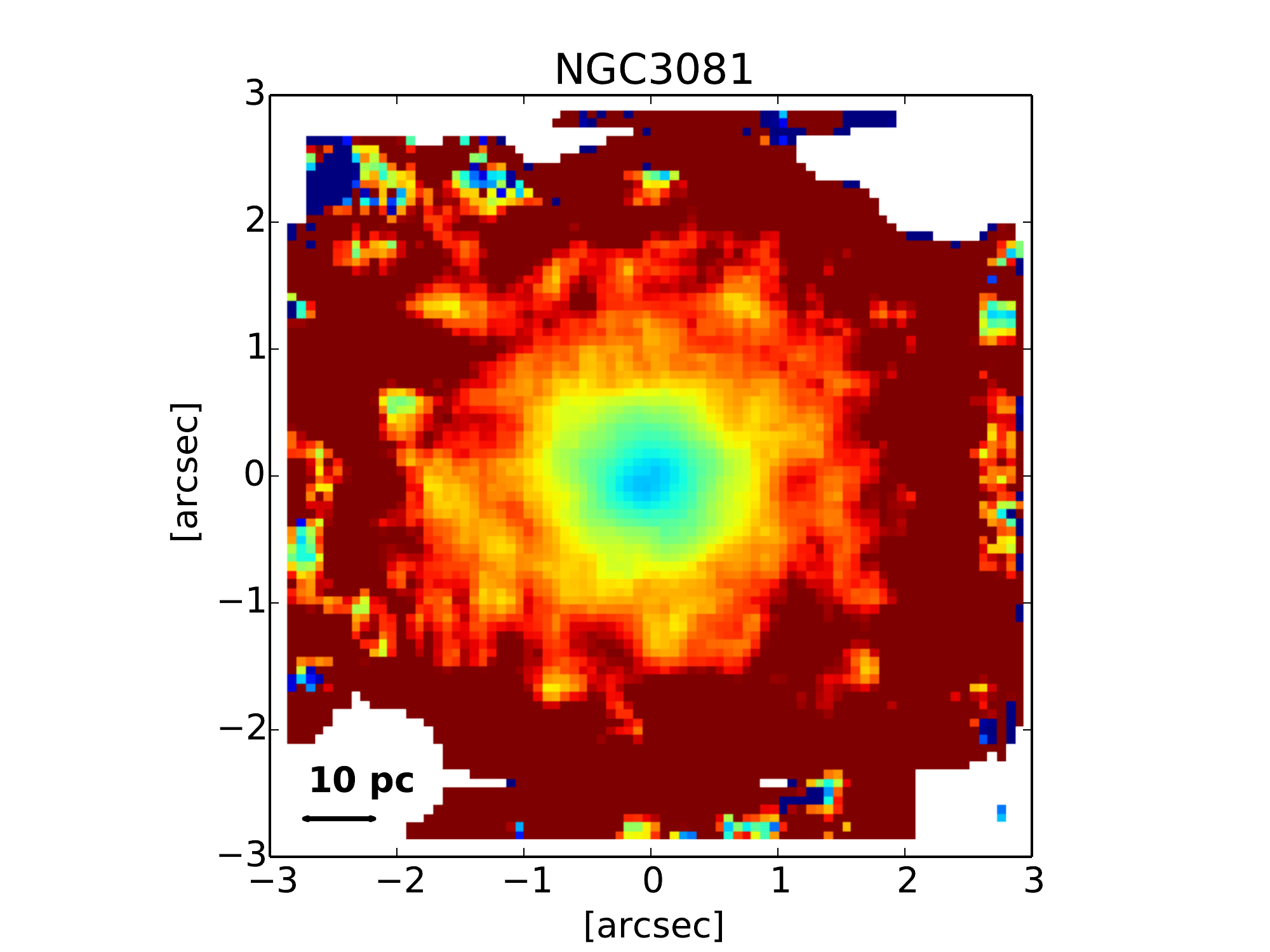}}\\
\subfloat{\includegraphics[trim=2cm 0.5cm 1.5cm 0.5cm, width=0.2\hsize]{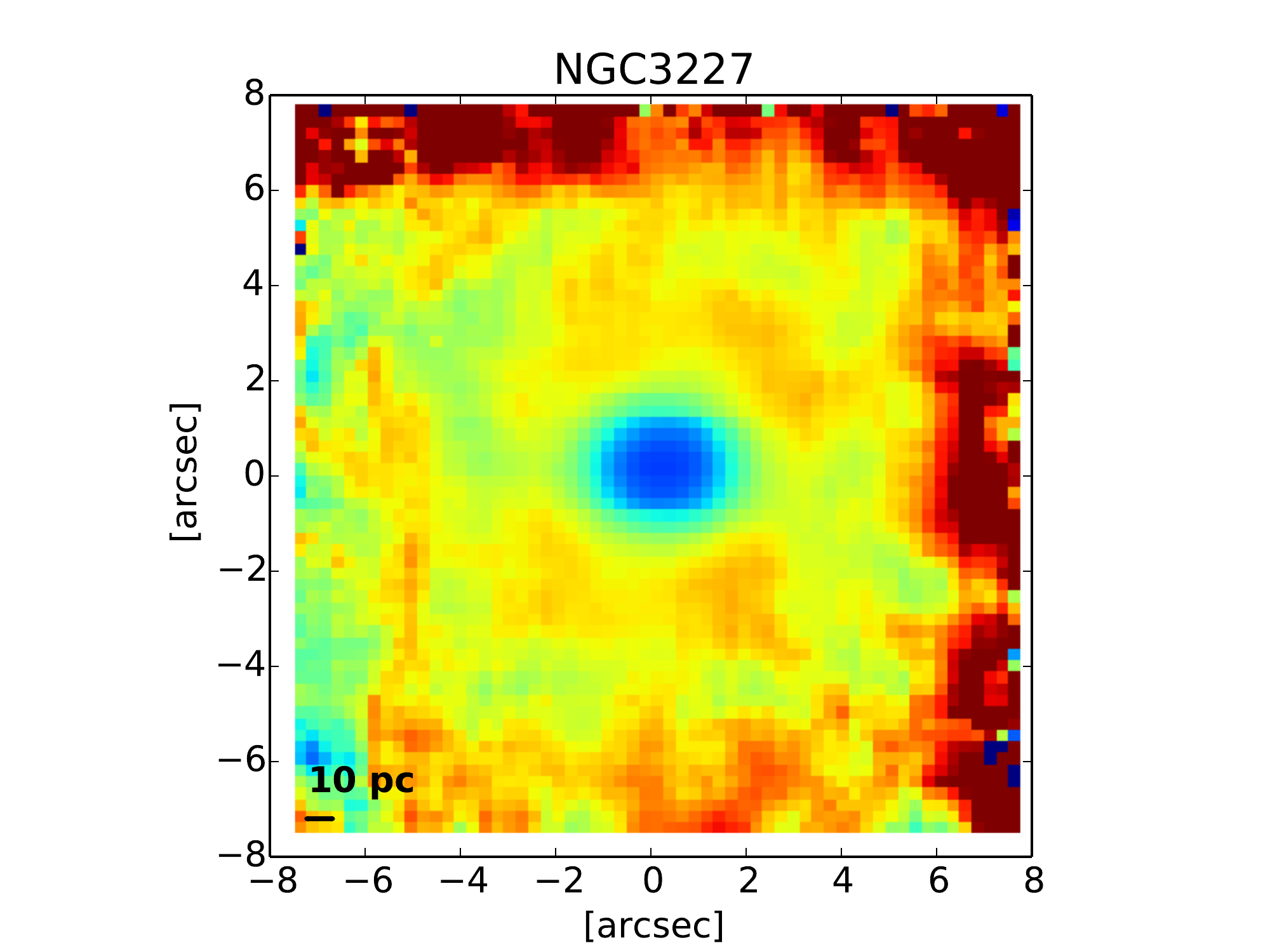}}
\subfloat{\includegraphics[trim=2cm 0.5cm 1.5cm 0.5cm, width=0.2\hsize]{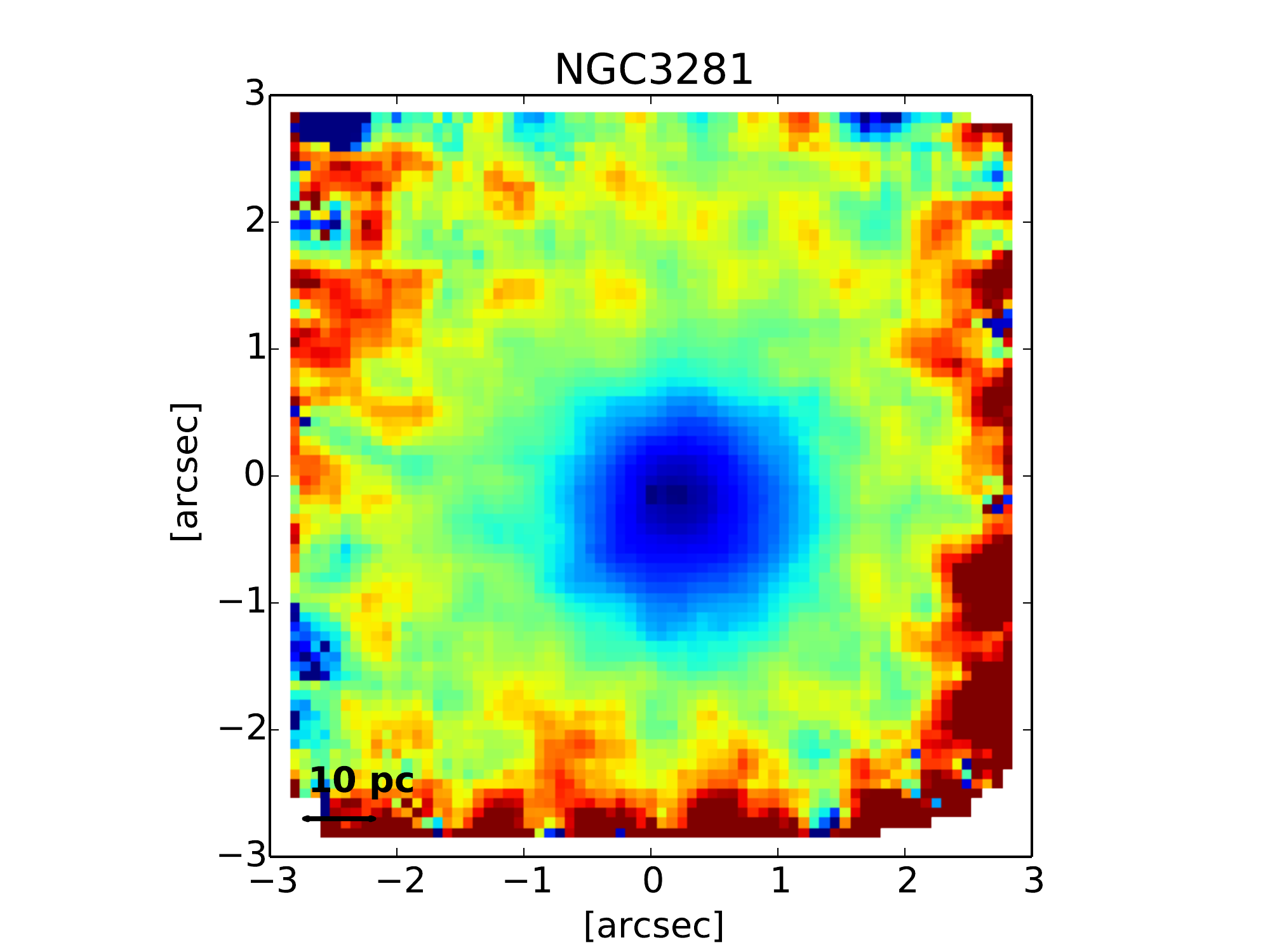}}
\subfloat{\includegraphics[trim=2cm 0.5cm 1.5cm 0.5cm, width=0.2\hsize]{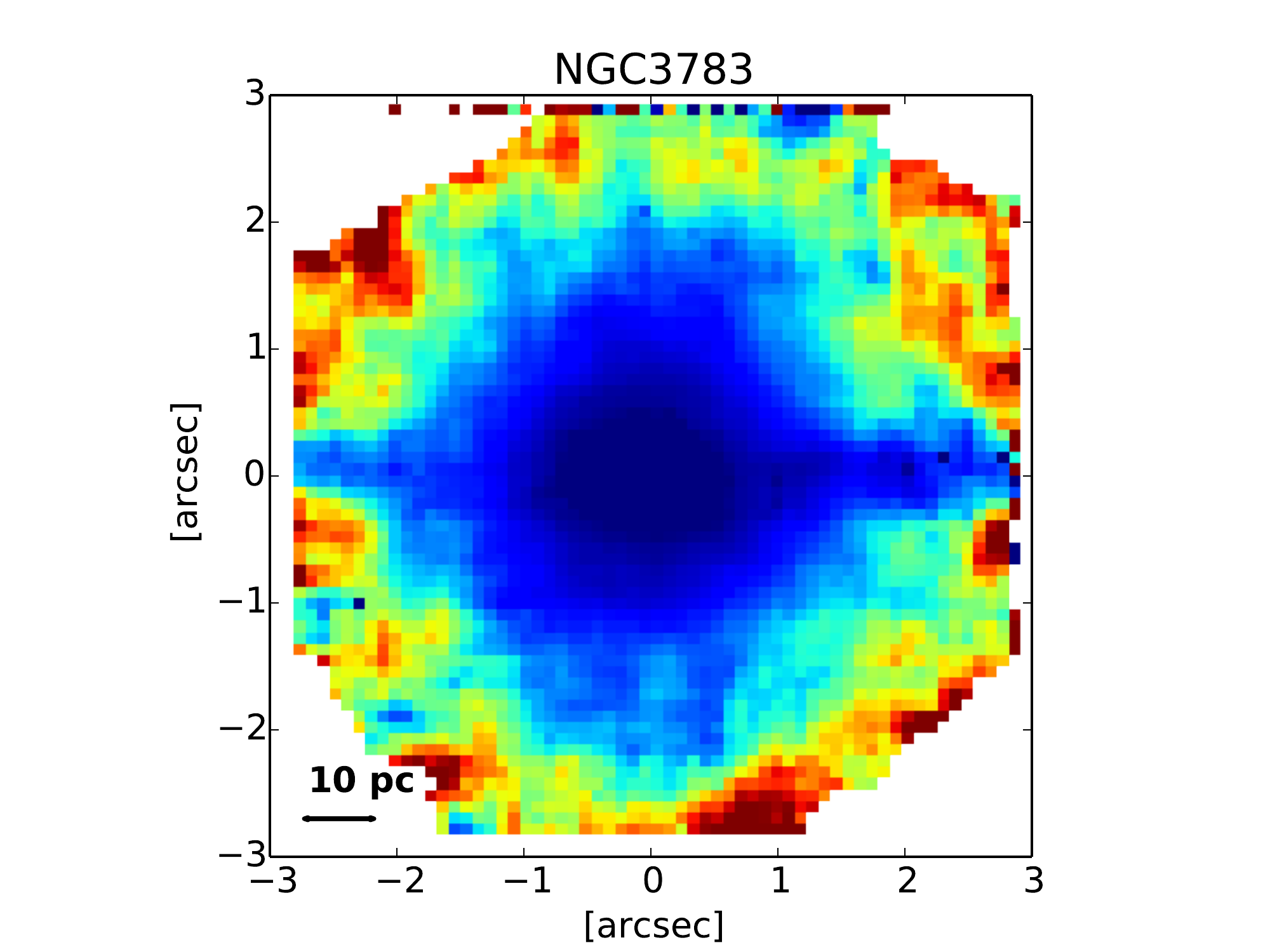}}
\subfloat{\includegraphics[trim=2cm 0.5cm 1.5cm 0.5cm, width=0.2\hsize]{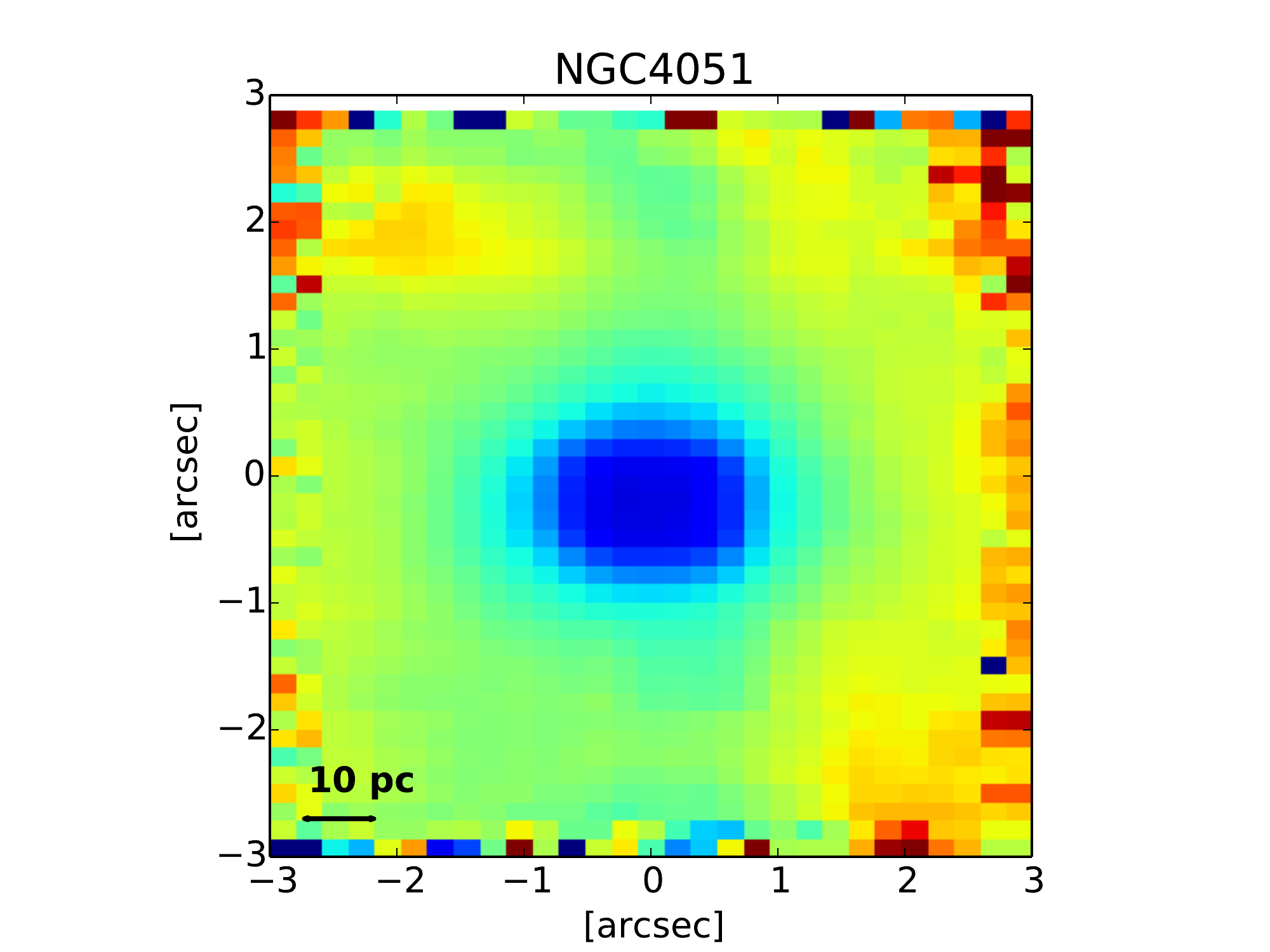}}
\subfloat{\includegraphics[trim=2cm 0.5cm 1.5cm 0.5cm, width=0.2\hsize]{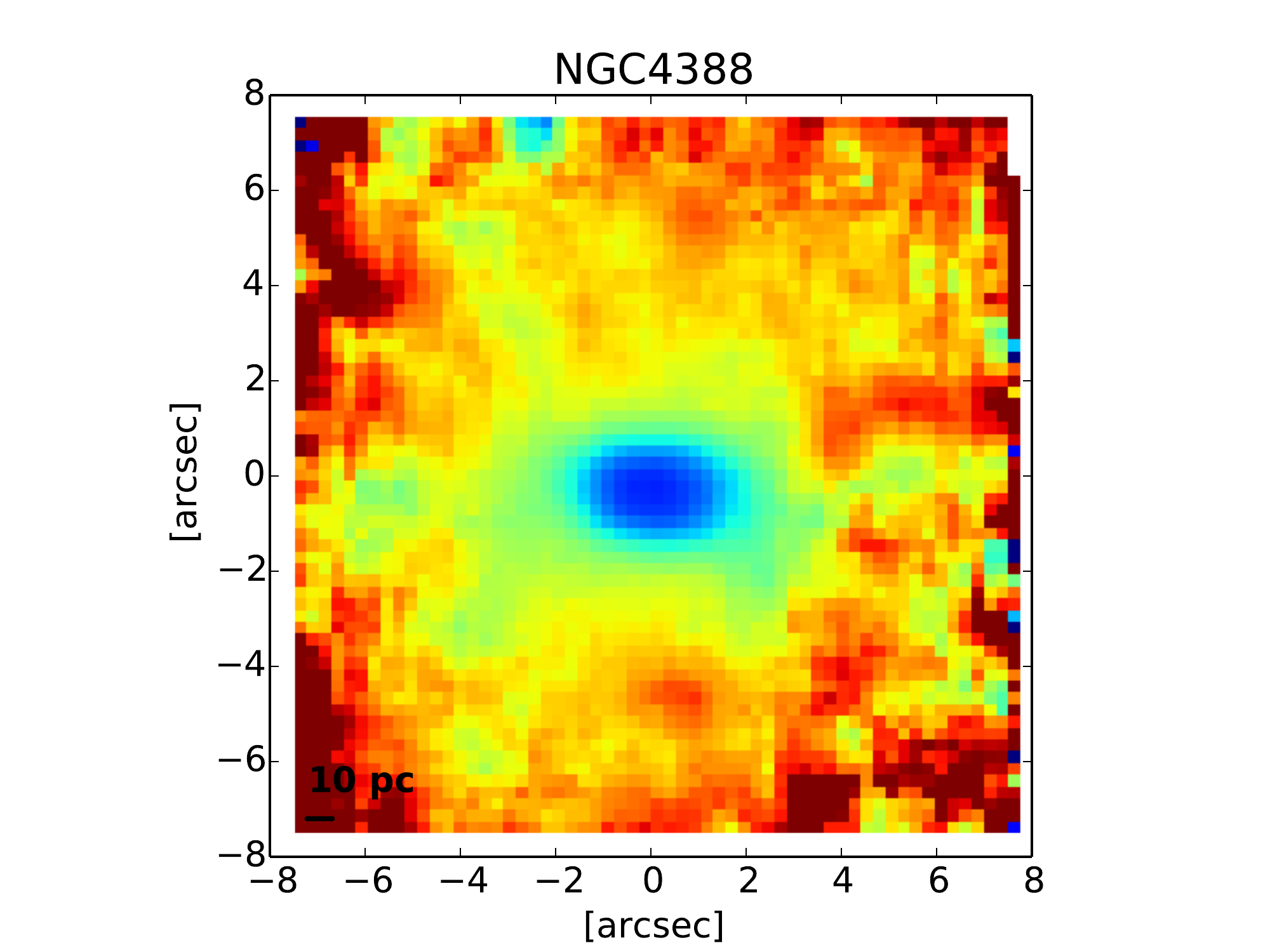}}\\
\subfloat{\includegraphics[trim=2cm 0.5cm 1.5cm 0.5cm, width=0.2\hsize]{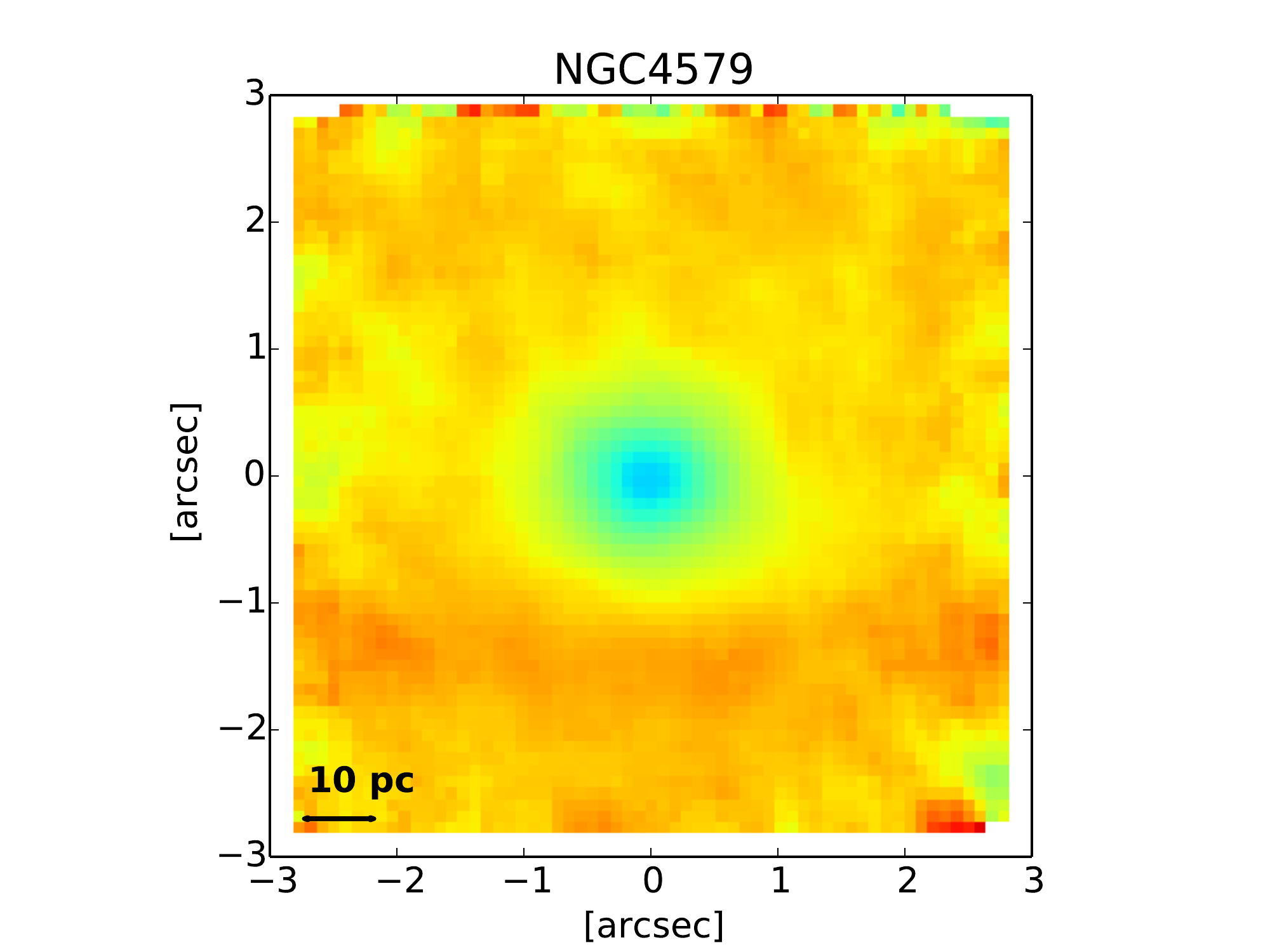}}
\subfloat{\includegraphics[trim=2cm 0.5cm 1.5cm 0.5cm, width=0.2\hsize]{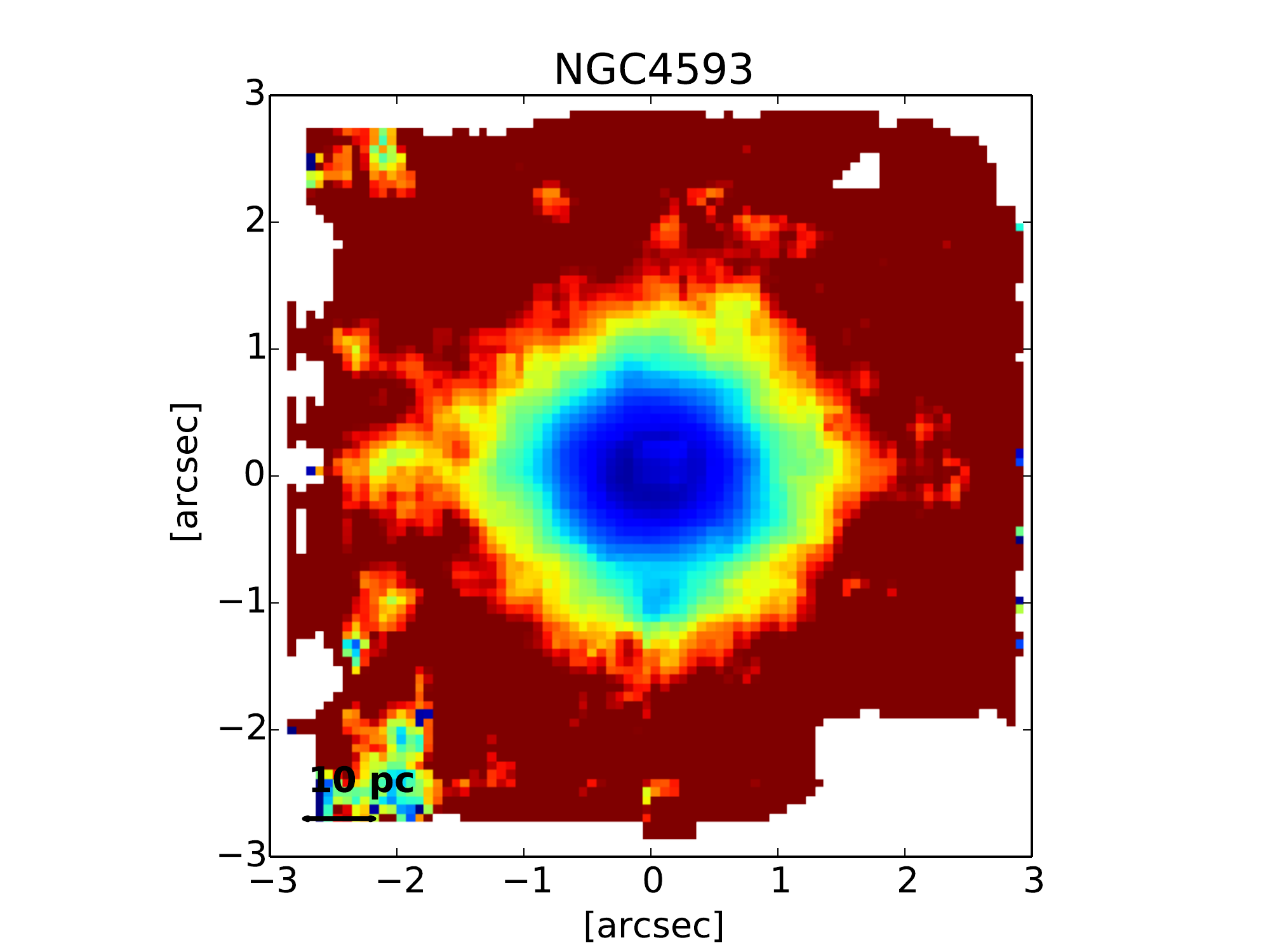}}
\subfloat{\includegraphics[trim=2cm 0.5cm 1.5cm 0.5cm, width=0.2\hsize]{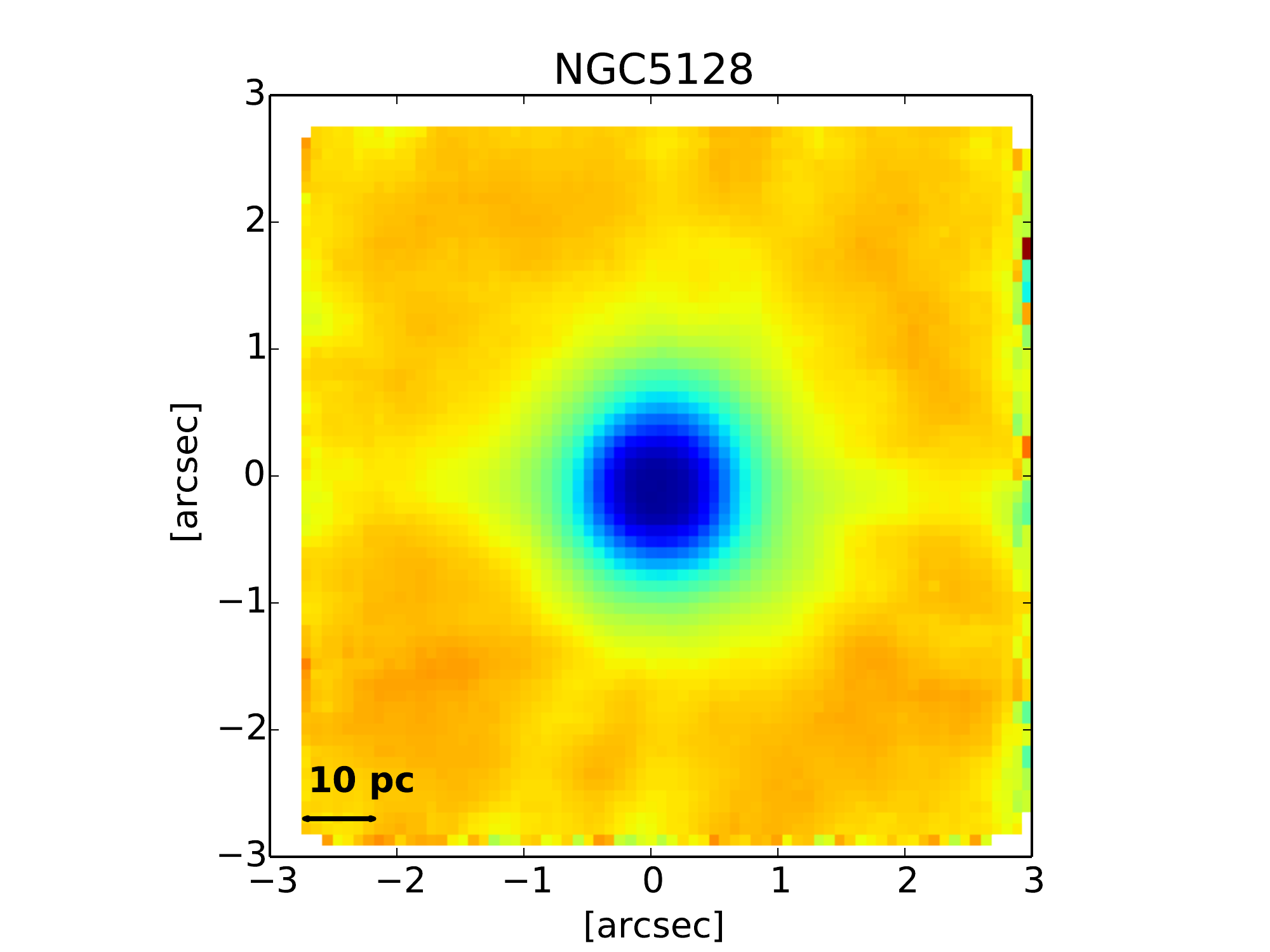}}
\subfloat{\includegraphics[trim=2cm 0.5cm 1.5cm 0.5cm, width=0.2\hsize]{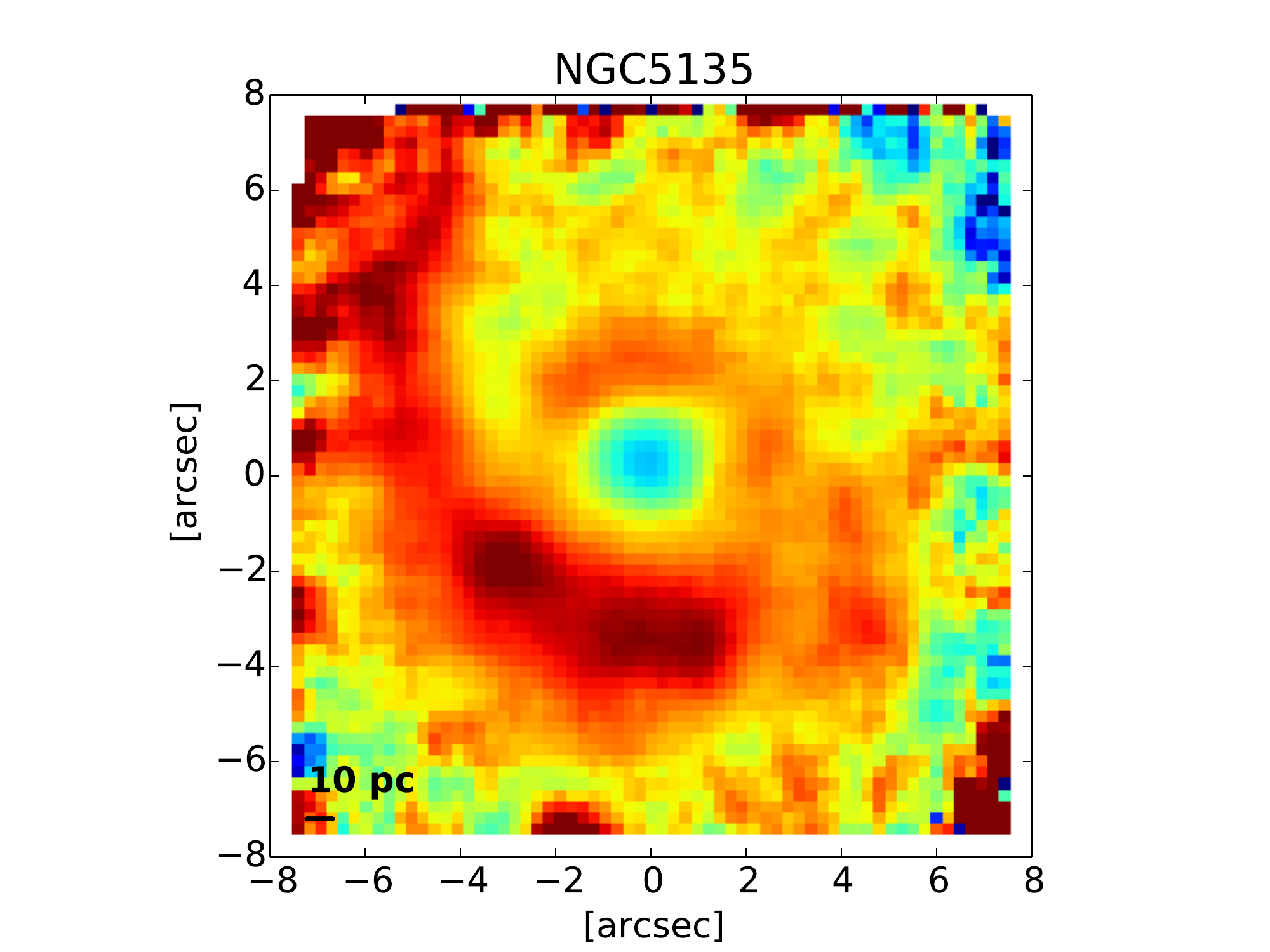}}
\subfloat{\includegraphics[trim=2cm 0.5cm 1.5cm 0.5cm, width=0.2\hsize]{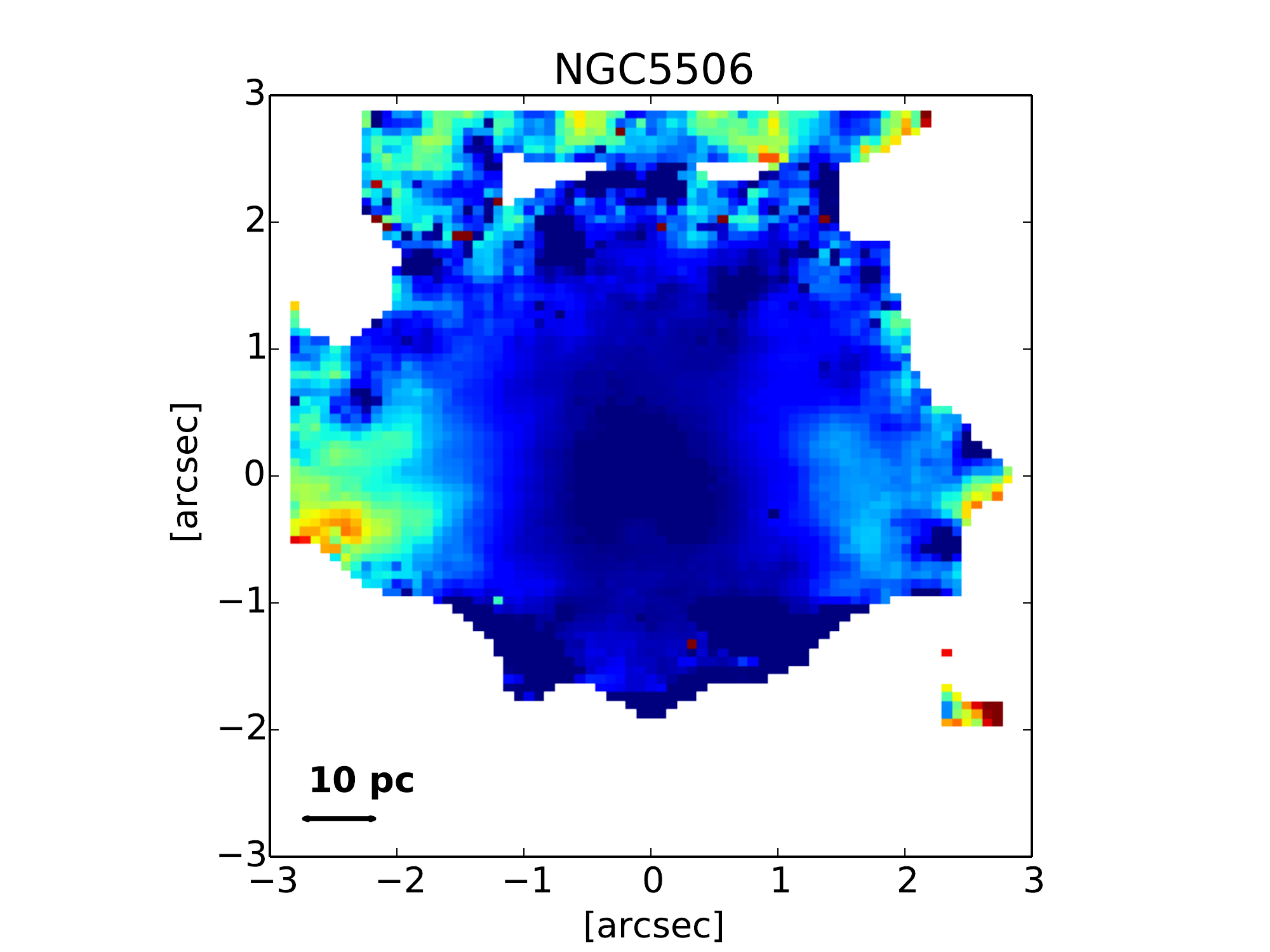}}\\
\subfloat{\includegraphics[trim=2cm 0.5cm 1.5cm 0.5cm, width=0.2\hsize]{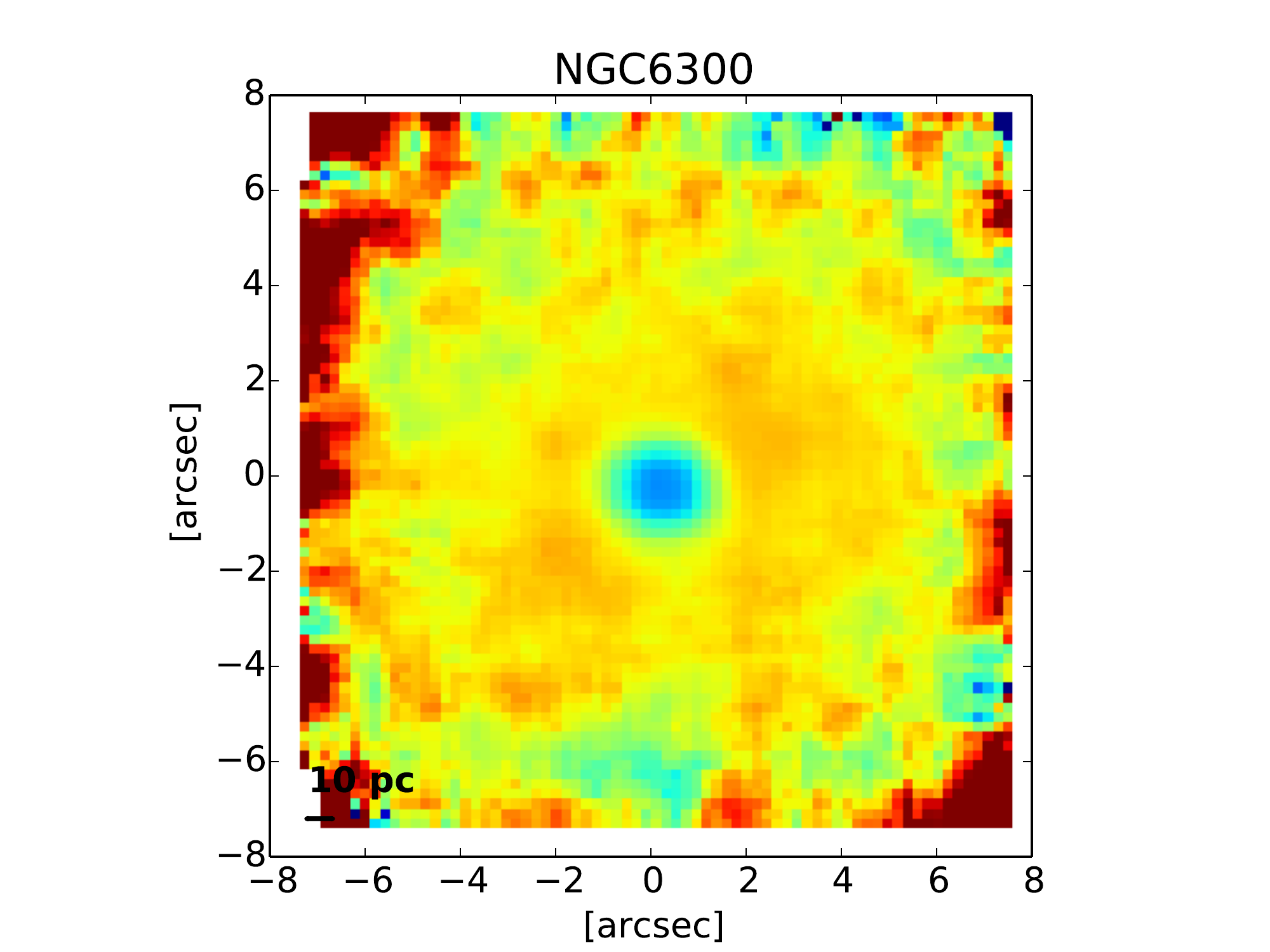}}
\subfloat{\includegraphics[trim=2cm 0.5cm 1.5cm 0.5cm, width=0.2\hsize]{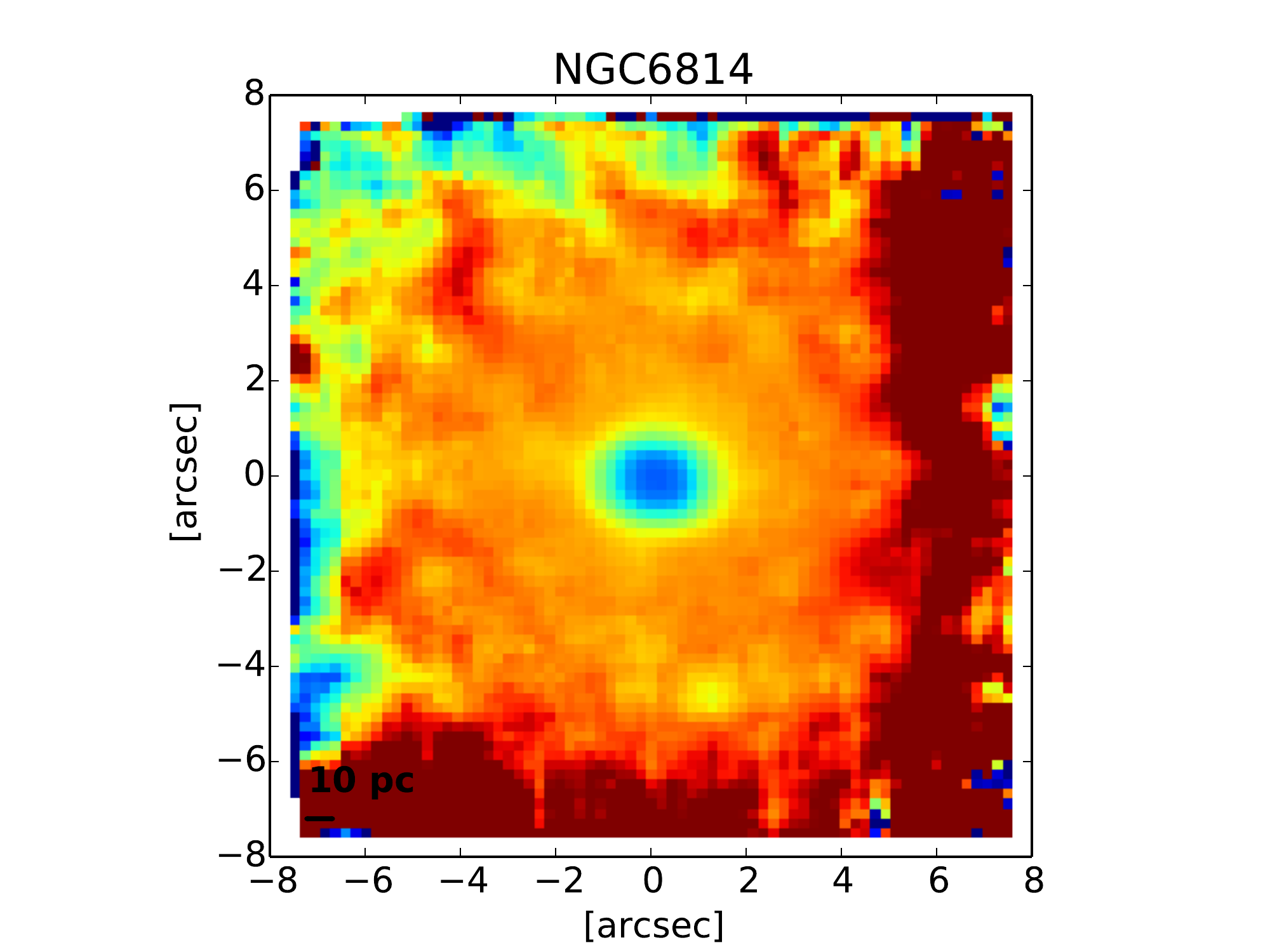}}
\subfloat{\includegraphics[trim=2cm 0.5cm 1.5cm 0.5cm, width=0.2\hsize]{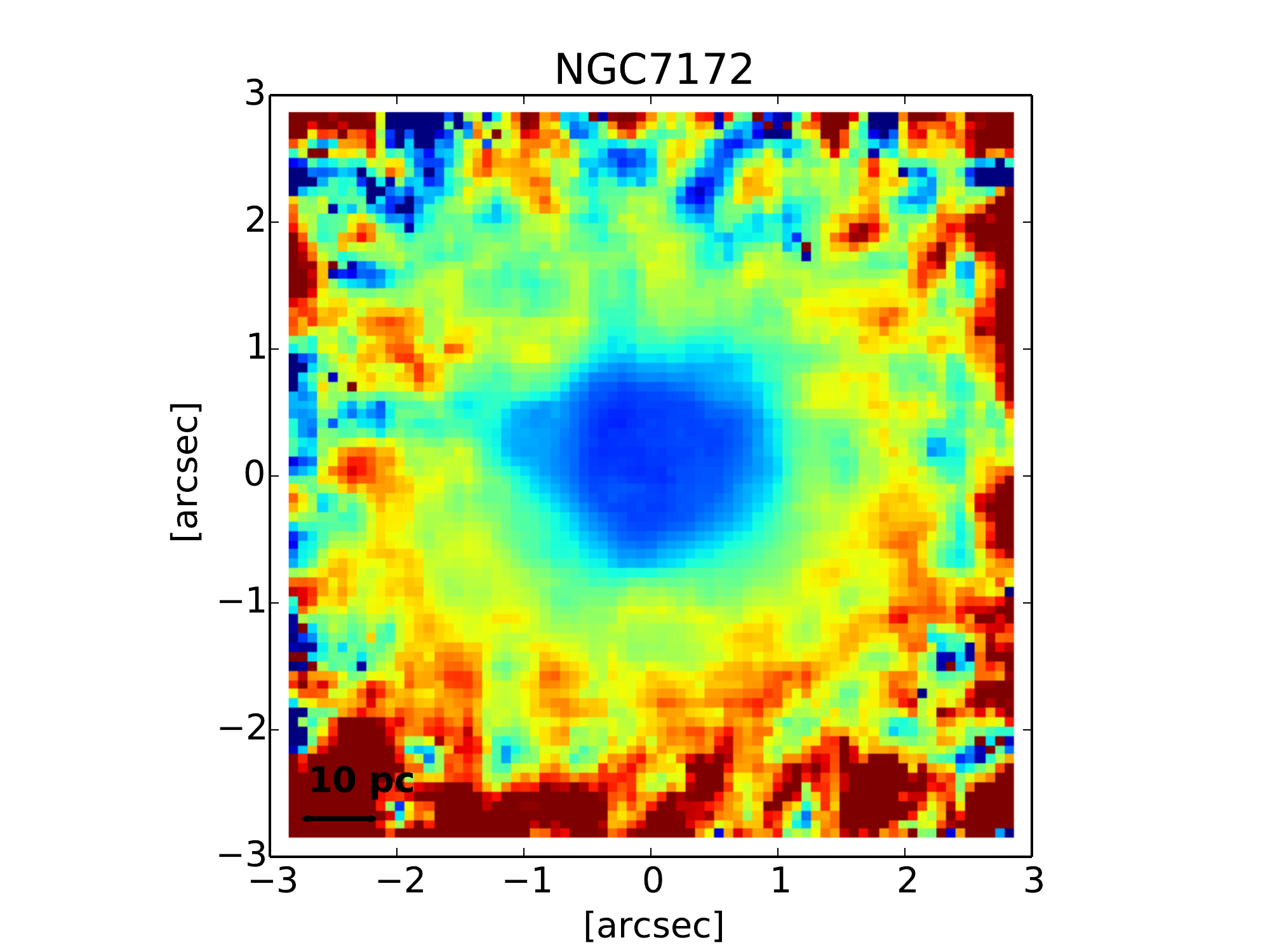}}
\subfloat{\includegraphics[trim=2cm 0.5cm 1.5cm 0.5cm, width=0.2\hsize]{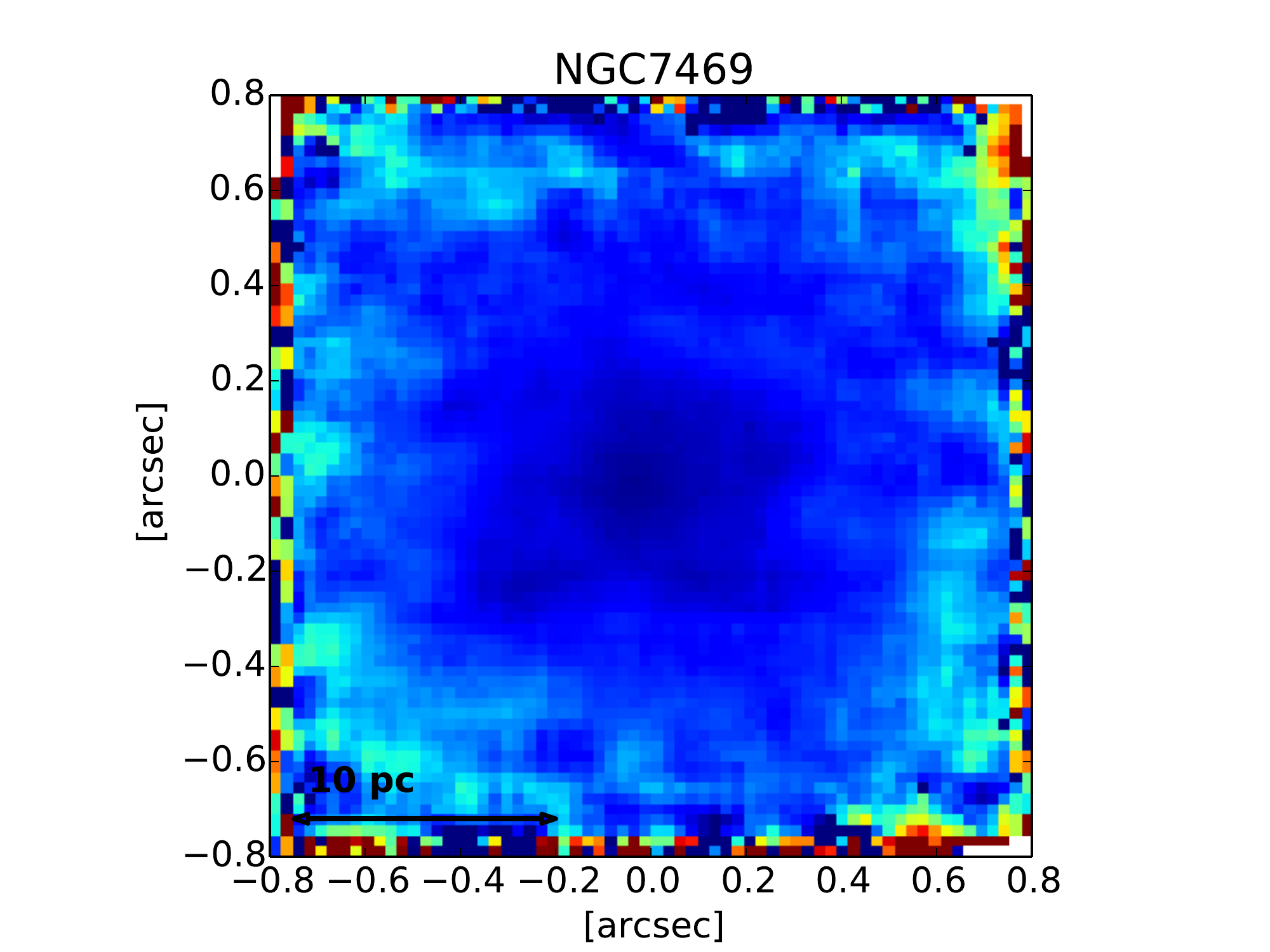}}
\subfloat{\includegraphics[trim=2cm 0.5cm 1.5cm 0.5cm, width=0.2\hsize]{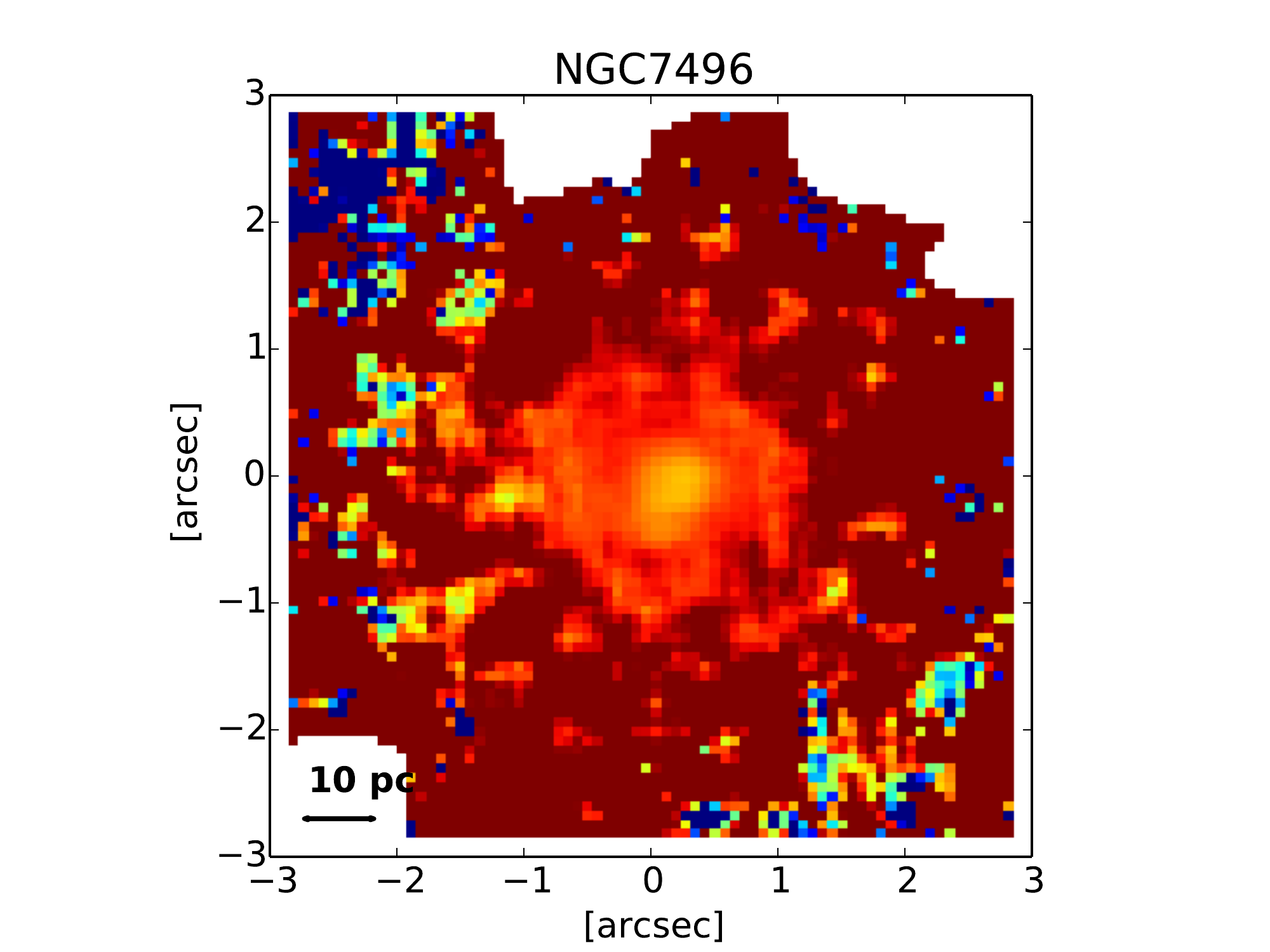}}\\
	\raggedright
\subfloat{\includegraphics[trim=2cm 0.5cm 1.5cm 0.5cm, width=0.2\hsize]{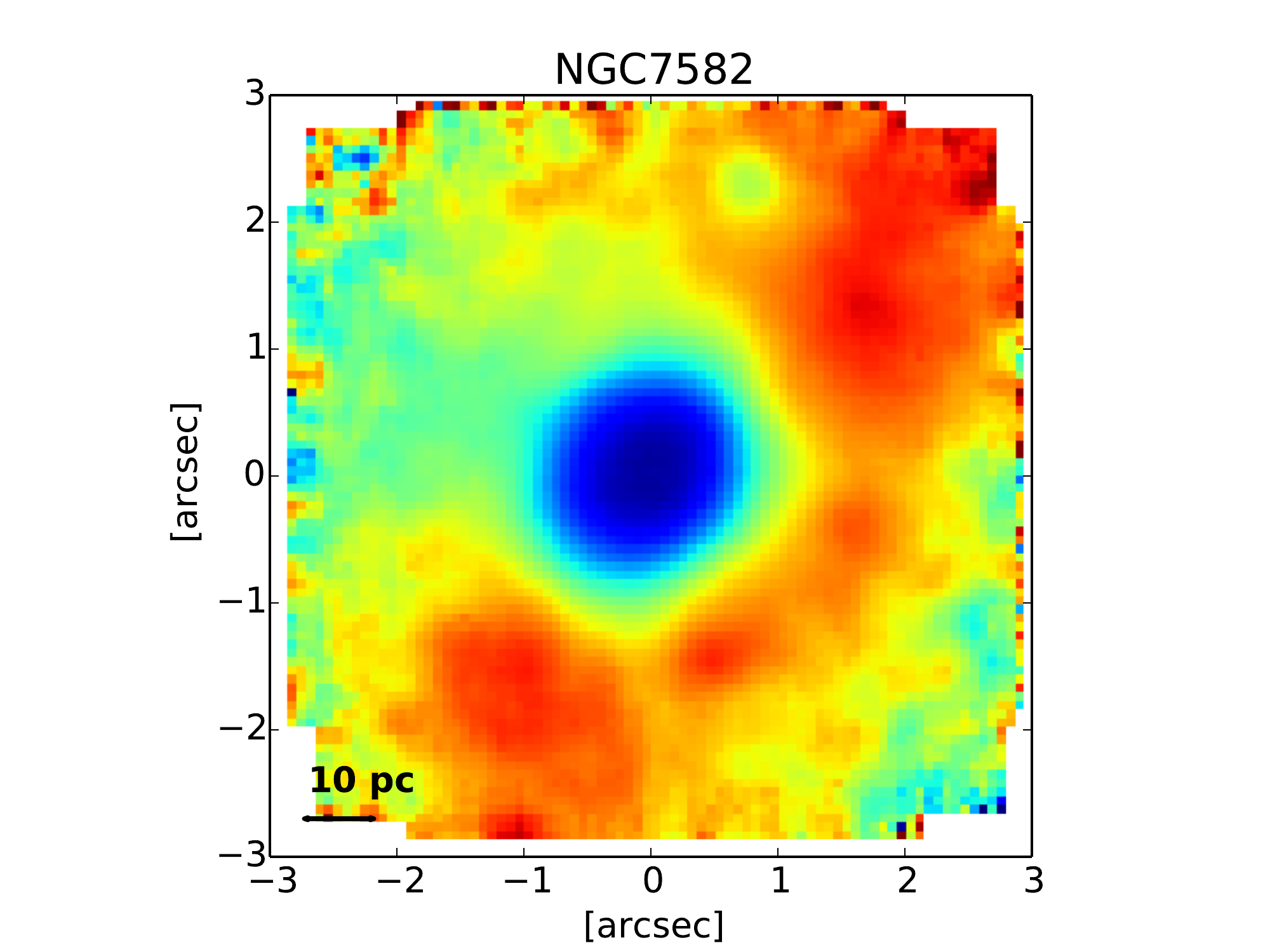}}
	\quad\quad\quad\quad
\subfloat{\includegraphics[height=0.12\vsize]{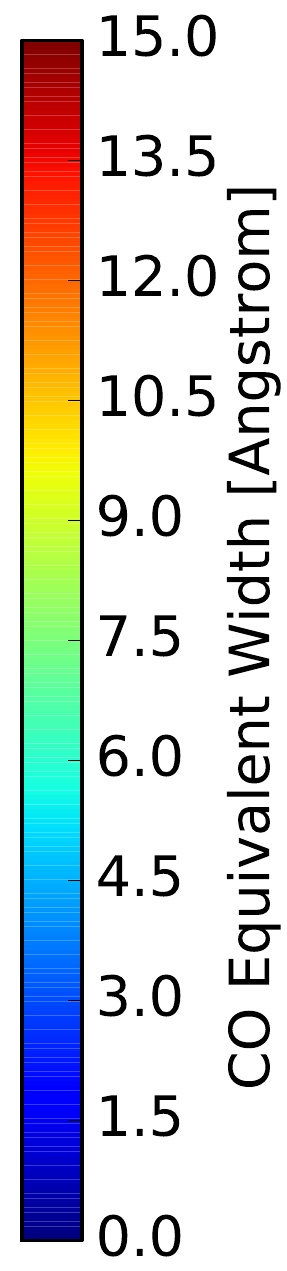}}
	\caption{\label{fig:ew:agns}EW maps of sources with unambiguous nuclear dilution. We note that some of these sources (e.g. NGC~5135) show additional structure in the EW map.}
\end{figure*}

\begin{figure*}
\subfloat{\includegraphics[trim=2cm 0.5cm 1.5cm 0.5cm, width=0.2\hsize]{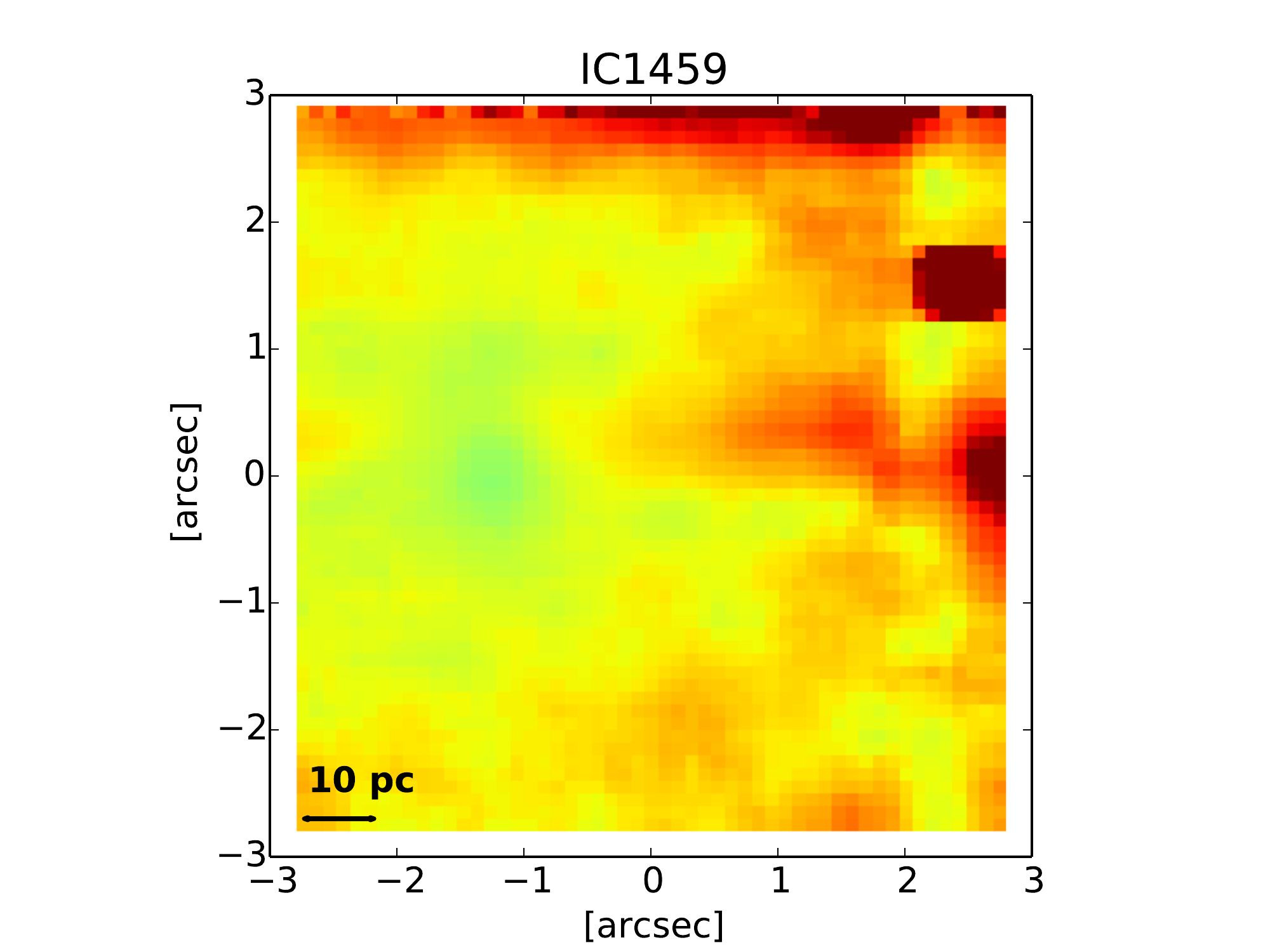}}
\subfloat{\includegraphics[trim=2cm 0.5cm 1.5cm 0.5cm, width=0.2\hsize]{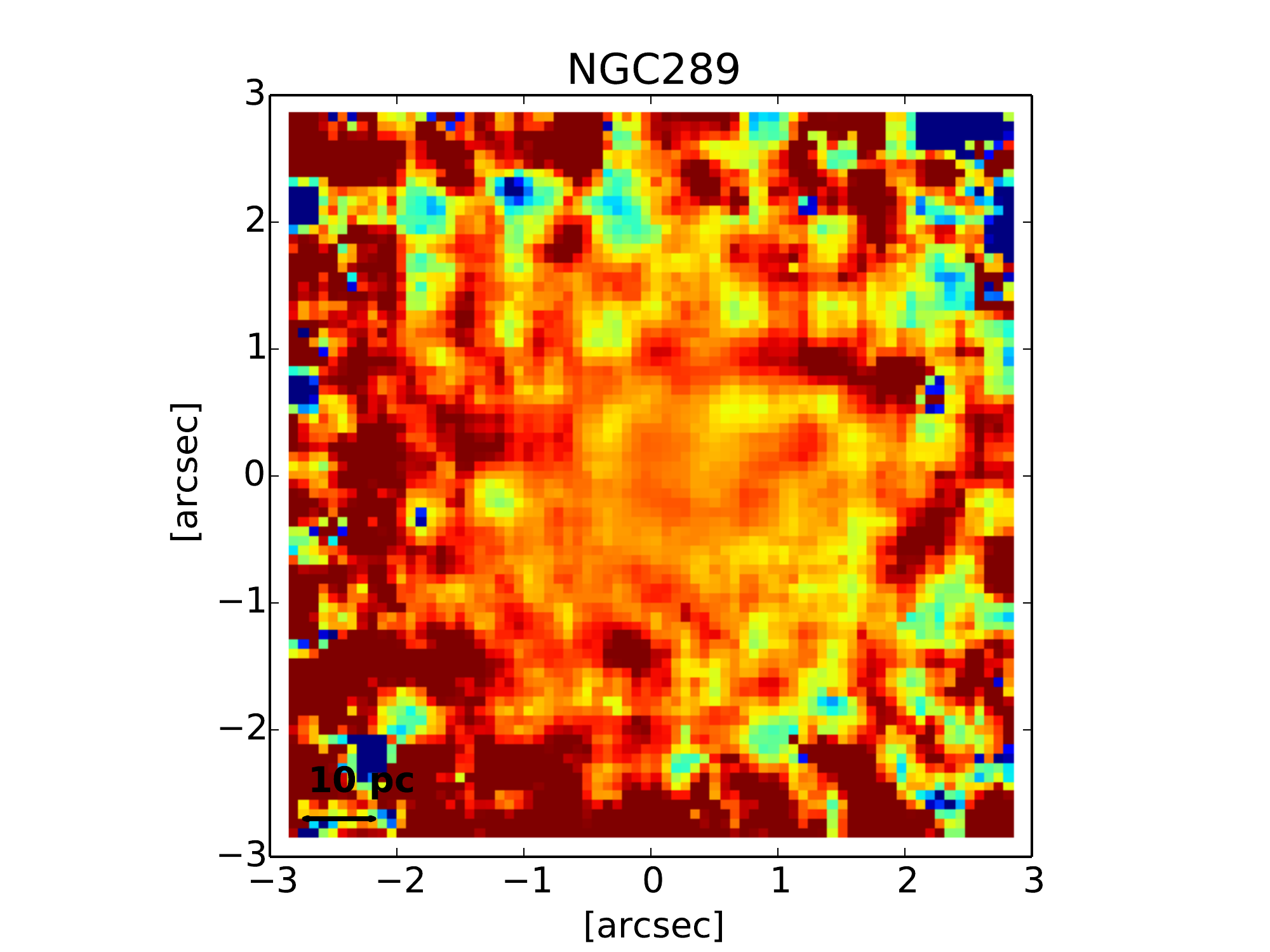}}
\subfloat{\includegraphics[trim=2cm 0.5cm 1.5cm 0.5cm, width=0.2\hsize]{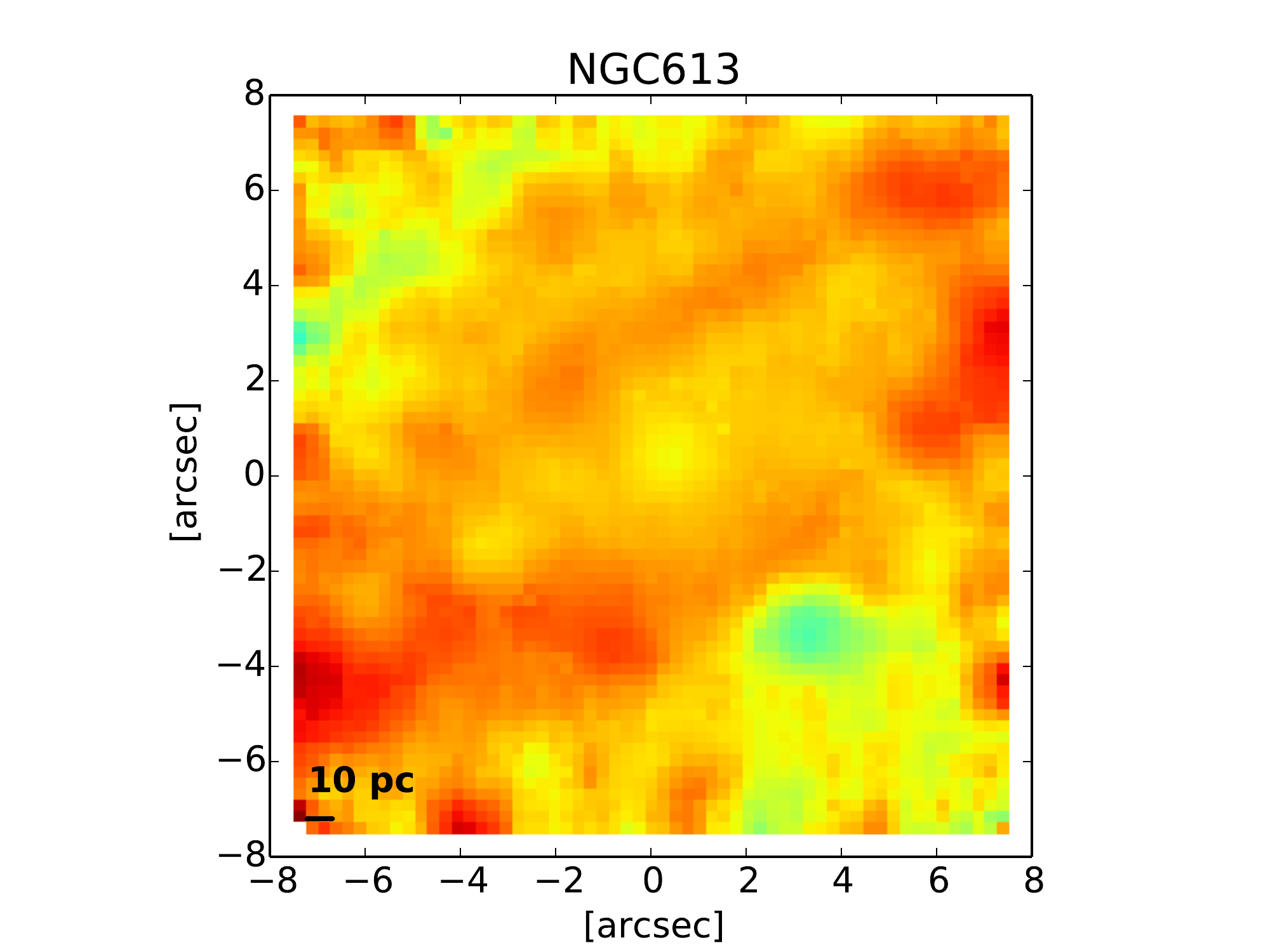}}
\subfloat{\includegraphics[trim=2cm 0.5cm 1.5cm 0.5cm, width=0.2\hsize]{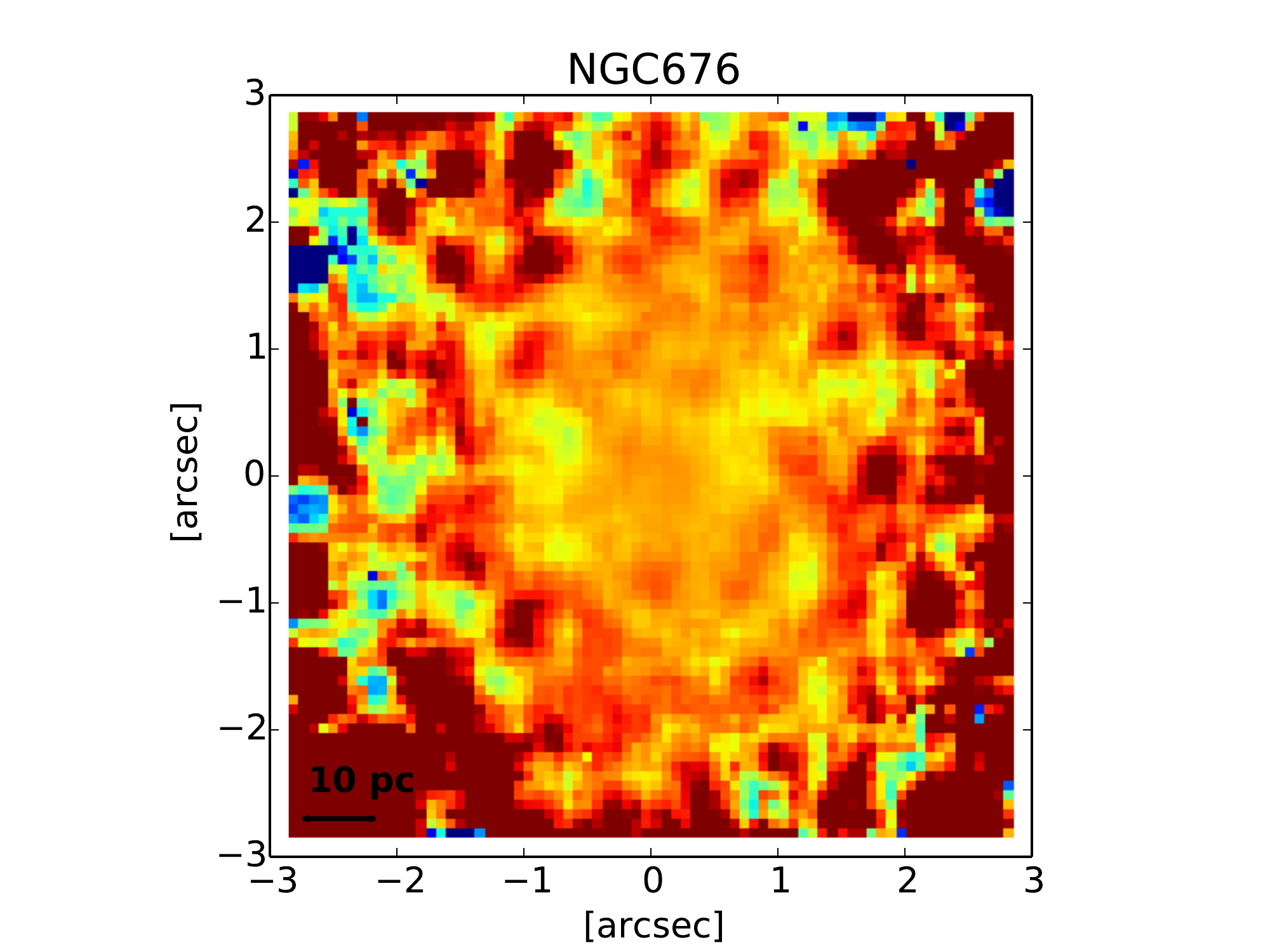}}\\
\subfloat{\includegraphics[trim=2cm 0.5cm 1.5cm 0.5cm, width=0.2\hsize]{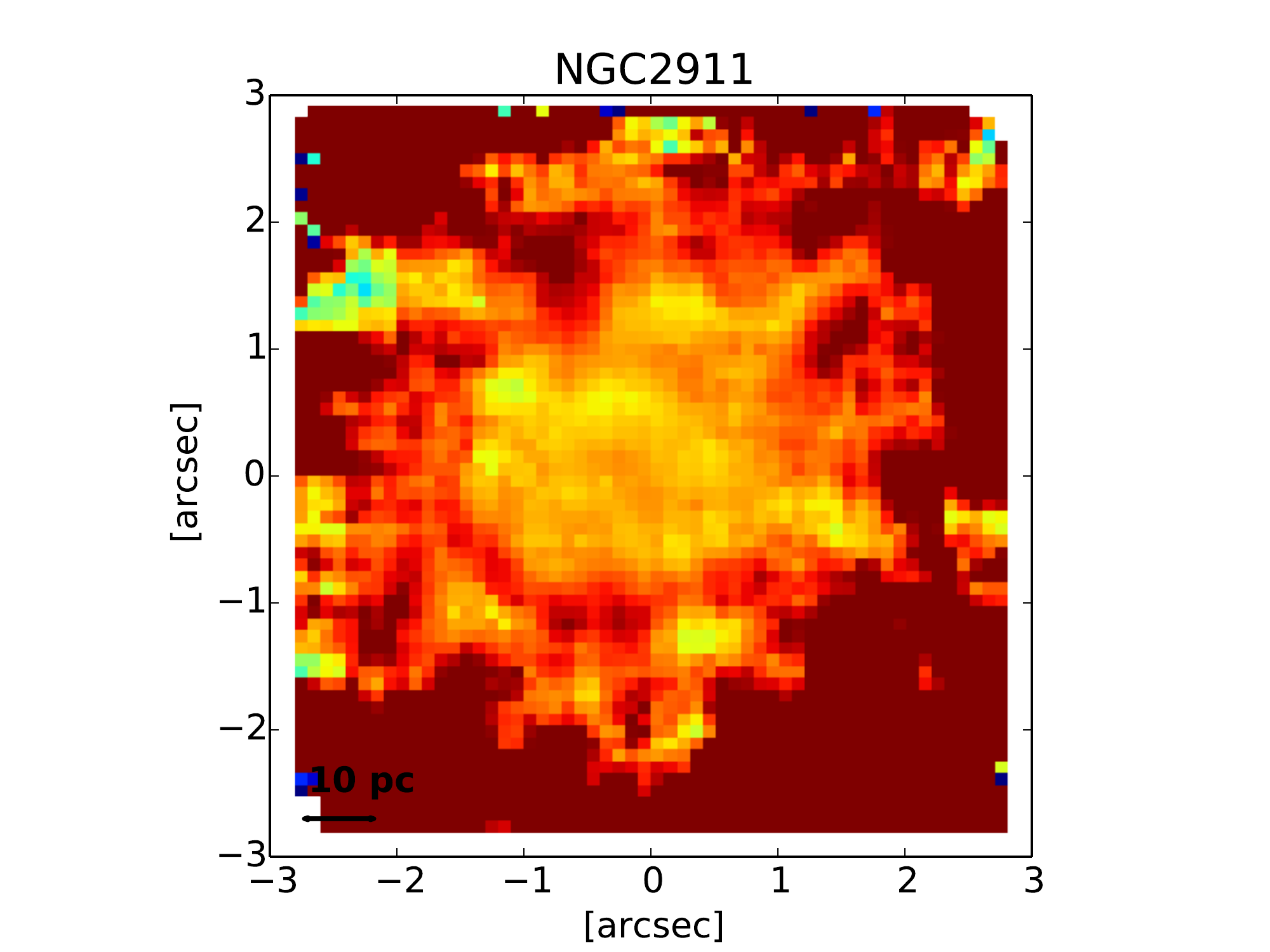}}
\subfloat{\includegraphics[trim=2cm 0.5cm 1.5cm 0.5cm, width=0.2\hsize]{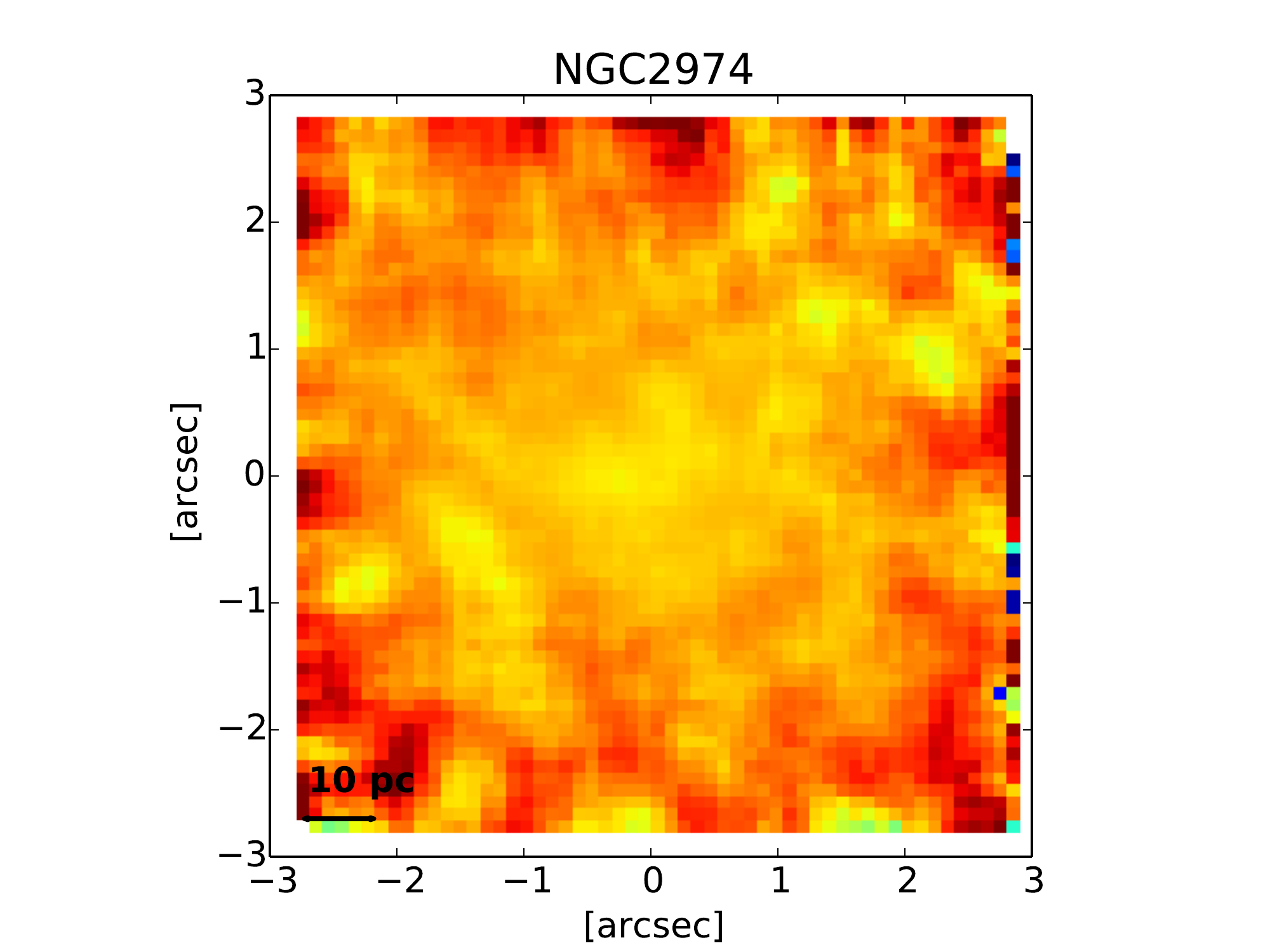}}
\subfloat{\includegraphics[trim=2cm 0.5cm 1.5cm 0.5cm, width=0.2\hsize]{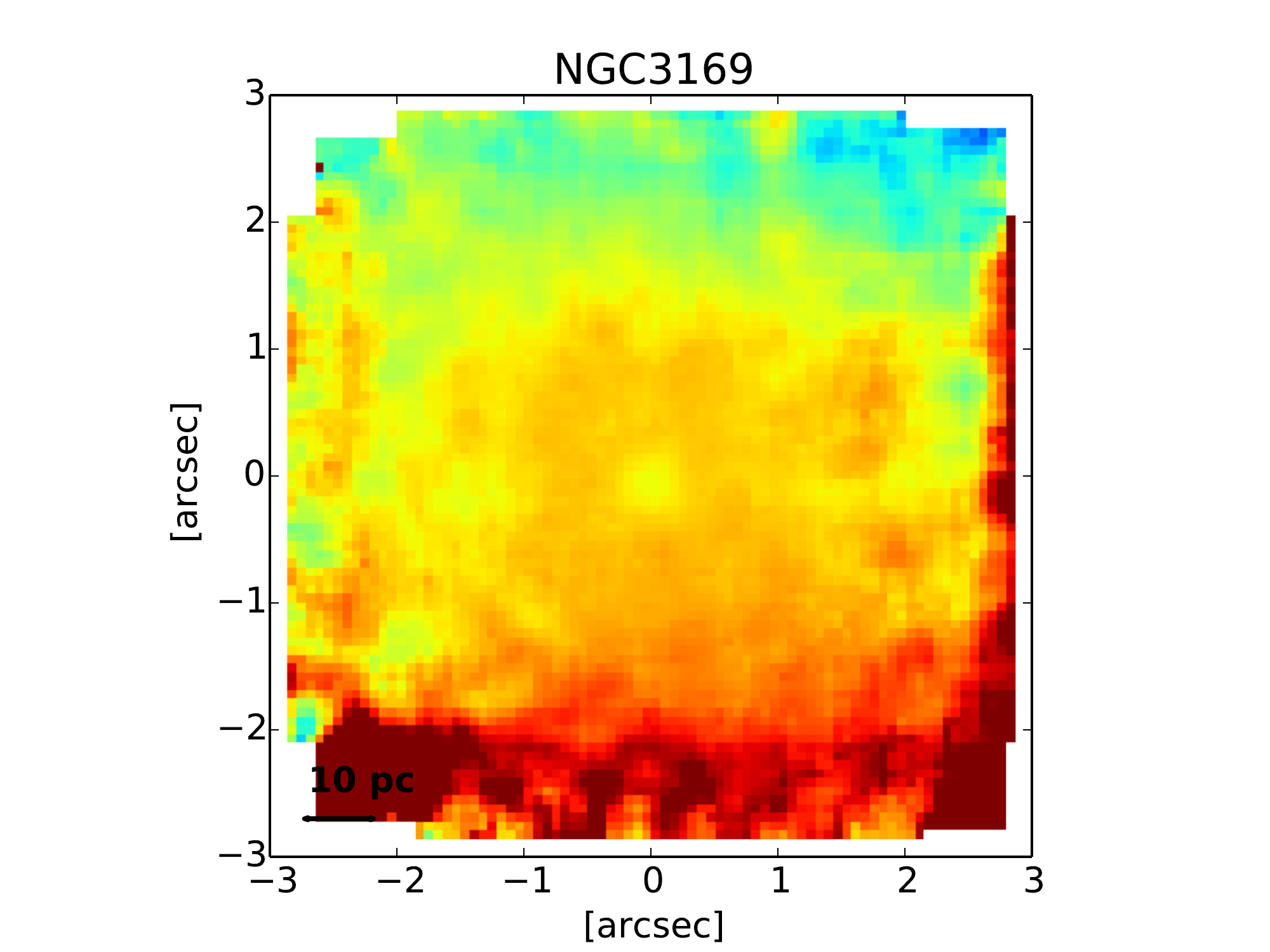}}
\subfloat{\includegraphics[trim=2cm 0.5cm 1.5cm 0.5cm, width=0.2\hsize]{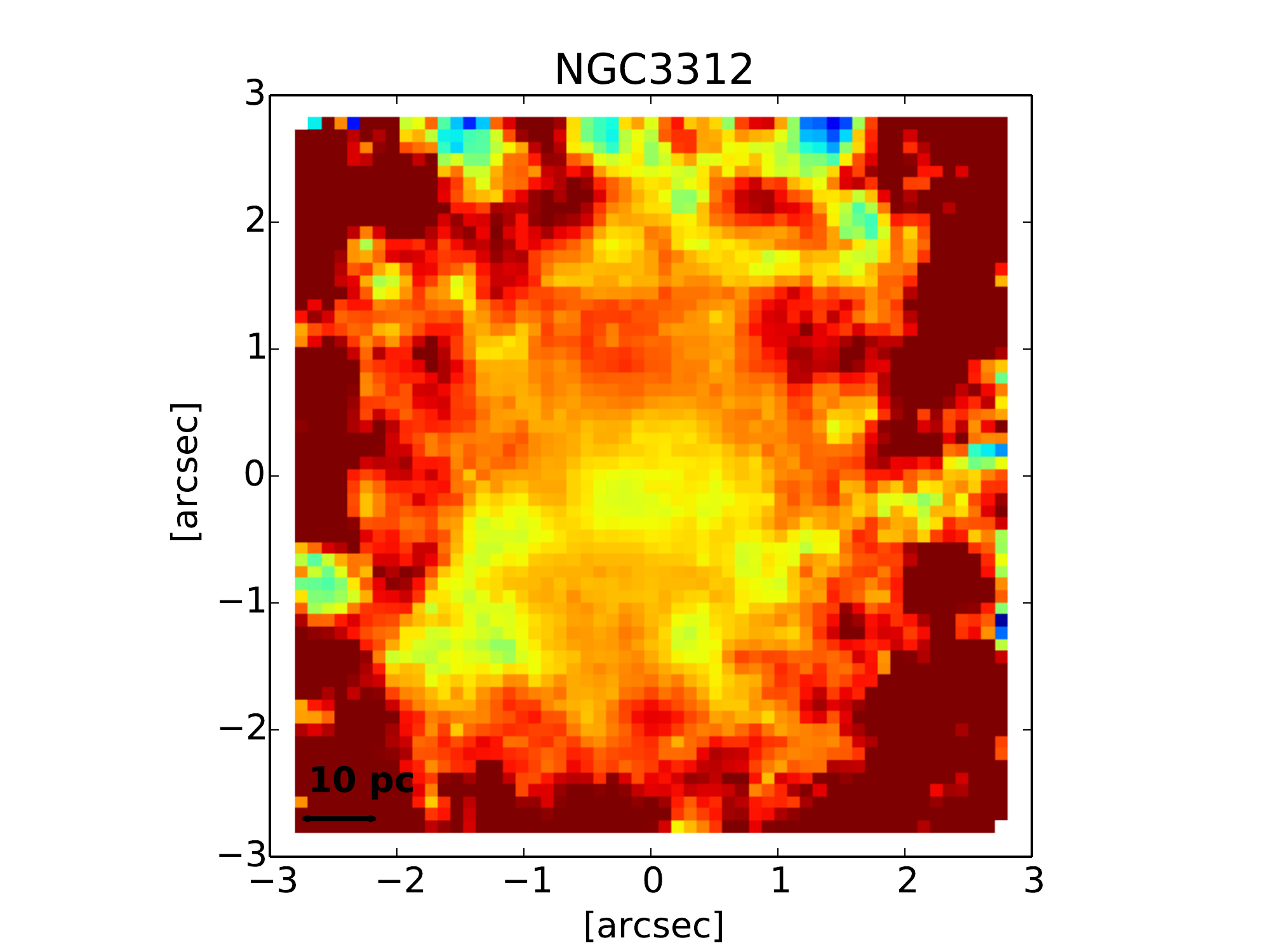}}\\
\subfloat{\includegraphics[trim=2cm 0.5cm 1.5cm 0.5cm, width=0.2\hsize]{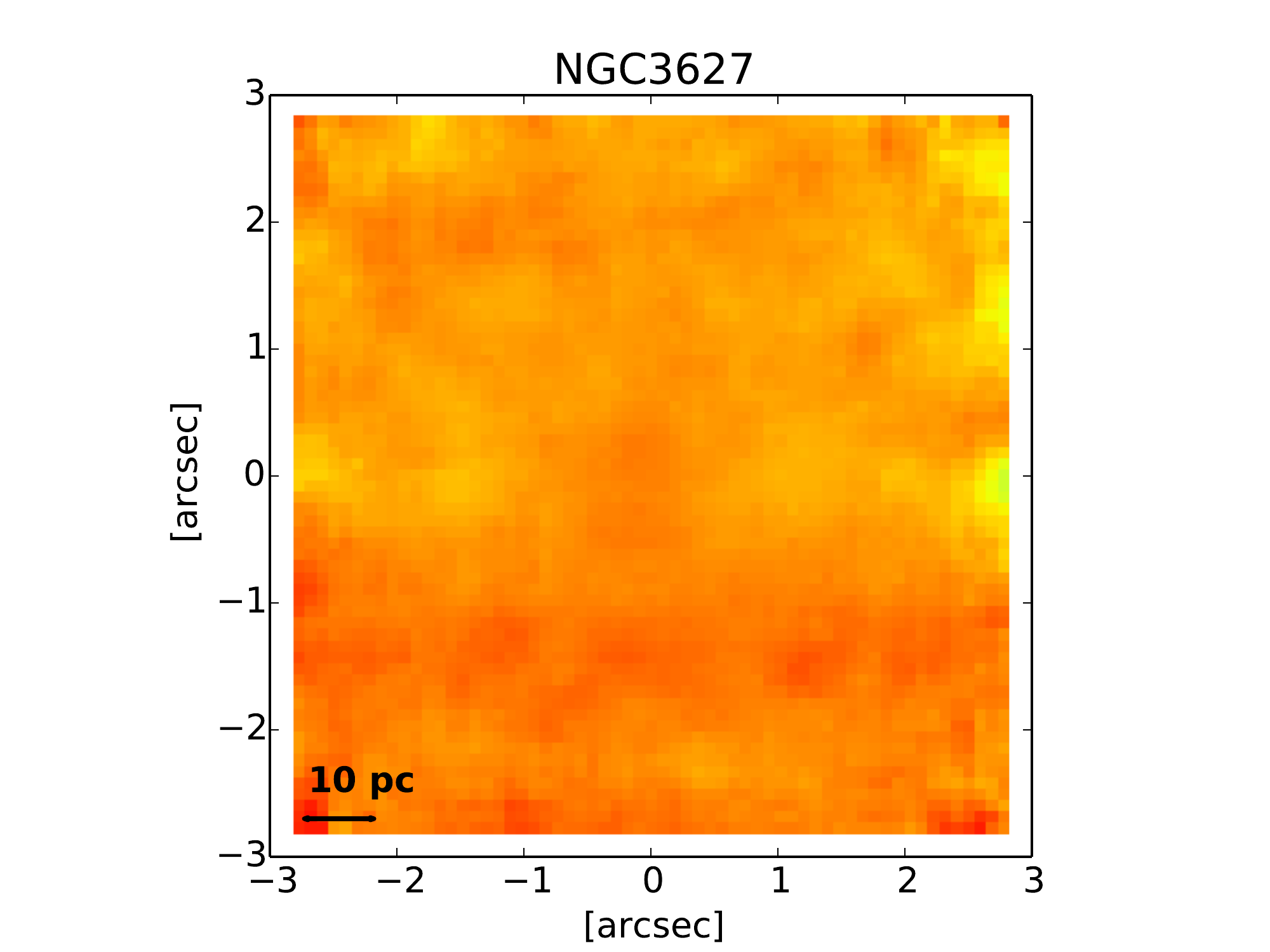}}
\subfloat{\includegraphics[trim=2cm 0.5cm 1.5cm 0.5cm, width=0.2\hsize]{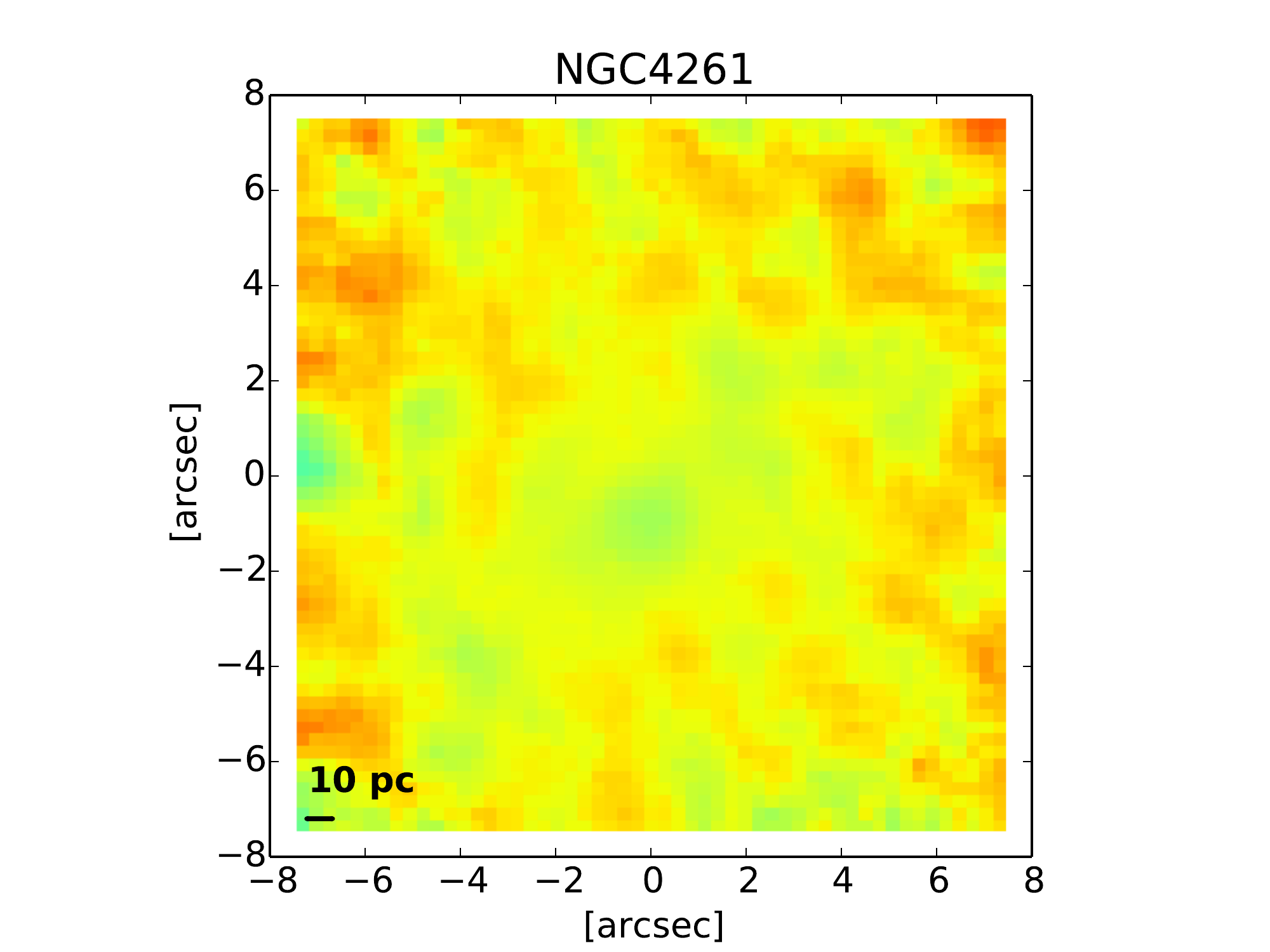}}
\subfloat{\includegraphics[trim=2cm 0.5cm 1.5cm 0.5cm, width=0.2\hsize]{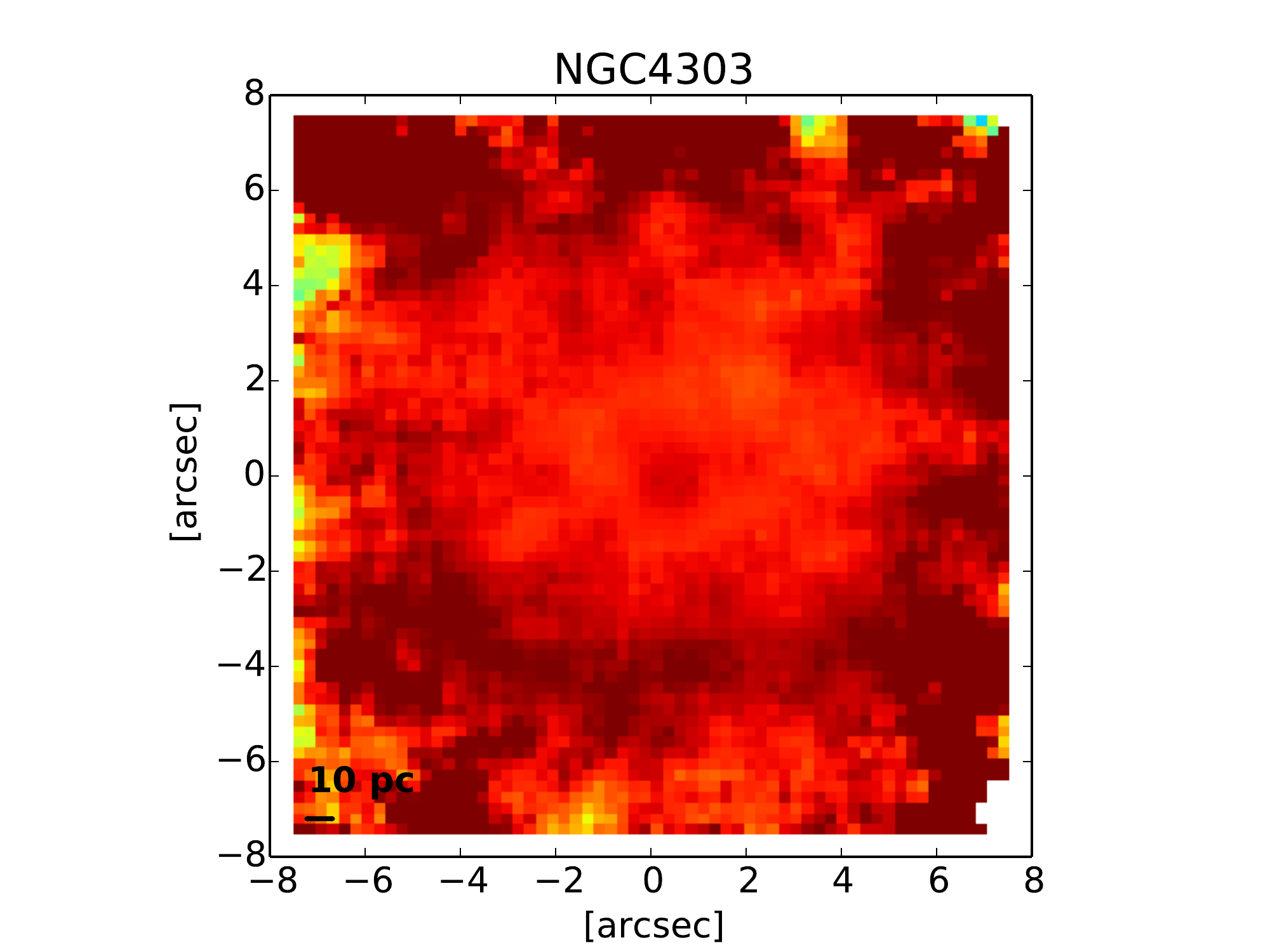}}
\subfloat{\includegraphics[trim=2cm 0.5cm 1.5cm 0.5cm, width=0.2\hsize]{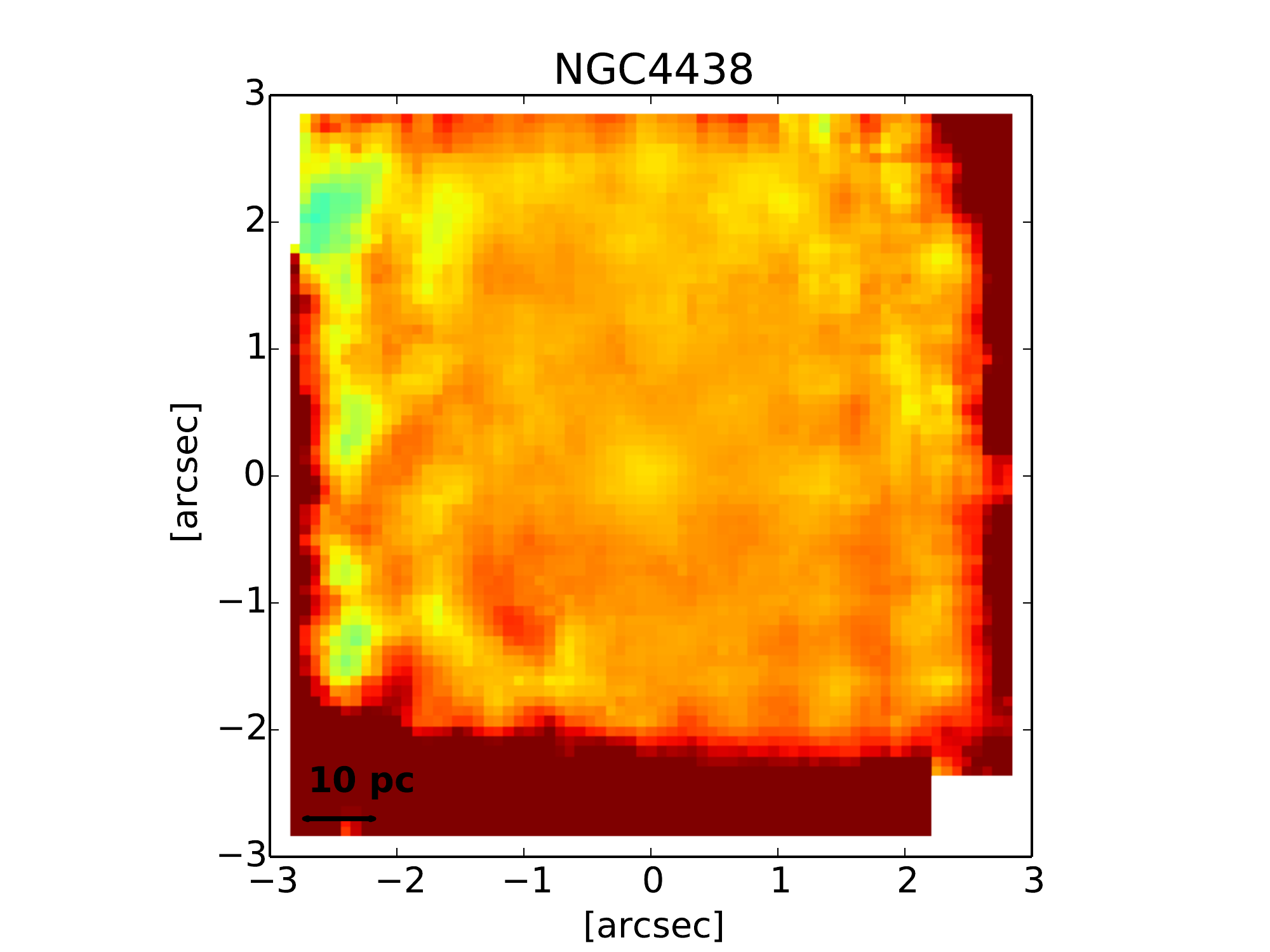}}\\
\subfloat{\includegraphics[trim=2cm 0.5cm 1.5cm 0.5cm, width=0.2\hsize]{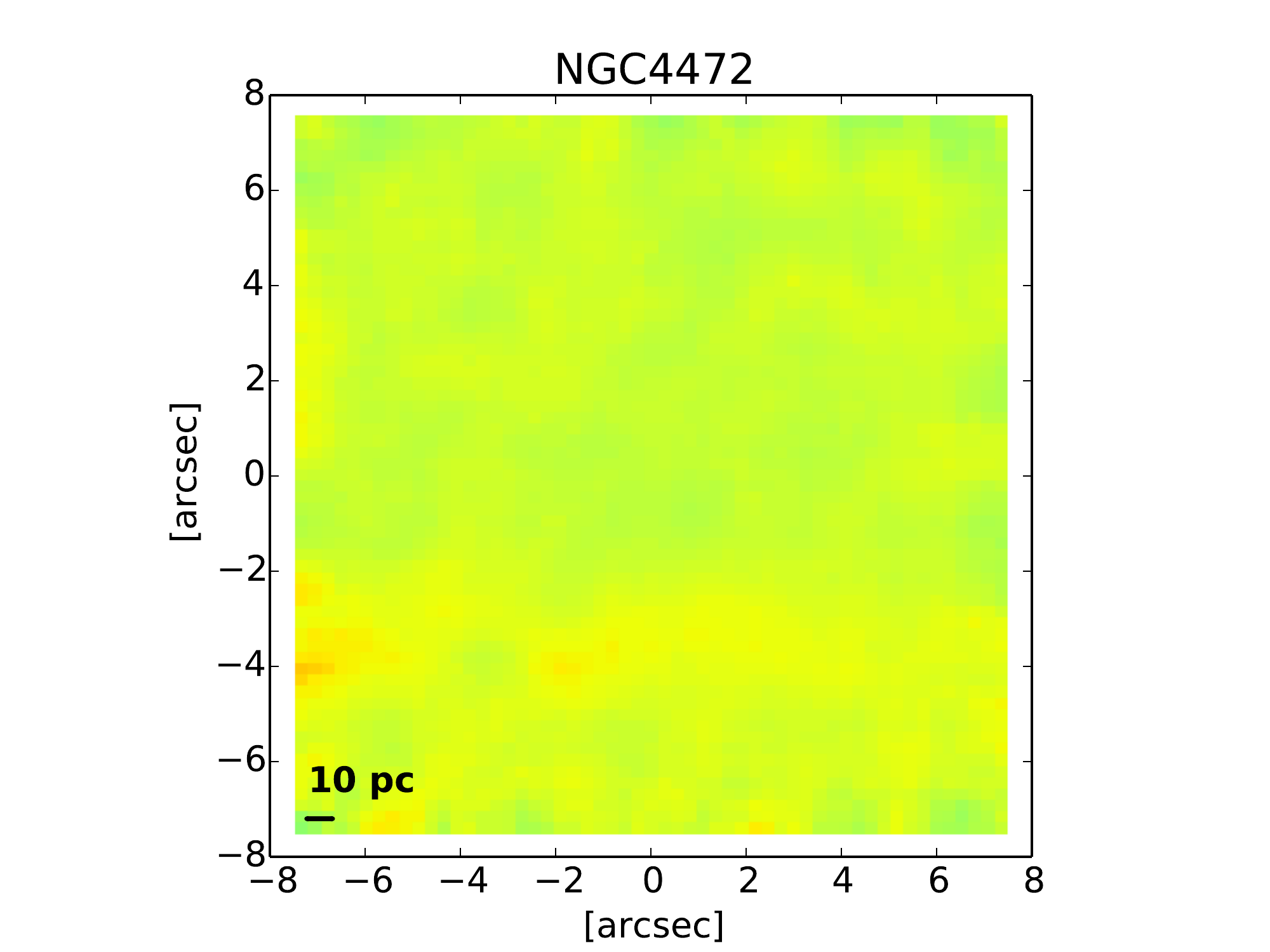}}
\subfloat{\includegraphics[trim=2cm 0.5cm 1.5cm 0.5cm, width=0.2\hsize]{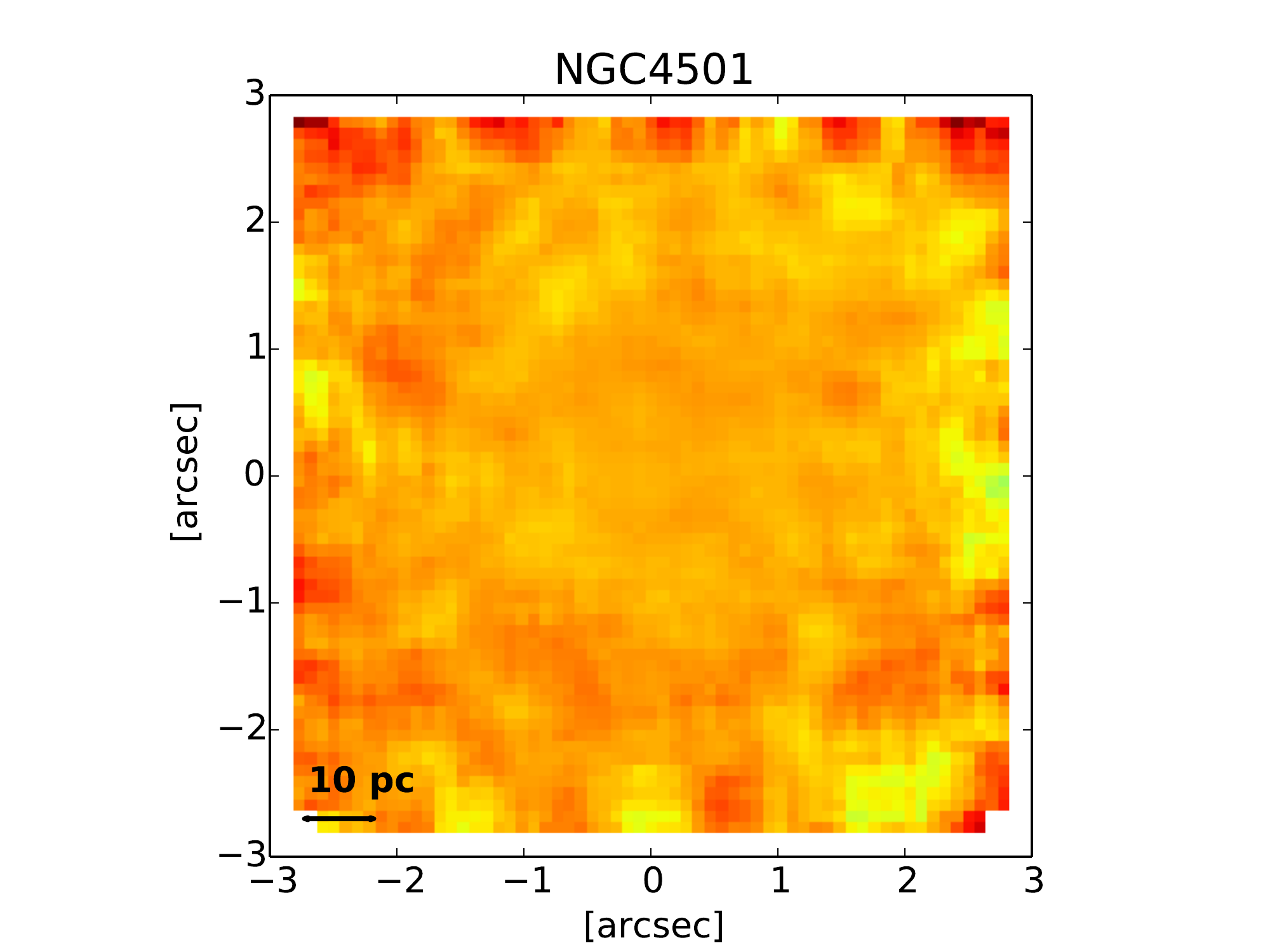}}
\subfloat{\includegraphics[trim=2cm 0.5cm 1.5cm 0.5cm, width=0.2\hsize]{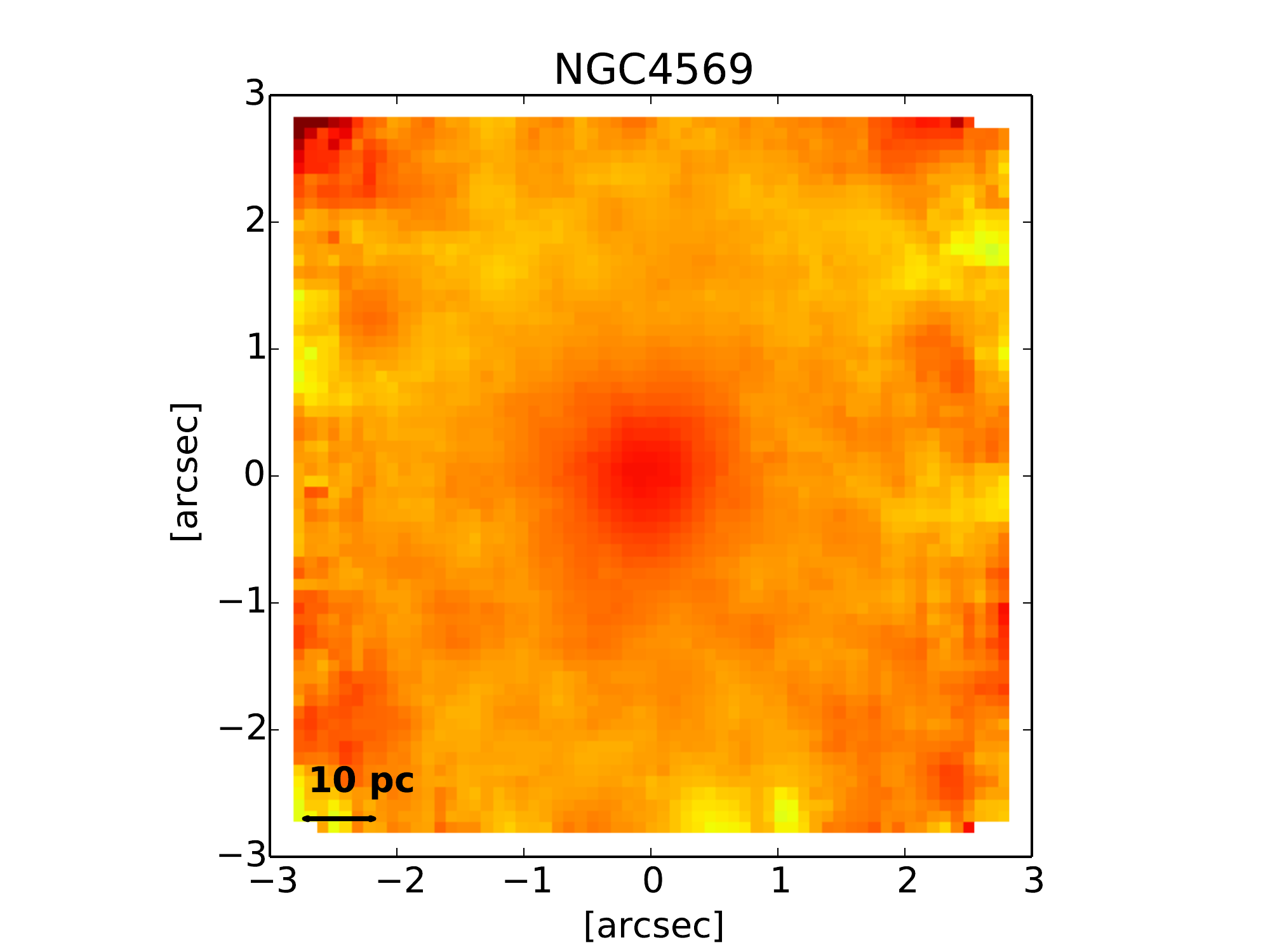}}
\subfloat{\includegraphics[trim=2cm 0.5cm 1.5cm 0.5cm, width=0.2\hsize]{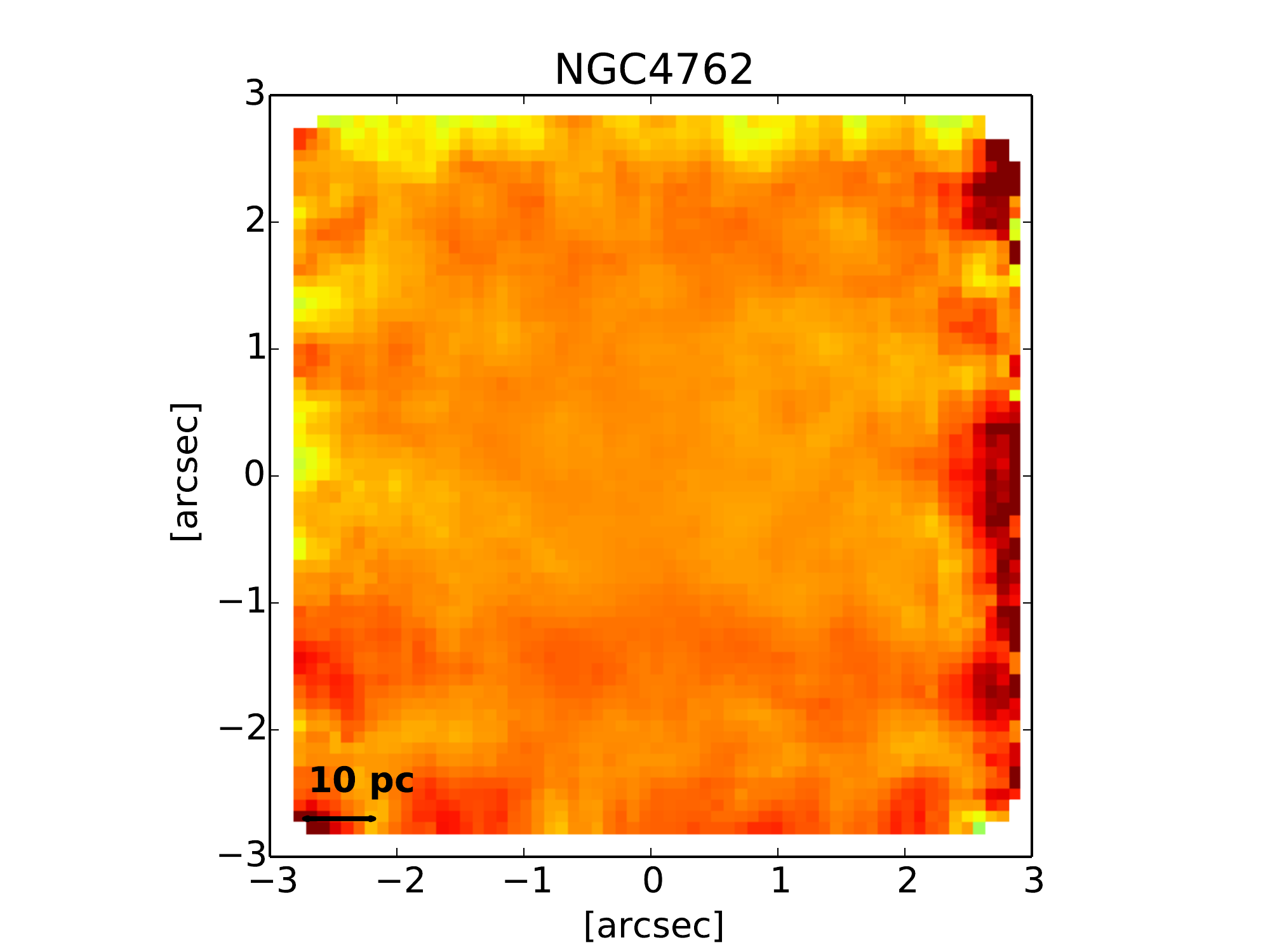}}\\
\subfloat{\includegraphics[trim=2cm 0.5cm 1.5cm 0.5cm, width=0.2\hsize]{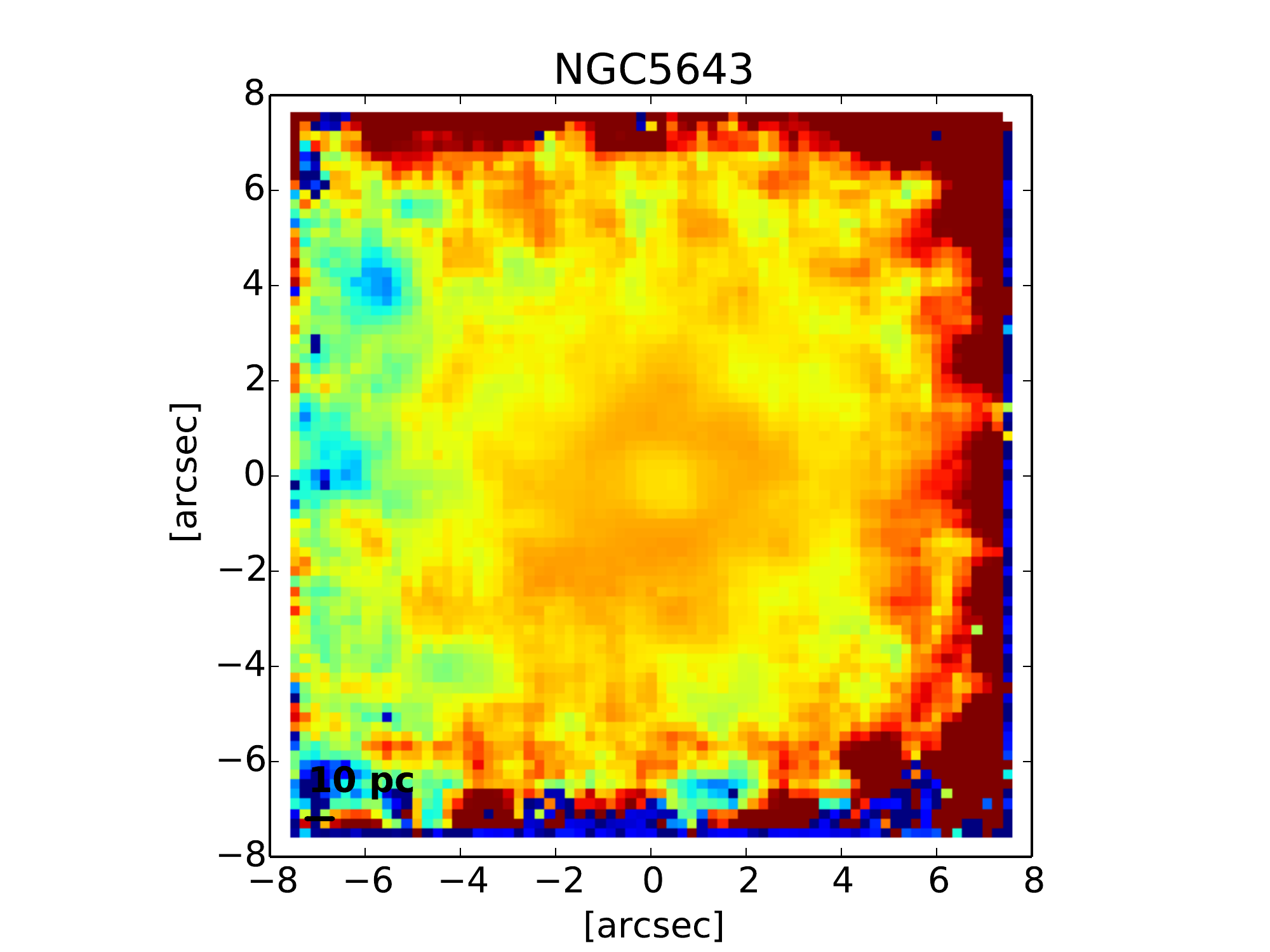}}
\subfloat{\includegraphics[trim=2cm 0.5cm 1.5cm 0.5cm, width=0.2\hsize]{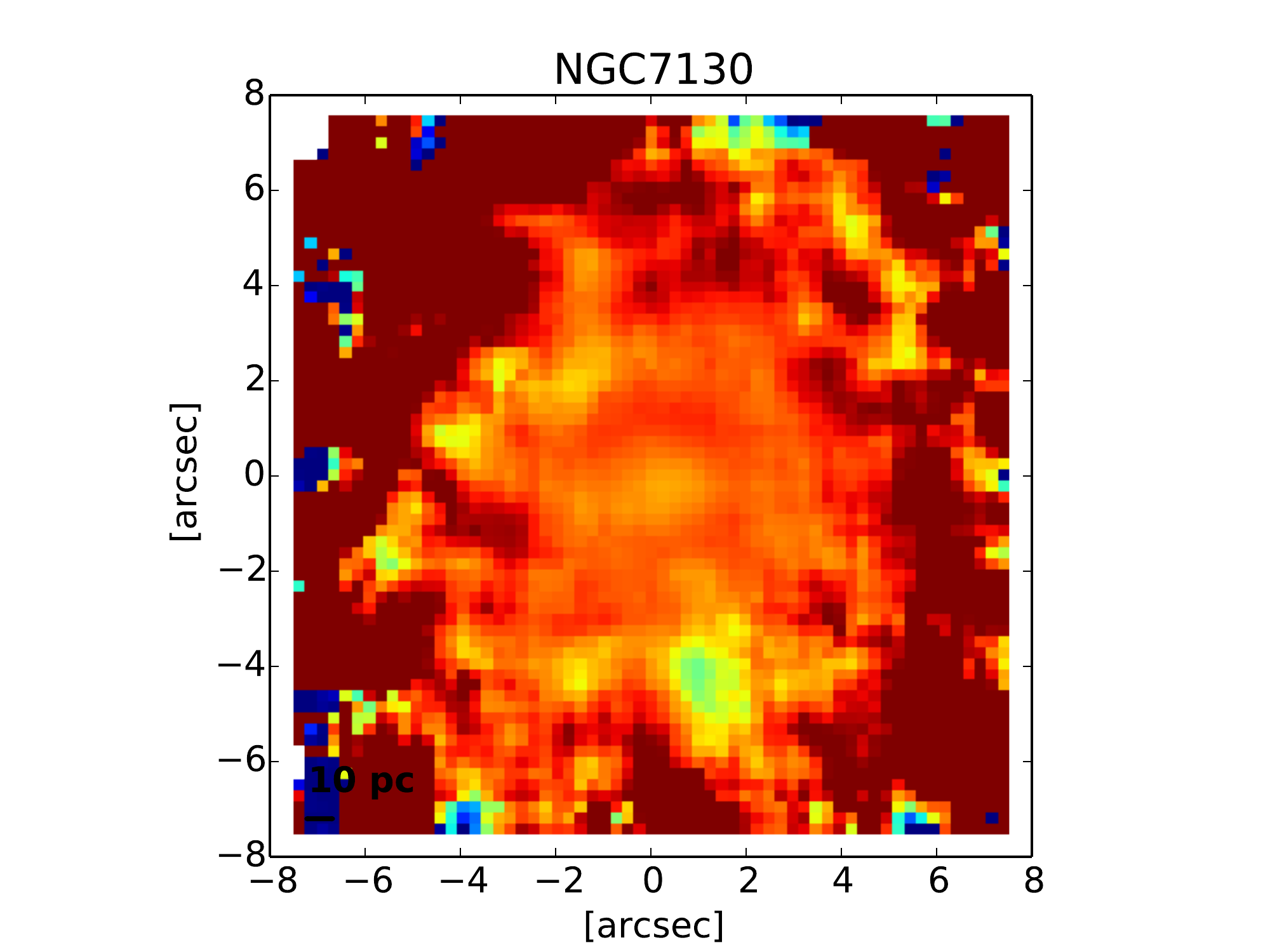}}
\subfloat{\includegraphics[trim=2cm 0.5cm 1.5cm 0.5cm, width=0.2\hsize]{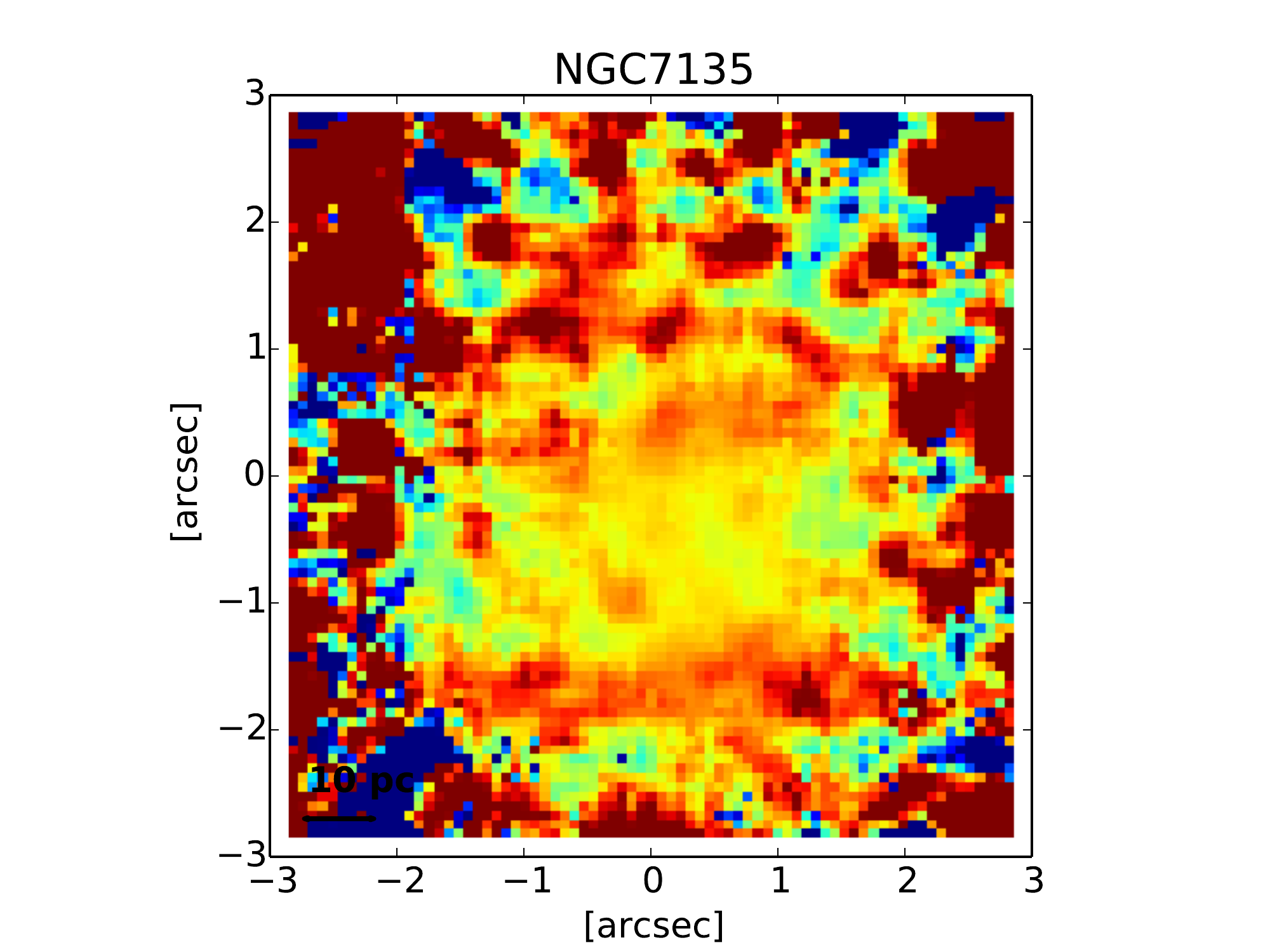}}
\subfloat{\includegraphics[trim=2cm 0.5cm 1.5cm 0.5cm, width=0.2\hsize]{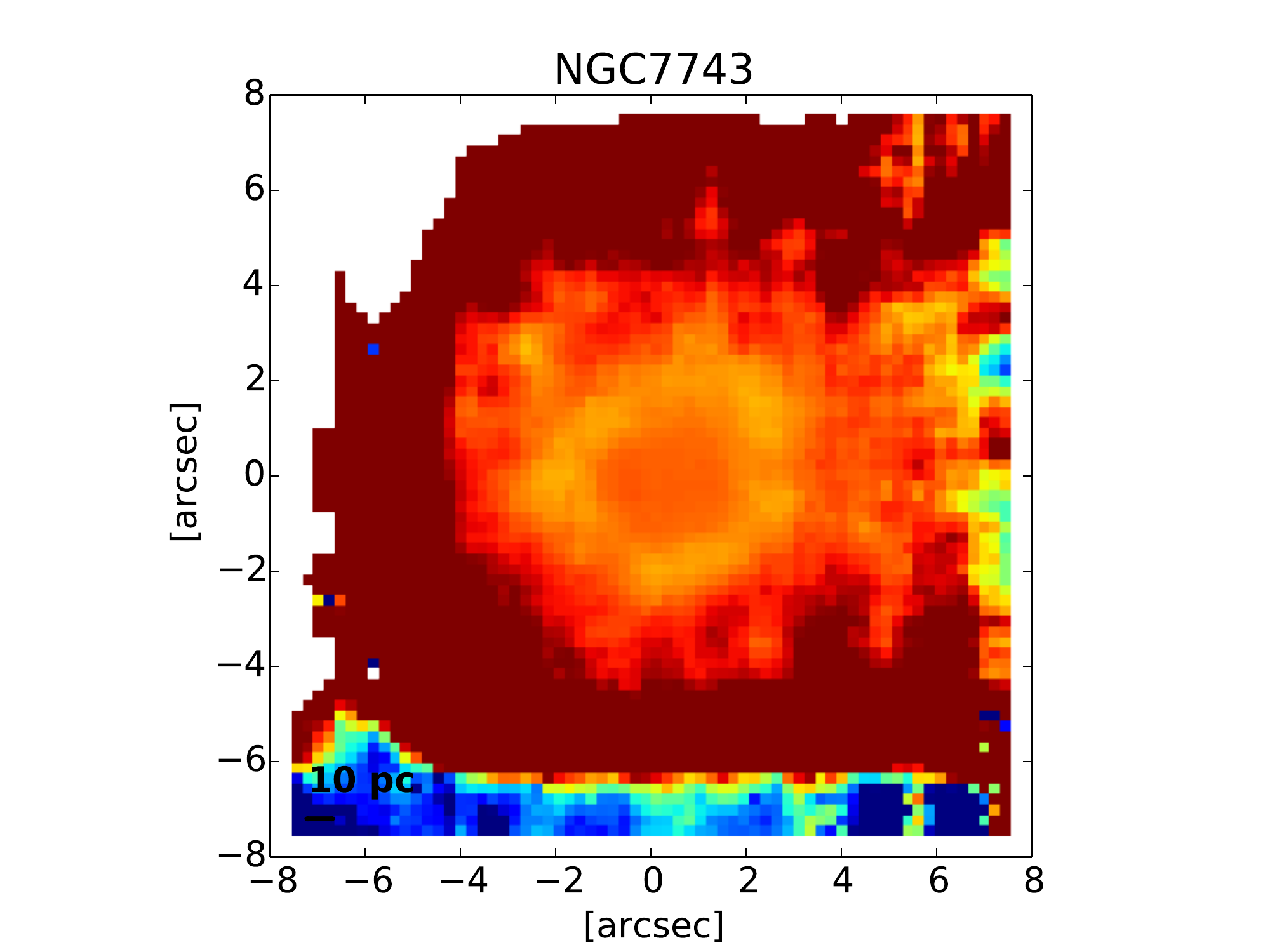}}
	\quad\quad\quad\quad
\subfloat{\includegraphics[height=0.12\vsize]{img/EW_maps/colorbar_vertical.pdf}}
	\caption{\label{fig:ew:noagns}EW maps of sources where nuclear dilution by AGN light cannot be safely established.}
\end{figure*}

The resulting EW maps are displayed in Figs.~\ref{fig:ew:agns}~and~\ref{fig:ew:noagns} for sources with and without detected dilution by non-stellar continuum respectively. Among the non-detections there are a few sources (NGC~613, NGC~3169, NGC~5643) which may show an AGN diluted core, but since the EW depression there is smaller than our adopted uncertainty we do not derive AGN luminosities from them.

Given the EW map at each pixel $x,y$, we calculate the diluting radiation, i.e. pure continuum emission $I_{\rm dilute}$ that does not show the CO absorption feature, using $f_{{\rm dilute}, x,y} = {\rm EW_{int}}/{\rm EW}_{x,y}$ where ${\rm EW_{int}}$ is the intrinsic CO equivalent width (see below) of the galaxy outside the influence of the AGN light. The diluting radiation is then given as

\begin{align}
I_{\rm dilute}(x,y) &= I_{\rm total}(x,y) - I_{\rm stars}(x,y) \notag\\
&= \left(1 - 1/f_{\rm dilute}(x,y)\right) \cdot I_{\rm total}(x,y)
\end{align}

The fraction of non-stellar light in the near-IR \fnir{} is then calculated as the fraction of diluting radiation (over total) within an aperture $A$ larger than the PSF:

\begin{equation}
\fnir = \frac{\int_A I_{\rm dilute}(x,y)}{\int_A I_{\rm total}(x,y)}
\end{equation}

We used an extraction aperture of 1\arcsec{} for all sources except for the Circinus~galaxy and NGC~7469 where 0.5 arcsec were used due to the smaller pixel scales (fields of view) of these observations. The full-width at half maximum (FWHM) of the point spread function (PSF) of these observations was 290 milli arcseconds (mas) and 181 mas, respectively.

The adopted intrinsic EW per source and \fnir{} are given in Tab.~\ref{tab:starfit} and its derivation is described in the next Section.

The isotropic AGN luminosity in the near-IR, \lnir, is then simply given by $I_{\rm dilute} \cdot 4 \pi D^2$ where $D$ is the distance of the galaxy. These luminosities can be compared to literature values, e.g. to \citet{prieto2010} who have compiled a number of nuclear photometries. We find their results for the K band photometry of the pure AGN source to be essentially identical to ours for most of the sources.\footnote{In NGC~1068, NGC~1566 and NGC~7582, however, our photometries are significantly (about two magnitudes or a factor 6) brighter than those of \citet{prieto2010}. This could indicate problems with the PSF correction in the very high spatial resolution data from \citet{prieto2010}. Since our star-formation corrected photometries are averaged over the central arcsecond, they should not be affected by PSF variations.}

\subsubsection{Determination of intrinsic CO equivalent width}

To determine the intrinsic, i.e. undiluted, equivalent width of the CO feature, we classify sources into three categories depending on the radial profile of their CO EW:

\begin{enumerate}
	\item Constant EW over the entire cube (i.e. undiluted by AGN light, $EW = EW_{\rm int}$),
	\item EW suppressed in the nucleus and more or less constant outside some radius (i.e. diluted by AGN light but intrinsic EW is observed),
	\item constantly increasing EW (i.e. diluted by AGN light and field of view too small to observe $EW_{\rm int}$.).
\end{enumerate}

The variety of EW profiles can be seen from the maps (Figs.~\ref{fig:ew:agns}--\ref{fig:ew:noagns}) and from the azimuthally-averaged radial profiles (Fig.~\ref{fig:radialplot}). One can see a clear bimodality in the nuclear value EW$_{\rm nuc}$: most sources are either highly diluted (EW$_{\rm nuc}$ $\lesssim$ 5 \AA{}) or essentially undiluted (EW$_{\rm nuc}$ $\sim$ EW$_{\rm int} \approx 11.1$). This is due to the fact that most of our sources have roughly the same stellar surface brightness (luminosity within our aperture), but we are probing a range of AGN luminosities (see Fig.~\ref{fig:lum_star_agn}). Depending on the AGN luminosity that we probe, the nucleus therefore appears either starlight dominated (undiluted CO feature) or AGN dominated (entirely diluted CO feature). The sources with intermediate values of EW$_{\rm nuc}$ have AGN luminosities that are similar to the stellar luminosity. This transition occurs at $\lx{} \approx 10^{41.5}$ erg/s which corresponds to $L_{\rm bol} \approx 10^{42.5}$ erg/s. This happens to be the luminosity below which the AGN torus is expected to disappear according to some models \citep{elitzur2006,hoenig2007}. Unfortunately we can however not constrain the torus disappearance with these measurements yet (we would need higher spatial resolution to probe lower AGN luminosities).

\begin{figure}[h]
   \centering
   \includegraphics[width=\hsize, trim=4cm 1cm 2cm 3cm]{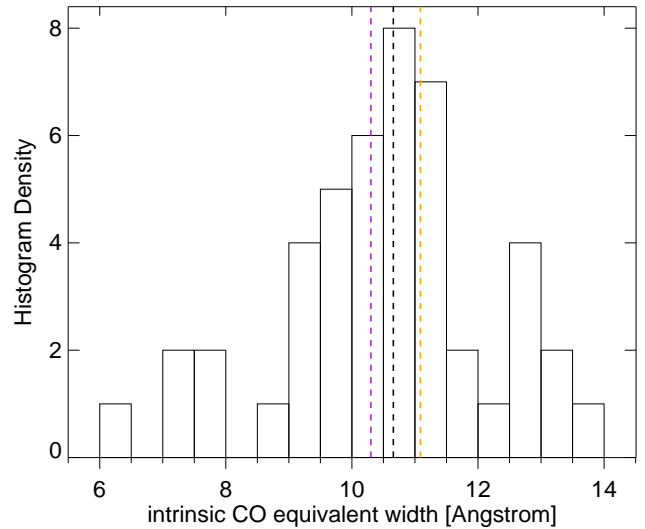}
   \caption{\label{fig:ew_int_hist}Histogram of intrinsic equivalent width as determined from Gaussian fits to the radial profile of those sources where a plateau level was seen. Black dashed line: median value of EW for all sources where the intrinsic EW can be measured (10.6 \AA{}); purple dashed line: median value for diluted sources only (10.3 \AA{}); orange dashed line: for non-diluted sources only (11.1 \AA{}).}
\end{figure}

\begin{figure}[h]
   \centering
   \includegraphics[width=\hsize, trim=5cm 3cm 5cm 3cm]{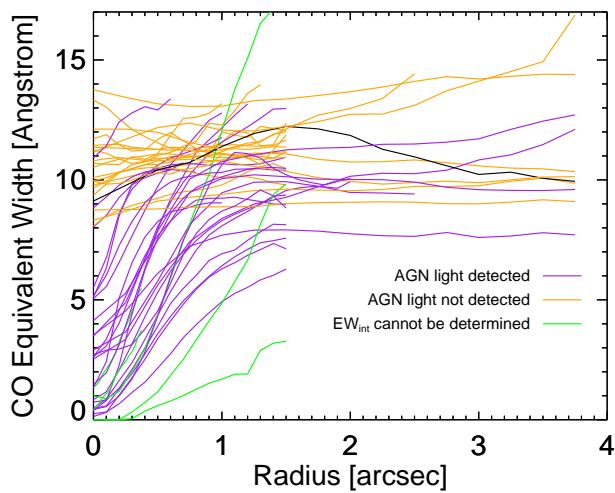}
   \caption{\label{fig:radialplot}EW averaged in annuli for all 51 sources. Only the good signal/noise part of the cube that was used to fit the EW profile is shown. Diluted sources are drawn in purple, undiluted in orange and sources where the intrinsic value could not be determined due to the limited field of view of the SINFONI data are shown in green. The black line denotes NGC~5135 which has a non-axisymmetric EW profile. This graph shows sources observed in all three SINFONI pixel scales (field of view: 0.8, 3 and 8 arcsec).}
\end{figure}

\begin{figure}[h]
   \centering
   \includegraphics[width=10cm, trim=5cm 3cm 5cm 3cm]{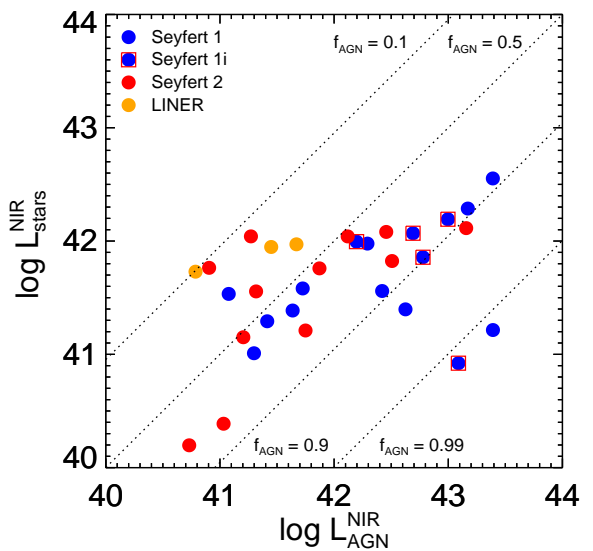}
   \caption{\label{fig:lum_star_agn}Distribution of AGN and stellar luminosities (surface brightnesses) within an aperture of 1 arc second; various levels of \fnir{} are indicated in the plot.}
\end{figure}

For all sources except five, we can determine the intrinsic value of EW from a fit to the radial profile. The median value is 10.7 \AA{} with a standard deviation of 1.6 \AA{} (see Fig.~\ref{fig:ew_int_hist}). There is a slight difference between the diluted sources (AGNs) and the undiluted sources with the former showing a median value of $10.3 \pm 1.8$ \AA{} and the latter $11.1 \pm 1.1$ \AA{}, but this difference is not significant given the relatively large errors.

For each source where we can determine the intrinsic value, we use that value as a reference to calculate the fraction of AGN light. The accuracy of this method is limited by our ability to resolve small-scale EW variations which we estimate to be $\sim$ 0.5 \AA{}. Thus we are effectively limited by spatial resolution, not by signal/noise.

For stellar-light-dominated sources (low \fnir{}) this converts to an uncertainty in \fnir{} of about 10\%. For AGN-dominated sources, the uncertainty in the intrinsic EW is irrelevant and the error is dominated by the photometric accuracy of our calibration which we also estimate to be about 10\%.

For sources where we cannot measure the intrinsic EW, we adopt a value of 11.1 \AA{}, the median of all undiluted sources, for the CO(2,0) equivalent width. This is slightly different from the value of 12.0 \AA{} that was suggested by \citet{davies2007}.

\subsection{Spectral decomposition and near-IR color temperatures}
\label{sec:gox_fit}

\begin{figure*}
	\sidecaption
	\includegraphics[width=12cm]{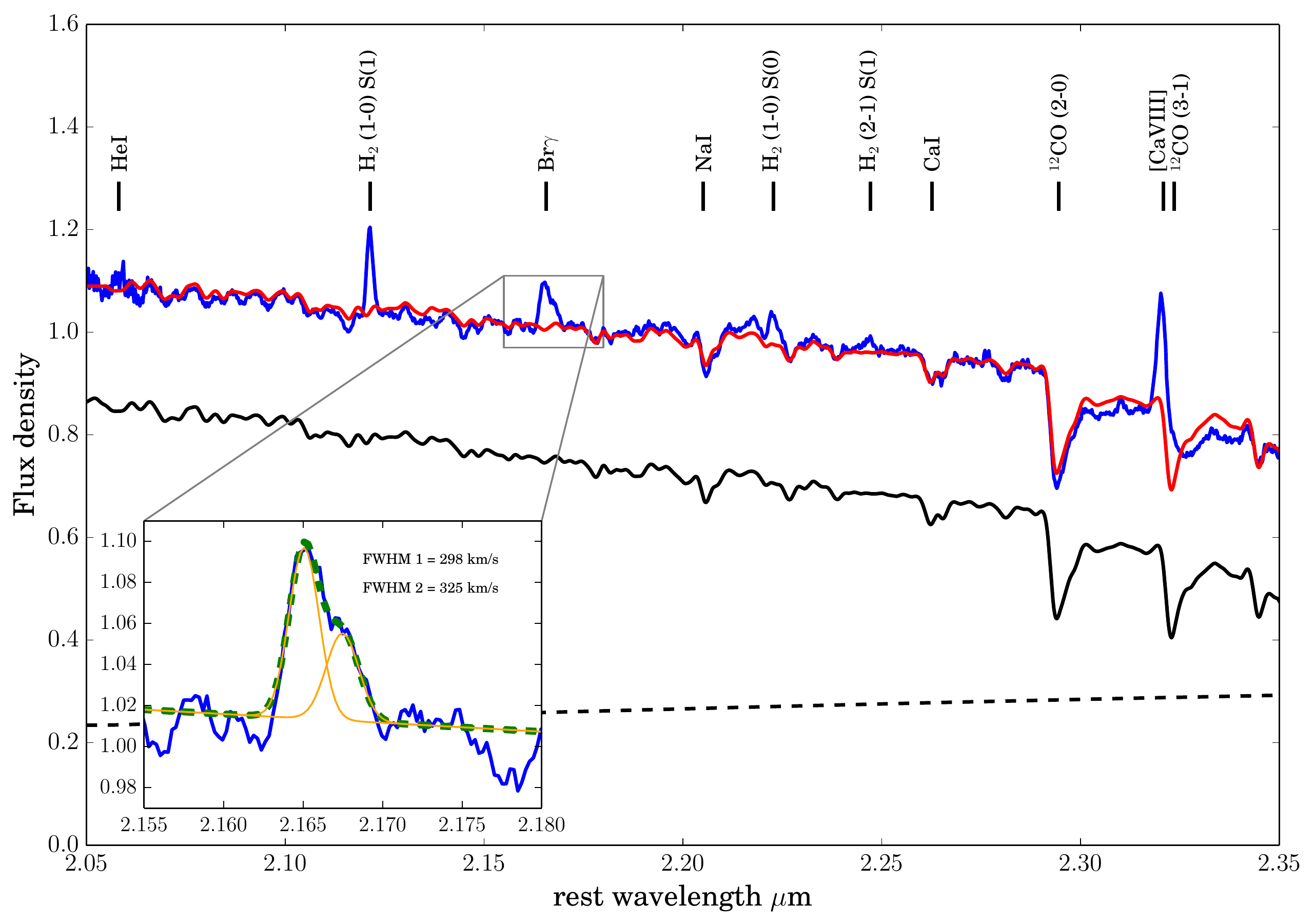}
	\caption{\label{fig:Tgox}Fit to the nuclear (aperture 1\arcsec) spectrum of NGC~1386. The observed spectrum is shown in blue, the continuum ($T$=982 K) and stellar template (HD~176617) are shown as dashed and straight black lines respectively and the total fit is shown in red. Prominent AGN emission lines and stellar absorption features are marked. In the inset we show our two-Gaussian fit to the \brg{} line of NGC~1386, demonstrating that the previously found ``broad'' line is actually a superposition of two narrow lines.}
\end{figure*}

In addition to the {\tt starfit} analysis, we also modeled the integrated nuclear spectrum (in an aperture of 1\arcsec{} for most galaxies, see above). We fitted a stellar template of a late-type star and a blackbody emitter (representing hot dust emission from the inner edge of the torus) to derive the color temperature of the continuum emission and an independent estimate of \fnir{}. An example of the spectral fit is shown in Fig.~\ref{fig:Tgox}. The resulting temperatures are mostly lower than the sublimation temperature of dust ($\approx$ 1200 -- 1800 K) and are given in Tab.~\ref{tab:starfit}. Only a few galaxies show very blue nuclear spectra that are not consistent with thermal radiation from dust (no temperatures are given for these sources in the table). Most of these sources show no AGN light in our analysis of the CO equivalent widths. ESO 428-G14 and NGC 1097 are the only exceptions showing clear AGN signatures both in the near-IR, in X-rays and in the mid-IR, but a very blue continuum that is not compatible with thermal radiation from hot dust.

The median temperature of Seyfert 1 (excluding 1i) is $1292 \pm 46$ K, while Seyfert 2s show lower temperatures of $887 \pm 68$ K. Sy 1i galaxies are in between at $1112 \pm 47$ K. The general trend of cooler K-band temperatures of Seyfert 2s has been seen before, but mostly with somewhat lower temperatures. \citet{alonsoherrero1996} for example require dust temperatures of 800--1200 K to account for the AGNs in near-IR color-color plots; \citet{riffel_r2009} report average temperatures of 1000 and 600 K for type 1 and type 2 objects, respectively. The reason for their lower numbers is likely that they use lower spatial resolution data which are contaminated by cooler dust from regions off the nucleus. This is consistent with the lower AGN fractions ($f_{\rm nuc}$) reported by \citet{alonsoherrero1996}. However, the hotter dust temperatures we find are probably more realistic given the fact the the hottest dust is indeed at the sublimation radius as seen by near-infrared interferometric and dust reverberation observations \citep[e.g.][]{kishimoto2011}. This innermost radius of dust is smaller than 1 mas for the K-band brightest local Seyfert galaxies \citep{kishimoto2011}, i.e. much smaller than our resolution limit of 25 mas. However, all of these interferometrically studied sources show very high visibility in the K band, implying that the the continuum in the central few ten mas is dominated by hot dust emission, i.e. we are able to isolate this emission.

\begin{figure}[h]
   \centering
   \includegraphics[width=\hsize, trim=6cm 4cm 6cm 4cm]{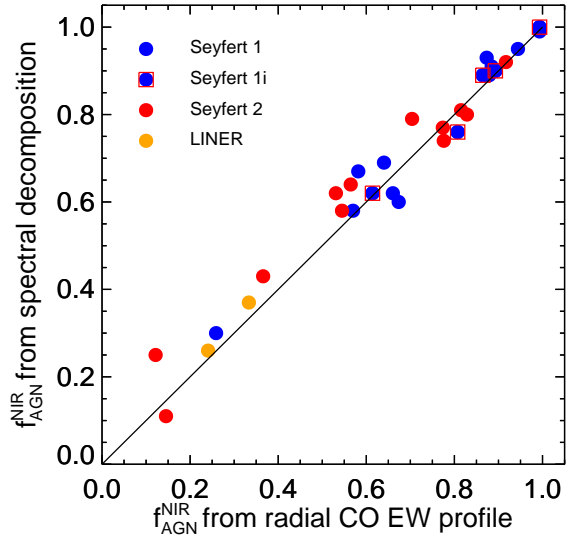}
   \caption{\label{fig:fagn_compare}Comparison of \fnir{} values (within 1 arcsecond for most sources) as derived from the radial profile of the CO EW and from the spectral decomposition.}
\end{figure}

An additional result of this analysis is a completely independent estimate of the fraction of AGN light in the near-IR \fnir{} -- on the same data and within the same aperture. We find very consistent results for most sources (Fig.~\ref{fig:fagn_compare}). Only in one source (NGC 1386), the results deviate substantially (\fnir{} = 12 \% in the radial decomposition, 25 \% in the spectral decomposition).

It is perhaps unexpected that these two methods work about equally well when one may have expected the radial decomposition to be a more solid analysis. The latter relies only on the  spatial invariance of the CO EW over the central 100 -- 200 pc (which, empirically, turns out to be the case) and not on the value of CO EW itself. It is helped by looking at the 2.3 \um{} CO bandhead, the depth of which does not vary much with star formation history for an evolving stellar population with only a few specific exceptions \citep[e.g.][]{davies2007}. The radial decomposition only uses the CO EW and is thus insensitive to variations in the spectral slope, i.e. extinction.

The spectral decomposition on the other hand has the observational advantage of not requiring spatial information at all, but relies on the invariance of both the spectral slope and the actual value of the CO EW (i.e. the stellar template chosen). It is sensitive to extinction, which can make the spectral slope redder (although that is mitigated by the fact we are looking in the K-band). This method will over-estimate the AGN fraction when there is significant reddening to the stellar population. This may be the reason why the \fnir{} values from the spectral decomposition tend to lie slightly above the value from the radial decomposition in Fig.~\ref{fig:fagn_compare}. In Section \ref{sec:twocomp} we discuss the relation between reddening of the AGN light and color temperature of the AGN heated dust in the context of a simple AGN obscuration model.

In conclusion it seems that both methods give similarly good results for the decomposition of stellar and AGN light. Given the fact that the dust temperatures derived from spectral decompositions are often higher than the dust sublimation temperature, however, we would still recommend to be cautious when only a spectral decomposition is used. If possible, it is advisable to use both the spectral and radial decomposition (on IFU data) to estimate \fnir{}.

\subsection{AGN classifications}

\begin{figure}[h]
   \centering
   \includegraphics[width=\hsize, trim=6cm 4cm 6cm 4cm]{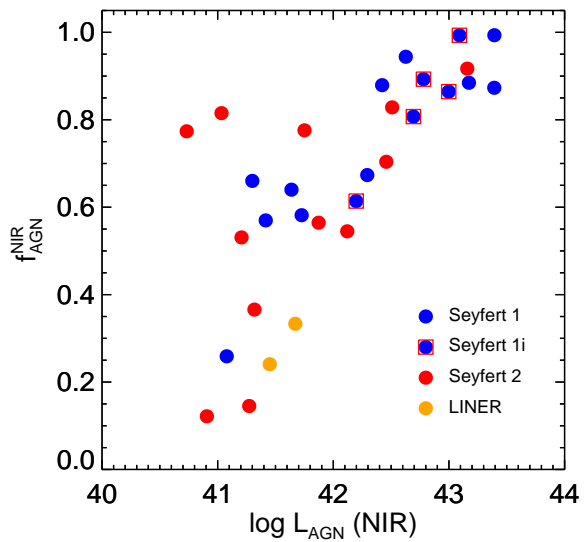}
   \caption{\label{fig:f_agn}Near-IR AGN fraction \fnir{} as a function of near-IR AGN luminosity for the various classes of AGNs. The apparent correlation between \fnir{} and $L_{\rm AGN}$ (NIR) is caused by the fact that the stellar luminosity is essentially constant within our aperture in all objects.}
\end{figure}

AGN fractions in the near-infrared (\fnir{}) are higher for optical type 1s than for optical type 2s, as expected, but this effect is entirely due to a luminosity mismatch between the two sub-samples as shown in Fig.~\ref{fig:f_agn}. There seems to be no difference between \fnir{} of type 1 and that of type 2 sources at any specific luminosity. The fact that \fnir{} can differ substantially for AGNs in our luminosity range means that one has to be very careful when constructing AGN SEDs from large-aperture photometry data in the near-infrared.

Here we classify AGNs as type 2 if they belong to the NED classes "Sy2" or "Sy1h" (Seyferts with hidden broad line regions that can be seen in polarized light). Classes Sy1, Sy1.5, Sy1.8, Sy1.9, Sy1n and Sy1i will be put into the type 1 bin -- but we will mark Sy1i specially to show how important it is to correctly classify these intermediate sources as type 1 in infrared studies.

For NGC~1386 the detection of a somewhat broad ($\sim 800$ km/s) \brg{} line has been reported by \citet{reunanen2002} and the source is therefore listed as an infrared type 1 Seyfert galaxy in NED (although most Sy1i have line widths $\gg$ 1000 km/s). With our higher spatial resolution and better S/N data, we cannot confirm this broad line, but instead find that the \brg{} line is composed of two narrow ($\sim 300$ km/s) components that are offset by 355 km/s (see Fig.~\ref{fig:Tgox}). We will therefore treat this galaxy as a Seyfert 2 galaxy in our analysis.

%\clearpage
%%
%% Table generated by IDL script tab_starfit, caption maintained by hand
%%
\begin{table*}[h]
\begin{center}
\caption{\label{tab:starfit}$EW_{\rm int}$: intrinsic equivalent width of the CO feature. For sources where it cannot be derived from the data due to the limited field of view, the median value for non-diluted cubes (11.1 \AA{}) is used. Luminosities are given in log(L/[erg/s]). Mid-IR, X-ray and [O IV] luminosities are only listed for sources where we can measure the NIR luminosity of the AGN. The K magnitude is extracted within 1 arcsecond, except for the two sources marked with $\star$ where the aperture is 0.5 arcseconds. References for mid-IR measurements: a: \citet{raban2008}, b: \citet{burtscher2013}, c: \citet{asmus2014}. References for X-ray luminosities: BAT: 70-month BAT survey \citep{baumgartner2013}, absorption-corrected 2-10 keV: d: \citet{levenson2006}, e: \citet{gandhi2009}, f: \citet{asmus2011}. References for [O IV] luminosities: g: \citet{diamondstanic2009}, h: \citet{weaver2010}, i: \citet{tommasin2010}}
\begin{tabular}{|l|l|l|l|l|l|l|l|l|}
Source id & K mag & $EW_{\rm int}$ [\AA{}] & \fnir [\%]& T [K] & $\log\left(\lnir\right)$ & $\log\left(\lmir\right)$&$\log\left(\lx\right)$&$\log\left(\loiv\right)$\\
\hline
Circinus $\star$&11.11&(11.1)& 77&$ 691 \pm    2$&40.73&42.73$^b$&41.76$^{\rm BAT}$&40.16$^g$\\
ESO428&11.75&10.5& 14&--&41.27&42.31$^c$&43.20$^{d}$&--\\
ESO548&10.30& 7.2& 87&$1236 \pm    3$&43.39&42.95$^c$&43.26$^{\rm BAT}$&--\\
IC1459&11.54& 9.6&< 10&--&$<$41.33&--&--&--\\
IC5063&12.89& 7.8& 56&$ 947 \pm   11$&41.87&43.63$^c$&43.18$^{\rm BAT}$&41.39$^h$\\
M87&12.56&12.8& 66&$1423 \pm    8$&41.30&41.27$^c$&41.43$^{f}$&--\\
MCG523&10.43& 8.0& 86&$1207 \pm    5$&43.00&43.53$^b$&43.58$^{\rm BAT}$&40.72$^h$\\
NGC289&13.17&11.8&< 10&$1006 \pm   61$&$<$40.44&--&--&--\\
NGC613&12.94&10.7&< 10&--&$<$40.46&41.53$^c$&--&--\\
NGC676&12.16&11.2&< 10&--&$<$40.75&--&--&--\\
NGC1052&11.43&11.2& 24&$1239 \pm   14$&41.45&42.22$^c$&42.15$^{\rm BAT}$&39.06$^h$\\
NGC1068& 7.90& 7.5& 91&$ 723 \pm    1$&43.16&44.00$^b$&41.94$^{\rm BAT}$&41.67$^g$\\
NGC1097&12.09&10.7& 25&--&41.08&41.17$^c$&41.50$^{e}$&39.33$^i$\\
NGC1365& 9.73& 6.4& 94&$1282 \pm    4$&42.63&42.56$^b$&42.39$^{\rm BAT}$&40.78$^g$\\
NGC1386&11.85&10.4& 12&$ 982 \pm   12$&40.91&42.48$^c$&42.80$^{f}$&40.54$^g$\\
NGC1566&11.38& 9.1& 56&$1479 \pm    8$&41.42&41.42$^c$&41.58$^{\rm BAT}$&39.20$^g$\\
NGC2110&10.89& 9.5& 80&$ 967 \pm    2$&42.69&43.08$^c$&43.69$^{\rm BAT}$&40.84$^h$\\
NGC2911&12.79&11.1&< 10&--&$<$41.35&--&--&--\\
NGC2974&11.80&11.0&< 10&$1106 \pm   32$&$<$41.10&--&--&--\\
NGC2992&11.57&12.6& 61&$1111 \pm    7$&42.20&42.76$^c$&42.51$^{\rm BAT}$&41.11$^g$\\
NGC3081&12.83&12.8& 36&$ 977 \pm   10$&41.32&42.53$^c$&42.85$^{\rm BAT}$&40.92$^g$\\
NGC3169&11.46&10.2&< 10&$1600 \pm   40$&$<$41.00&40.93$^c$&--&--\\
NGC3227&10.53& 9.7& 67&$1543 \pm   15$&42.29&42.62$^b$&42.76$^{\rm BAT}$&40.47$^g$\\
NGC3281&12.06& 9.2& 82&$ 732 \pm    3$&42.51&43.36$^b$&43.39$^{\rm BAT}$&41.60$^g$\\
NGC3312&12.44&11.4&< 10&$ 518 \pm   15$&$<$41.36&41.68$^c$&--&--\\
NGC3627&11.56&11.3&< 10&$1433 \pm   39$&$<$40.06&40.23$^c$&--&--\\
NGC3783&10.00&(11.1)& 99&$1244 \pm    3$&43.39&43.67$^b$&43.70$^{\rm BAT}$&40.88$^g$\\
NGC4051&11.14& 8.8& 64&$1305 \pm    4$&41.64&42.39$^c$&41.92$^{\rm BAT}$&39.75$^g$\\
NGC4261&12.99& 9.6&< 10&$1455 \pm   22$&$<$40.78&41.53$^c$&--&--\\
NGC4303&12.39&13.8&< 10&$ 702 \pm   22$&$<$40.44&40.62$^c$&--&--\\
NGC4388&12.09&10.7& 77&$ 887 \pm    4$&41.75&42.41$^c$&43.18$^{\rm BAT}$&41.15$^g$\\
NGC4438&11.63&11.0&< 10&$1185 \pm   33$&$<$40.68&40.77$^c$&--&--\\
NGC4472&12.88& 9.0&< 10&$1603 \pm   28$&$<$40.35&--&--&--\\
NGC4501&11.75&10.9&< 10&--&$<$40.79&40.53$^c$&--&--\\
NGC4569&10.92&11.2&< 10&--&$<$40.86&--&--&--\\
NGC4579&11.34&10.3& 33&$1374 \pm   12$&41.67&42.00$^c$&41.80$^{e}$&39.18$^g$\\
NGC4593&11.02&(11.1)& 87&$1292 \pm    4$&42.42&42.72$^b$&42.87$^{\rm BAT}$&40.05$^g$\\
NGC4762&12.09&11.4&< 10&--&$<$40.65&--&--&--\\
NGC5128&10.20&10.3& 81&$ 796 \pm    1$&41.03&41.84$^b$&42.39$^{\rm BAT}$&39.24$^g$\\
NGC5135&13.03&13.0& 54&$1033 \pm    9$&42.12&43.15$^c$&43.70$^{e}$&41.40$^g$\\
NGC5506& 9.25&(11.1)& 99&$1223 \pm    5$&43.09&43.22$^b$&43.22$^{\rm BAT}$&41.18$^g$\\
NGC5643&12.08&10.1&< 10&$ 900 \pm   30$&$<$40.66&42.34$^c$&41.79$^{\rm BAT}$&--\\
NGC6300&12.05& 9.9& 53&$ 677 \pm    2$&41.21&42.48$^c$&42.34$^{\rm BAT}$&39.82$^g$\\
NGC6814&11.98&11.7& 58&$1049 \pm    4$&41.73&42.18$^c$&42.68$^{\rm BAT}$&40.12$^g$\\
NGC7130&13.06&12.7&< 10&$1061 \pm   56$&$<$41.42&43.11$^c$&42.89$^{\rm BAT}$&--\\
NGC7135&12.20&10.5&< 10&--&$<$41.24&--&--&--\\
NGC7172&11.22& 9.6& 70&$ 822 \pm    3$&42.46&42.80$^c$&43.37$^{\rm BAT}$&40.83$^g$\\
NGC7469 $\star$&10.80&(11.1)& 88&$1355 \pm    6$&43.17&43.78$^b$&43.41$^{\rm BAT}$&41.15$^g$\\
NGC7496&12.62&13.1&  0&$ 667 \pm   53$&38.75&--&--&--\\
NGC7582& 9.63&10.5& 89&$1082 \pm    2$&42.78&42.95$^a$&42.63$^{\rm BAT}$&41.07$^g$\\
NGC7743&12.15&12.1&< 10&$1563 \pm   82$&$<$40.75&--&--&--\\
\end{tabular}
\end{center}
\end{table*}%
%\clearpage

%%%%%%%%%%%%%%%%%%%%%%%%%%%%%%%%%%%%%%%%%%%%%%%%%%%%%%%%%%%%%%%%%%%%%%%%%%%%%%%%%%%%%%%%%%%
%%%%%%%%%%%%%%%%%%%%%%%%%%%%%%%%%%%%%%%%%%%%%%%%%%%%%%%%%%%%%%%%%%%%%%%%%%%%%%%%%%%%%%%%%%%
%%%%%%%%%%%%%%%%%%%%%%%%%%%%%%%%%%%%%%%%%%%%%%%%%%%%%%%%%%%%%%%%%%%%%%%%%%%%%%%%%%%%%%%%%%%
%%%%%%%%%%%%%%%%%%%%%%%%%%%% DISCUSSION: NIR - X-RAY RELATIONS %%%%%%%%%%%%%%%%%%%%%%%%%%%%
%%%%%%%%%%%%%%%%%%%%%%%%%%%%%%%%%%%%%%%%%%%%%%%%%%%%%%%%%%%%%%%%%%%%%%%%%%%%%%%%%%%%%%%%%%%
%%%%%%%%%%%%%%%%%%%%%%%%%%%%%%%%%%%%%%%%%%%%%%%%%%%%%%%%%%%%%%%%%%%%%%%%%%%%%%%%%%%%%%%%%%%
%%%%%%%%%%%%%%%%%%%%%%%%%%%%%%%%%%%%%%%%%%%%%%%%%%%%%%%%%%%%%%%%%%%%%%%%%%%%%%%%%%%%%%%%%%%

%\clearpage
\section{Discussion}
\label{sec:discussion}

\subsection{AGN luminosity relations}

\begin{figure*}
   \centering
	% trim does l b r t
   \subfloat{\includegraphics[trim=7cm 4cm 7cm 4cm, width=0.5\hsize]{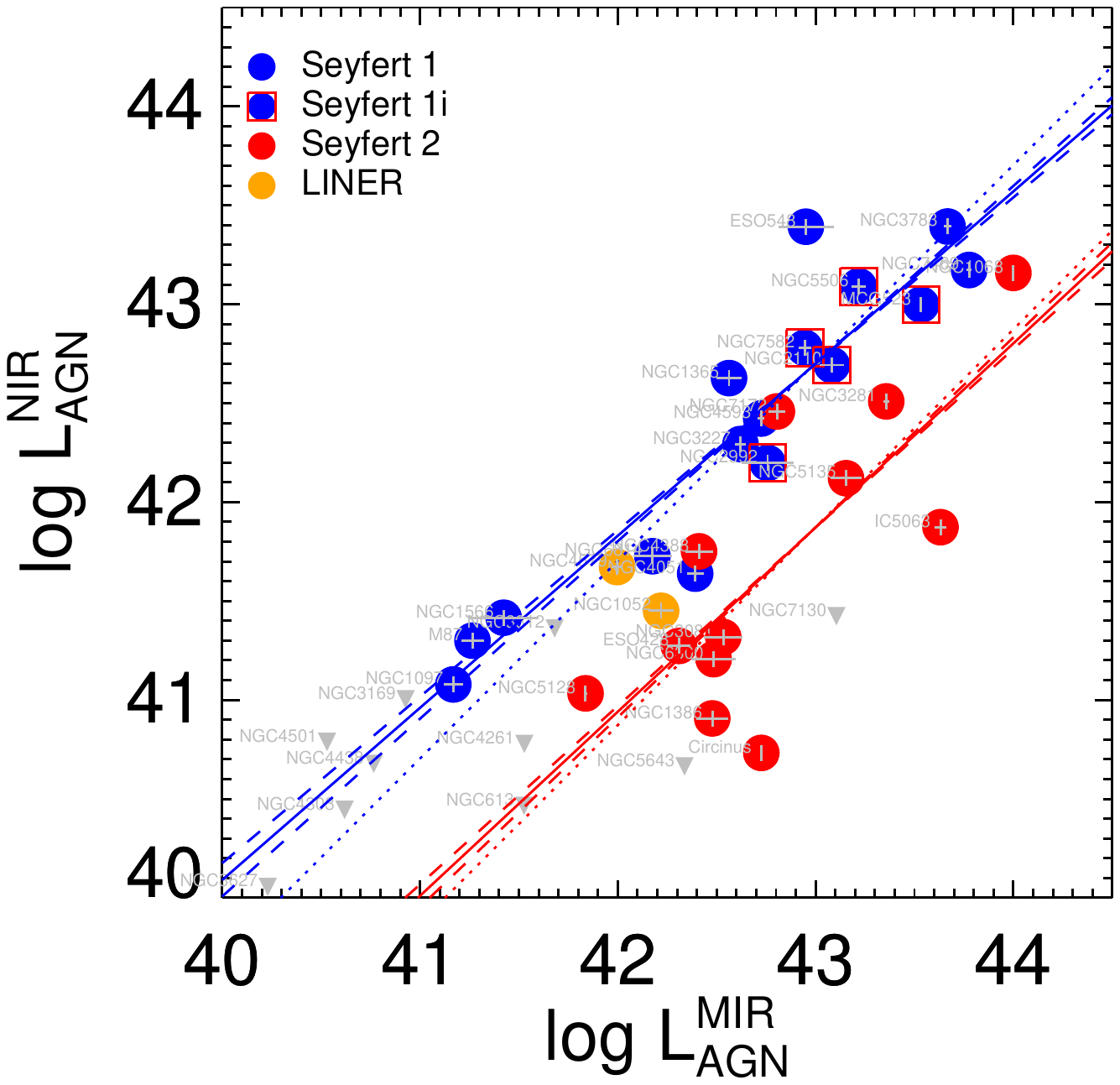}}
   \subfloat{\includegraphics[trim=7cm 4cm 7cm 4cm, width=0.5\hsize]{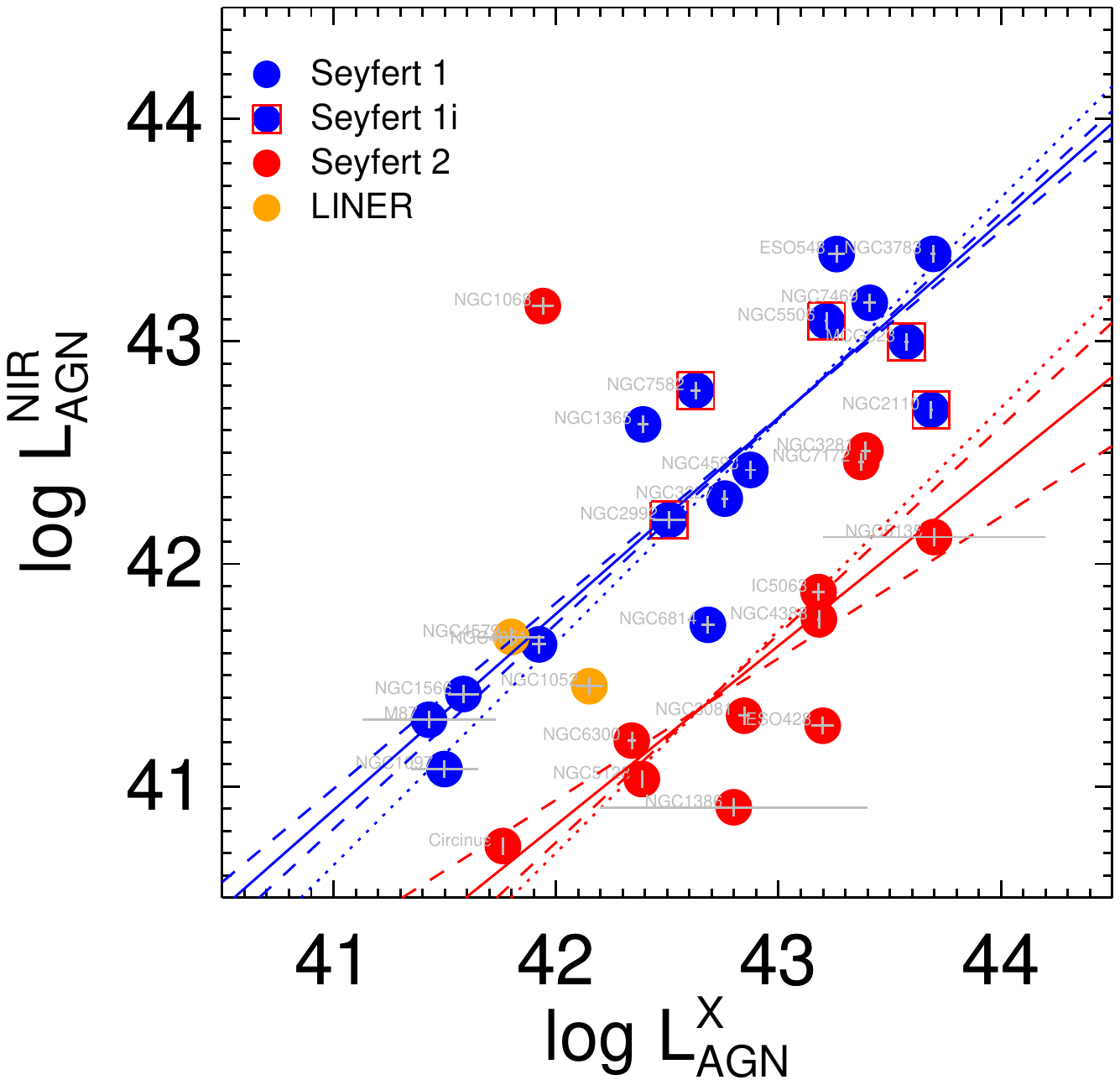}}\\
   \subfloat{\includegraphics[trim=7cm 3cm 7cm 4cm, width=0.35\hsize]{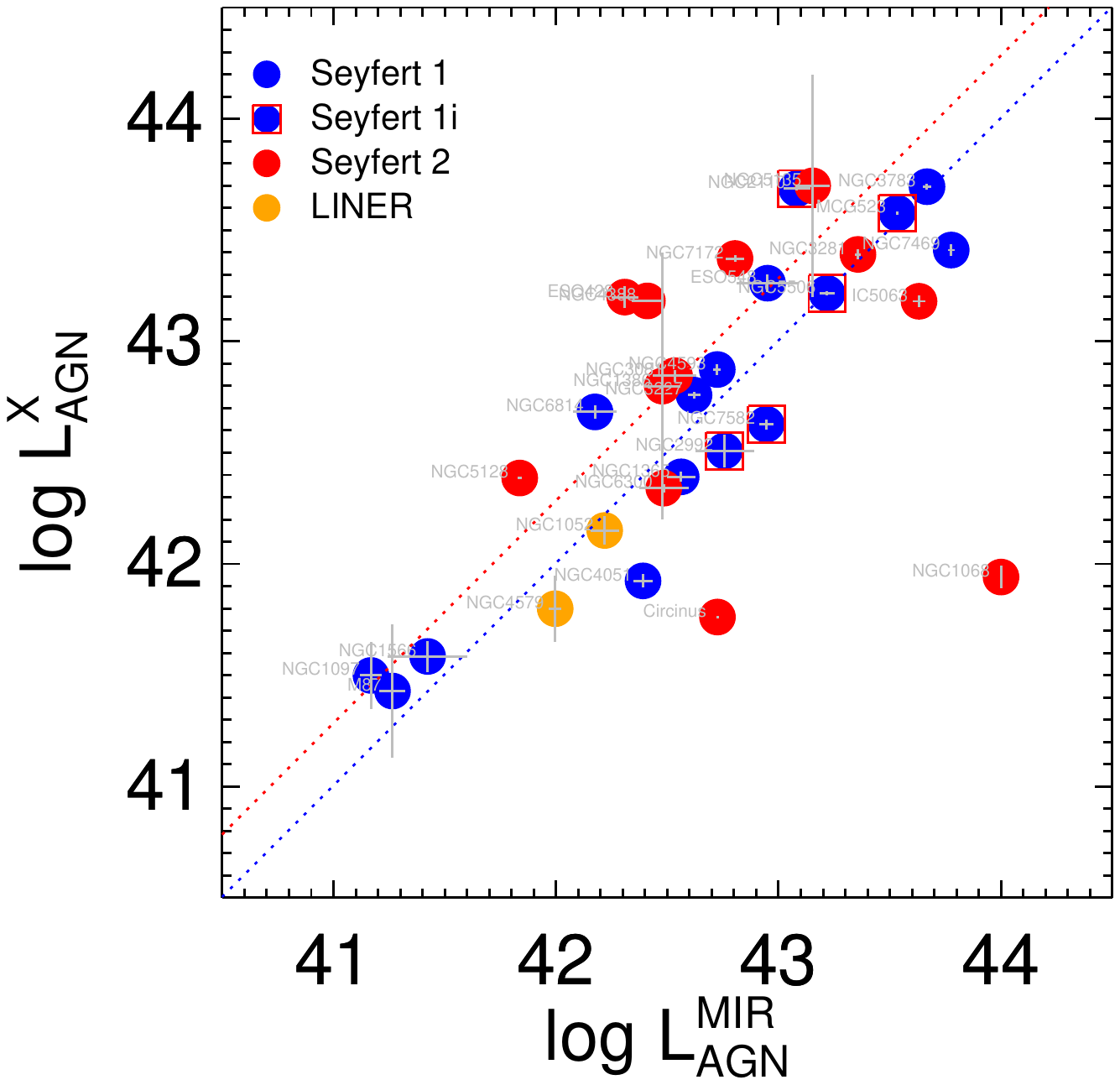}}
   \subfloat{\includegraphics[trim=7cm 3cm 7cm 4cm, width=0.35\hsize]{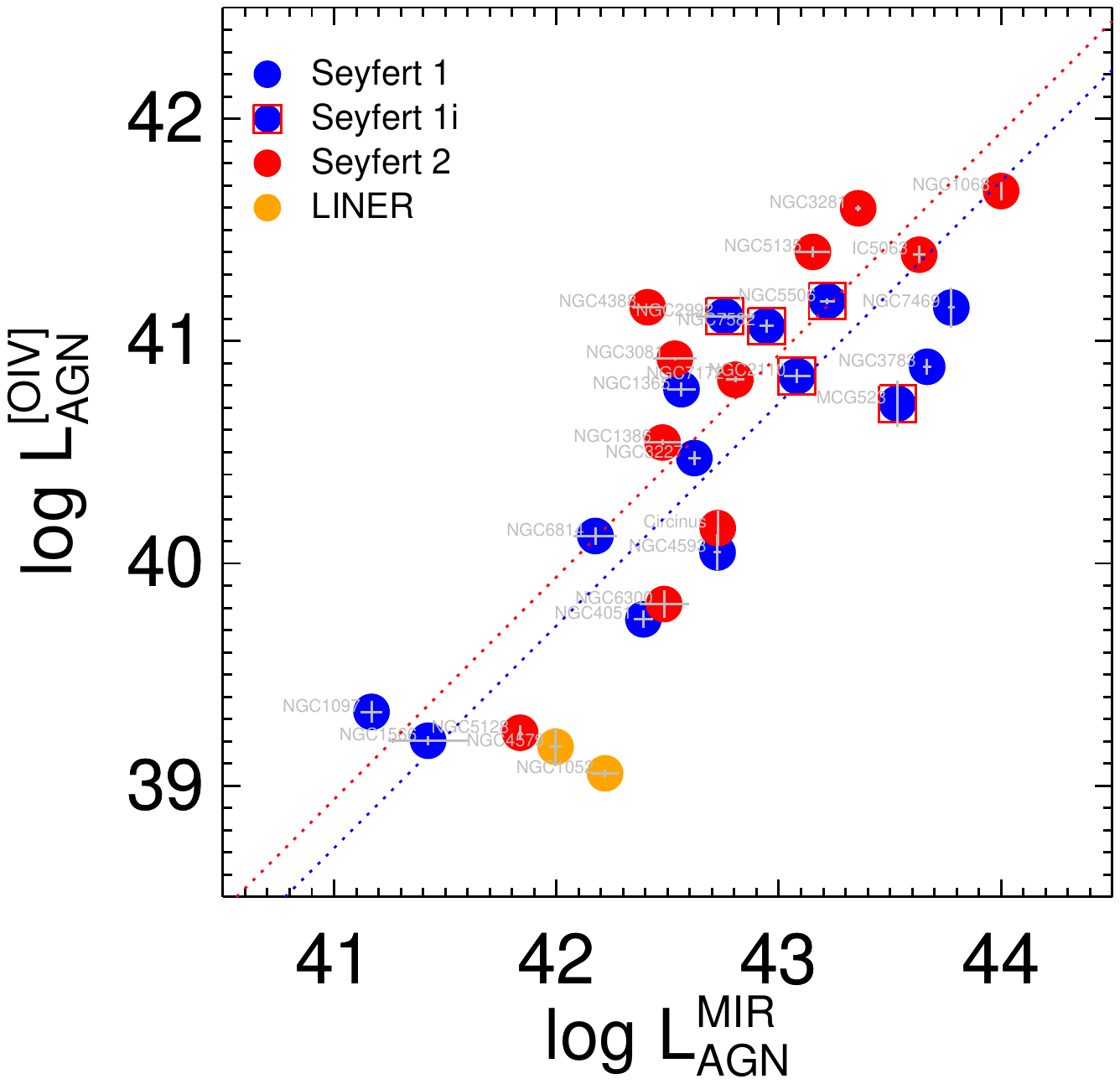}}
   \subfloat{\includegraphics[trim=7cm 3cm 7cm 4cm, width=0.35\hsize]{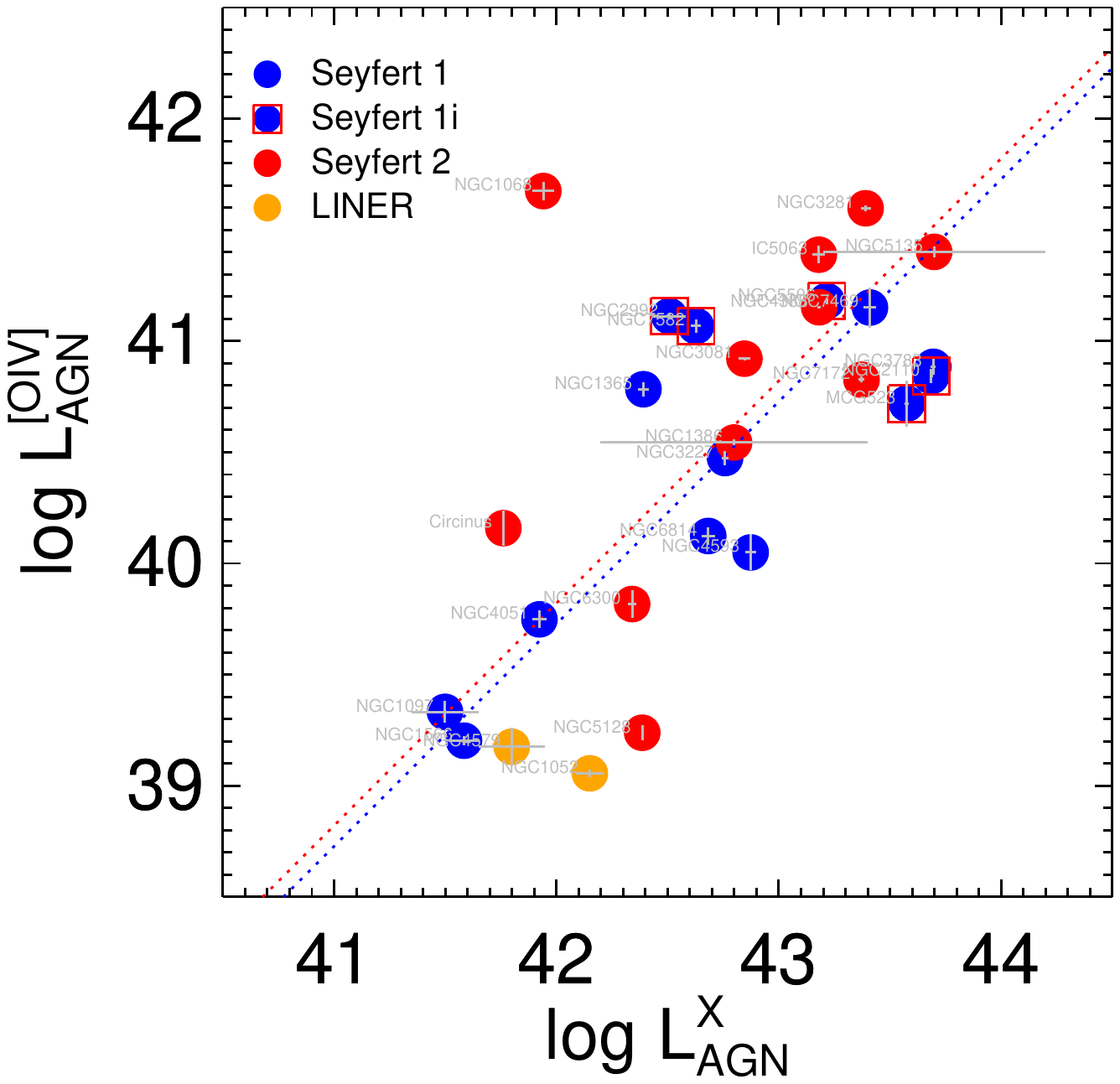}}
   \caption{\label{fig:relations}AGN luminosity relations. From top to bottom, left to right: (1) mid-IR -- near-IR, (2) hard X-rays -- near-IR, (3) mid-IR -- hard X-rays, (4) mid-IR -- [OIV], (5) hard X-rays -- [OIV]. Luminosities are given in $\log\left(L/{\rm [erg/s]}\right)$. The straight lines indicate the best fit, the dashed lines the $\approx 1 \sigma$ uncertainties and the dotted lines the best fit with slope fixed at 1; see text for details of the fitting procedure.}
\end{figure*}

% the data rows in this table have been generated by IDL procedure bootstrap_all
\begin{table*}[h]
\begin{center}
\caption{\label{tab:relations}Fit results for various AGN luminosity relations. The intercept and slope for e.g. the NIR--MIR relation is defined as: $\log\lnir$ = intercept + slope $\times \log\lmir{}$. For each relation, the K-S test is computed from the ratio of luminosities for type 1 vs. type 2 AGNs.}
\begin{tabular}{|l|l|l|l|l|l|l|}
\hline
relation & intercept type 1 & slope type 1 & intercept type 2 & slope type 2 & K-S distance & K-S prob \\
\hline
\hline
\multicolumn{7}{|c|}{type 1i treated as type 1}\\
NIR--MIR & $ 5.287^{+ 1.189}_{- 1.238}$&$ 0.870^{+ 0.029}_{- 0.028}$&$ 1.844^{+ 1.247}_{- 1.244}$&$ 0.931^{+ 0.029}_{- 0.029}$& 0.854&2.5e-05\\
NIR--X-rays & $ 4.748^{+ 1.745}_{- 1.968}$&$ 0.882^{+ 0.046}_{- 0.040}$&$ 6.990^{+ 5.483}_{- 7.227}$&$ 0.806^{+ 0.169}_{- 0.129}$& 0.875&2.5e-05\\
X-rays--MIR & $ 5.308^{+ 2.367}_{- 2.314}$&$ 0.877^{+ 0.054}_{- 0.055}$&$21.113^{+ 4.468}_{- 4.404}$&$ 0.511^{+ 0.103}_{- 0.105}$& 0.449&1.0e-01\\
OIV--MIR & $ 8.587^{+ 1.299}_{- 1.289}$&$ 0.747^{+ 0.030}_{- 0.030}$&$-2.597^{+ 1.630}_{- 1.655}$&$ 1.012^{+ 0.038}_{- 0.038}$& 0.292&5.9e-01\\
OIV--X-ray & $10.673^{+ 1.144}_{- 1.130}$&$ 0.697^{+ 0.026}_{- 0.027}$&$ 0.968^{+ 5.795}_{- 7.250}$&$ 0.926^{+ 0.170}_{- 0.136}$& 0.286&6.5e-01\\
\hline
\hline
\multicolumn{7}{|c|}{type 1i treated as type 2}\\
NIR--MIR & $ 4.624^{+ 1.228}_{- 1.292}$&$ 0.886^{+ 0.030}_{- 0.029}$&$-5.062^{+ 1.344}_{- 1.385}$&$ 1.097^{+ 0.032}_{- 0.031}$& 0.610&7.3e-03\\
NIR--X-rays & $ 0.586^{+ 2.438}_{- 2.726}$&$ 0.980^{+ 0.064}_{- 0.057}$&$ 5.467^{+ 5.439}_{- 7.362}$&$ 0.848^{+ 0.172}_{- 0.127}$& 0.722&9.2e-04\\
X-rays--MIR & $ 6.518^{+ 2.459}_{- 2.381}$&$ 0.848^{+ 0.055}_{- 0.057}$&$17.611^{+ 3.471}_{- 3.499}$&$ 0.592^{+ 0.081}_{- 0.081}$& 0.375&2.5e-01\\
OIV--MIR & $10.115^{+ 1.326}_{- 1.302}$&$ 0.708^{+ 0.031}_{- 0.031}$&$ 2.813^{+ 1.717}_{- 1.722}$&$ 0.886^{+ 0.040}_{- 0.040}$& 0.340&4.3e-01\\
OIV--X-ray & $ 6.932^{+ 1.520}_{- 1.531}$&$ 0.782^{+ 0.036}_{- 0.036}$&$13.078^{+ 2.617}_{- 3.566}$&$ 0.645^{+ 0.083}_{- 0.061}$& 0.422&2.0e-01\\
\hline
\end{tabular}
\end{center}
\end{table*}

\paragraph{Linear fits and correlation analysis}
Having compiled a large sample of nuclear AGN luminosities in a number of bands, we will now discuss their relations (Fig.~\ref{fig:relations}). For each relation (in log-log space), we generate $10^5$ re-sampled data points according to their two-dimensional (assumed Gaussian) error distribution and compute linear fits using a simple $\chi^2$ metric. We use the median of the resulting distributions to represent the best fit relation and cast our lower and upper uncertainties in terms of the $16^{\rm th}$ and $84^{\rm th}$ percentiles of the distributions. The results are given in Tab.~\ref{tab:relations}.

NGC~1068 has been excluded from all fits involving X-ray measurements since it is heavily absorbed \citep{bauer2015}. This affects even the very high energy band used by Swift/BAT as can easily be seen from its large offset from the respective relations.

\subsubsection{Near-IR luminosity relations}
We find that a strong correlation exists between the nuclear near-IR and mid-infrared luminosities of local AGNs (Fig.~\ref{fig:relations}, top left) as well as between the near-IR and hard X-ray luminosities (Fig.~\ref{fig:relations}, top right). The best fit is slightly different from a linear relation which is indicated as a dotted line in the fits. However, we consider the differences marginal and note that there is no indication in the literature of a non-linear behavior in the luminosity range that we cover \citep{lutz2004,gandhi2009}. Only at higher luminosities, the X-ray obscured fraction decreases \citep{merloni2014} which can be very well explained in the context of the receding (in fact: sublimating) torus \citep{davies2015}. This may change the slope (especially for the mid-IR -- X-ray relation, see below) at X-ray luminosities $>10^{44}$ erg/s.

Therefore, and since the scatter is dominated by intrinsic variations rather than observational errors, we compute the two-sided Kolmogorov-Smirnov statistic of the ratio of the luminosities to determine the significance of the offset between type 1 and type 2 AGNs. Both near-IR relations show a significant offset between type 1 and type 2 sources. At \lmir{} (\lx{}) = 42.5, type 1 sources are about 7 (10) times brighter in the near-IR than type 2 sources with very high significance (see Tab.~\ref{tab:relations}).

It is important to note that infrared Seyfert 1 galaxies \citep[e.g.][]{lutz2002}, where broad lines are visible only in infrared bands, lie in the same locus as optical type 1 sources and should therefore be treated as type 1 objects in infrared studies. This is perhaps not surprising since our analysis is based upon infrared emission as well, but it is an often overlooked issue. The K-S distance and probability indicate a better separation and higher significance of the separation if type 1i objects are treated as type 1 AGNs (see Tab.~\ref{tab:relations}). We have therefore marked these sources as type 1 objects (blue) with a red box around the symbol. In our type 1 sample, five out of 16 sources belong to the 1i class.

Our results can be compared to the finding from \citet{hoenig2011} who estimated the anisotropy in the infrared emission in a sample of powerful quasars and radio galaxies and find an anisotropy factor of about eight at 2.3 \um{} which is consistent with the offset we find in the near-IR -- mid-IR relation for our sample of much lower luminosity AGNs.

%\citet{ivanov2000} have also used the equivalent width of the near-IR stellar CO features to derive the fraction of AGN light. However, likely due to the much lower spatial resolution of their data, they do not find differences between type 1 and type 2 sources.
% referee reply: In discussion with colleagues we were pointed at an earlier paper (Ivanov et al. 2000) that had used a similar method but with different results. We refer to this study briefly in Section 4.1.

\subsubsection{Isotropic AGN luminosity relations}
From the same sample, we also construct the relations between mid-IR, hard X-ray and [O IV] luminosities (Fig.~\ref{fig:relations}, bottom row), all of which are supposed to be indicators of the isotropic luminosity of AGNs \citep[e.g.][]{lamassa2010}. In our compilation we indeed do not see significant differences between type 1 and type 2 sources in these wavebands. The best fits to the relations indicate that some of them are highly non-linear (see Tab.~\ref{tab:relations}). Due to the larger scatter, however, we do not further discuss these non-linearities. In the plots we show the best fit for an assumed linear relation.

For the well-known mid-IR -- hard X-rays relation there is a marginal indication (on a 10\% significance level) that type 2 sources are brighter by a factor of about 2 in \lx{} at \lmir{}= 42.5 than type 1 sources. This indicates that the mid-IR emission from the AGN ``torus'' is indeed slightly anisotropic as suggested by both smooth and clumpy torus models. \citet{nenkova2008}, for example, find a factor 2 increase in flux between edge-on and face-on clumpy tori and \cite{snyder2013} find the least obscured sources to be more than a factor of ten more luminous at 12 \um{} than the most obscured sources. In most studies of this relation, however, no difference has been seen \citep[e.g.][]{lutz2004,gandhi2009,asmus2011}. We believe that the marginal difference we see arises due to two improvements over previous studies:

\begin{enumerate}
	\item Absorption corrected 2-10 keV luminosities are a less reliable indicator of the hard X-ray radiation of an AGN than the Swift/BAT fluxes from the 70-month data release that we mostly use (for 24 out of 30 sources). The Swift/BAT fluxes are essentially unaffected by absorption (the attenuation is less than 2 for $\log(N_H/{\rm cm^2}) \lesssim 24.3$). Due to the long-term nature of the BAT survey, variability effects are also reduced.
	\item The small anisotropy in the hard X-ray -- mid-IR relation is washed out if type 1i sources are not classified as type 1 (``face-on'') sources.
\end{enumerate}

%A similar effect is also seen in the [O IV] -- mid-IR relation. There, type 2 sources are brighter than type 1 sources by a factor of 1.3 on average at a given mid-IR luminosity. This is nicely consistent with the recent result of \citet{yang2015} who find a factor 2.6 anisotropy in the mid-IR emission using [O IV] as an indicator of the bolometric luminosity of the AGN. Using our multiband analysis, we show that this effect is dominated by an anisotropy in the mid-IR radiation since there is nearly no difference between type 1 and type 2 objects in the [O IV] -- hard X-ray relation (Fig.~\ref{fig:relations}, bottom right). 

Most recently, \citet{mateos2015} studied the relationship between the 6 \um{} and 2-10 keV luminosities of a sample of relatively powerful AGNs and also found an anisotropy in the sense that, at a given 2-10 keV luminosity, type 2 AGNs are $\sim$ 1.3 -- 2 times fainter at 6 \um{}. This indicates that the 6 \um{} emitting region is more closely related to the 12 \um{} emitting region (which has a similar anisotropy) than to the hottest dust, emitting at 2.3 \um{}, which has a much larger anisotropy of a factor of about 7. This is encouraging for the upcoming VLTI instrument MATISSE \citep{lopez2014} which will open the 3-5 \um{} window for imaging interferometry. The small anisotropy at 6 \um{} suggests that at the wavebands of MATISSE, one will be able to resolve the relatively large warm body of the torus and not the innermost hot dust (which is essentially unresolved with the currently available baselines).

There is a slight offset between type 1 and type 2 sources in the [O IV] -- mid-IR relation (Fig.~\ref{fig:relations}, bottom center) that would also indicate that the mid-IR emission is anisotropic (as also seen by \citet{yang2015}). However, this offset is not significant. There is nearly no difference between type 1 and type 2 objects in the [O IV] -- hard X-ray relation (Fig.~\ref{fig:relations}, bottom right). 

%While type 1 and type 2 AGNs differ significantly in the near-IR relations according to our fits (see Tab.~\ref{tab:relations}), the intrinsic scatter in the relations is so large that the probability of both type 1 and type 2 relations being drawn from the same parent distribution is still very high.

\subsection{Two component model}
\label{sec:twocomp}
\begin{figure*}
\sidecaption
   \includegraphics[trim=5cm 4cm 5cm 4cm, width=12cm]{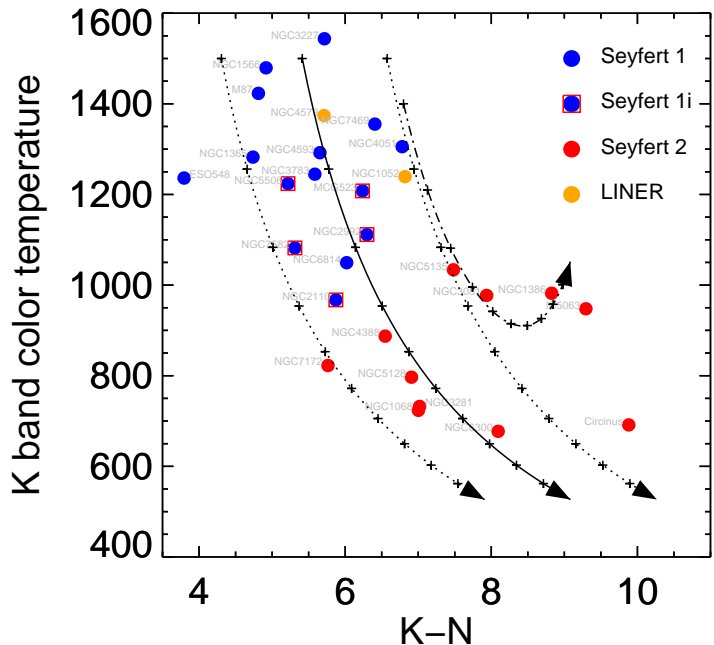}
   \caption{\label{fig:color_T}Near-IR color temperature vs. NIR-MIR color. Red infrared colors are to the right. The arrows are extinction vectors from our two blackbody model (crosses are at $A_{\rm V}^{\rm hot} = 0, 5, 10, \ldots$ mag). Solid arrow: $T_{\rm hot}$ = 1500 K, $T_{\rm warm}$ = 300 K, $f_{\rm area}$ = 1000. The dotted vectors are for a third (left) and three times (right) $f_{\rm area}$. The curled (dash-dotted) vector is with a scattering fraction of 10\% (for details see text).}
\end{figure*}

If both type 1 and type 2 sources are drawn from the same population -- the central premise of unification models -- then the clear offset of the type 1 and type 2 sources in the near-IR -- mid-IR luminosity relation has to be explained simply by inclination, i.e. different amounts of obscuration of the central source.

In order to test the effect of obscuration, we plot the near-IR color temperatures that we derive from our spectral fitting against the offset from the near-IR -- mid-IR luminosity relation (Fig.~\ref{fig:color_T}). Here we have converted this offset to an astronomical ``color'' so that it can be easier compared to other observations. We refer to the near--to--mid-infrared dust color as ``K-N'' although the magnitudes are actually not drawn from band-averaged fluxes, but from 2.3\um{} and 12.0\um{} monochromatic fluxes, respectively.

The first thing to note in our temperature--color plot is that AGNs of all optical types have very red near--to--mid-infrared colors: $K-N \gtrsim 4$. As expected, type 1s are less red and hotter than type 2s, but both have near-infrared temperatures that are indicative of hot dust. Again we see that infrared type 1 Seyfert galaxies are closer to optical type 1 than type 2 galaxies in this plot with temperatures intermediate between the two classes.

To explain both the hot near-IR temperatures and the red near--to--mid-infrared color, we require a hot and a warm component and only the hot one can be significantly inclination dependent (obscured) because of the near isotropy of the mid-IR radiation. Such a two-component model is also observationally motivated as AGN tori are known to consist of two spatially distinct components of hot and warm dust \citep[e.g.][]{burtscher2009}.

The intensity in this model is then given by

\begin{align}
	I_{\lambda} &=& I_{\lambda}^{\rm hot} + f_{\rm area} \cdot I_{\lambda}^{\rm warm},\\
\intertext{where $f_{\rm area}$ is the area ratio between the warm and the hot emitter and the two components are given by}\\
I_{\lambda}^{\rm hot} &=& B_{\lambda}(\lambda,{\rm T_{hot}}) \cdot \exp(-\tau_{\rm V} (\lambda) \cdot A_{\rm V}^{\rm hot}/1.09),\\
I_{\lambda}^{\rm warm} &=& B_{\lambda}(\lambda,{\rm T_{warm}}).
\end{align}

Here $B_{\lambda}$ is the Planck function for the spectral radiance in wavelength space, $T_{\rm hot}$ and $T_{\rm warm}$ are the temperatures of the respective components and $\tau_V(\lambda)=\tau_{\lambda}/\tau_{\rm V}$ is the optical depth of the dust screen at a given wavelength. We interpolate it logarithmically from the extinction curve of \citet{schartmann2005} and give it relative to the optical depth at V band (5500 \AA{}) where it is normalized to 1. $A_{\rm V}^{\rm hot}$ is the screen extinction towards the hot component at V band in magnitudes. We do not require an extinction to the warm component.

The temperatures of our simple two-component ``torus'' model are set to $T_{\rm hot} = 1500$ K and $T_{\rm warm} = 300$ K. The hot temperature is chosen according to the hottest temperatures that we measure from the near-IR spectra. The warm temperature is not very well defined by this analysis, but since the maximum of the re-emitted AGN radiation is at $\approx$ 10 \um{}, the temperature of the warm component must be at about 300 K (see e.g. Figs. 3 and 4 of \citet{prieto2010}).

The K-N color of the type 1 galaxies in our temperature-color plot is set by $f_{\rm area} \approx 1000$. This value can be compared to interferometrically measured radii for the warm and hot dust, the ratio of which is 10 \ldots 30 (see Fig. 36 of \citet{burtscher2013}), i.e. $f_{\rm area} = 100 \ldots 1000$, in reasonable agreement with our temperature--color plot, Fig.~\ref{fig:color_T}, where the horizontal range covered by the dotted lines corresponds to $f_{\rm area} = 300 \ldots 3000$. The modeled range of $f_{\rm area}$ is about three times larger than what one may expect from the interferometric observations. This difference cannot be explained by extinction to the mid-IR component, since an obscuration of about 2 mag (at 12 \um{}) would have the same effect as {\em reducing} $f_{\rm area}$ by a factor of 3. However, a slight variation of $T_{\rm warm}$ by a factor of $3^{1/4} \approx 1.3$ suffices to explain the small discrepancy. 

The only other free parameter of our model is the extinction to the hot component $A_{\rm V}^{\rm hot}$ which may be related to the inclination of the ``torus'' axis out of our line of sight. This parameter takes values of $0 \ldots 5$ mag for optical type 1 Seyferts, $5 \ldots 15$ mag for type 1i Seyferts and $15 \ldots 35$ mag for type 2 Seyferts. Since $\tau_V \approx 13 \cdot \tau_{2.3 \um}$ \citep{schartmann2005}, the extinction for type 1i Seyferts can also be expressed as $A_{2.3 \um{}}^{\rm hot} = 0.4 \ldots 1.2$ mag. This is consistent with their classification which is based on the fact the broad lines are visible in the infrared. Probably these sources host an intrinsic type 1 nucleus but are slightly obscured by foreground dust in their host galaxy.

Even this simple model is degenerate: a variation in $T_{\rm warm}$ has nearly the same effect as a variation in $f_{\rm area}$ and instead of obscuring the hot component by small amounts of dust (i.e. moderate inclinations?), we could also attribute the range of Seyfert 1 temperatures to intrinsic variations. But the point of this model is that with only two free parameters ($f_{\rm area}$, $A_{\rm V}^{\rm hot}$) as well as two observationally determined parameters ($T_{\rm hot}, T_{\rm warm}$) we can explain almost the entire range of near-IR temperatures and near-to-mid-infrared colors in our sample of local Seyfert galaxies.

There are four outliers that cannot be explained by a variable amount of extinction of the hot component and small variations of $f_{\rm area}$. On the blue side there is one source, ESO 548-G81 which may just have a lower fraction of warm to hot dust compared to the bulk of Seyfert 1 sources. On the red side there are three outliers: NGC~1386, IC~5063 and the Circinus~galaxy. In the reddest of all, the Circinus~galaxy with K-N $\approx$ 10, the hot component seems to be intrinsically much weaker than in the other sources. This has been seen in high-resolution SED models \citep{prieto2004} and recently been confirmed by a detailed analysis of the highest resolution interferometric data \citep{tristram2014}. Although the Circinus~galaxy is so nearby that the sublimation radius is nearly resolved, there is no evidence for hot dust in the interferometric data. This is in contrast to NGC~1068, where a similar resolution is reached in infrared interferometric observation and a hot component is clearly detected \citep{lopezgonzaga2014} -- consistent with the ``bluer'' color of NGC~1068 compared to the Circinus~galaxy.

The other two red outliers, NGC~1386 and IC~5063, are harder to explain as they are both red {\em and} hot\footnote{NGC~1386 should perhaps be treated with caution as it is the only source with a factor two difference in \fnir{} between our two analyses. We have plotted the K-N color derived from the radial profile decomposition, but note that the source would move to the left by about 0.8 mag if we had taken the value derived from the spectral decomposition. We have, however, no reason not to trust the radial decomposition in this source.}. There is no extinction vector that would connect the locus of these sources with a plausible ``starting'' position among the type 1 sources, but there are a few possible ways of increasing the color temperature in the $K$ band while keeping the red colors.

A significant contribution from a non-thermal source with a flat spectrum ($F_{\nu} \propto \nu^0$) would increase the modeled temperature, but most Seyfert galaxies are not dominated by non-thermal radiation in the infrared, as evidenced for example by the fact that the infrared emission is resolved in most objects with interferometry \citep{burtscher2013}.

Alternatively, if some part of the hot dust radiation could escape unreddened, it would also affect the K band color in the desired way, i.e. if we replace $I_{\lambda}^{\rm hot}$ by

\begin{equation}
\widehat{I}_{\lambda}^{\rm hot} = I_{\lambda}^{\rm hot} + f_{\rm scat} \cdot I_{\lambda}(\lambda,{\rm T_{hot}}),
\end{equation}

where $f_{\rm scat}$ is the fraction of hot blackbody emission that is scattered into our line of sight without obscuration.

One possibility to achieve this would be via elastic scattering off free electrons in the narrow-line region. This is perhaps attractive since IC~5063 is observed to have a hidden broad-line region as seen in polarized light \citep{inglis1993}. This indicates that a scatterer exists in this source. The scattered radiation should be measurable via a high degree of polarization also in the near-infrared as the Thomson cross-section for scattering off free electrons is wavelength-independent. For an extinction vector including scattering to match the red and hot outliers in Fig.~\ref{fig:color_T} (the curled dot-dashed line), we require a scattering fraction of 10 \% which corresponds to an intrinsic polarization of about 60 \% in IC~5063 since the direct radiation would be attenuated by $A_{2.3 \um{}} \approx 3$. For this source, very detailed polarization observations are available by \citet{lopez_rodriguez2013} who measure a degree of polarization of 8 \% at 2.3 \um{}. Using our own \fnir{} as well as an aperture correction factor of 1.6, this corresponds to an intrinsic polarization of 23 \%. Scattering therefore seems to be an unlikely explanation. \citet{lopez_rodriguez2013} also considered synchrotron radiation as the origin of the polarized light in IC~5063, but since the source shows no variability, it is unlikely to host a strong non-thermal source.

Another way of ``by-passing'' the obscuring torus would be to give up the assumption of uniform screen absorption of the hot component. This could be through ``holes'' in the clumpy torus or because only a part of the region hosting the hottest dust is covered by colder torus clumps or because we are seeing the source at a grazing inclination angle. This may imply that these sources are good candidates for ``changing-look'' AGNs which are in the process of changing between obscured and unobscured classes \citep[e.g.][]{risaliti2010}. However, the red outliers in the temperature-color plot are all (moderately) Compton-thick sources, where one may not expect holes in the torus. In conclusion, most the sources are explained well with this very simple two-component model and the few outliers probably represent the intrinsic diversity in AGN nuclear dust structures.

\section{Conclusions}
\label{sec:conclusions}

For an archival sample of 51 local Seyfert galaxies, we analyze high-resolution near-IR integral field spectroscopic observations, mostly from VLT/SINFONI observations. We use these data to produce equivalent width maps of the stellar CO $\lambda\, 2.29\um{}$ absorption feature. From the dilution of this feature, we infer the AGN fraction and the AGN luminosity in the near-IR \lnir{}, i.e. hot dust luminosity. We find a significant fraction of nonstellar light in 31 sources. From the literature, we additionally collect measurements of the nuclear mid-IR radiation from high-resolution ground-based observations, hard X-ray fluxes from the Swift/BAT all-sky survey and other observations as well as \oiv{} fluxes from Spitzer observations. These measurements are considered isotropic indicators for the AGN luminosity. Equipped with this collection of high resolution data, our findings are:

\begin{itemize}
	\item The near-IR AGN fraction \fnir{} (within 1 arcsec) varies quite significantly among AGNs. In our sample, type 1 Seyfert galaxies have higher \fnir{} than type 2s only because of a luminosity mismatch between the two subsamples. AGNs of both types can take on fractions from near 0 to 100 \%. It is therefore important to take into account contamination by stellar light even when observing with small apertures.
	\item There is an apparent bimodality in the nuclear equivalent width of the stellar CO feature in nearby galaxies. Inactive galaxies and weak AGNs show an almost undiluted feature while powerful AGNs are strongly diluting the stellar light. The transition occurs at the luminosity where the torus is expected to vanish. However this is a coincidence that is caused by the fact that the stellar surface brightness is nearly constant for our sample of galaxies over a wide range of AGN luminosities.
	\item Using our estimate of the near-IR AGN luminosity with high-resolution mid-IR and hard X-ray measurements, we find a strong correlation with significant offsets between optical (and infrared) type 1 AGNs and type 2 AGNs. They are typically 7 (10) times brighter in the NIR at a typical luminosity of $\log$ \lmir{} = 42.5 ($\log$ \lx{} = 42.5).
	\item We also find a marginally significant offset between type 1 and type 2 sources in the hard X-rays--mid-IR relation in the sense that type 1 sources are about two times brighter at a typical hard X-ray luminosity and attribute this to anisotropy in the mid-IR emission of the torus. The relation between [O IV] luminosity and hard X-ray luminosity, on the other hand, shows almost no offset between the two classes of AGNs. This suggests that [O IV] and hard X-rays are equally isotropic indicators of the AGN luminosity and the anisotropy in the other relations arises in the near- and mid-IR emitting regions.
	\item We invoke a simple two component model that consists of a warm ($T_{\rm warm} \approx 300$ K) and a hot ($T_{\rm hot} \approx 1500$ K) blackbody and an area scaling factor between them. We find that the scaling factor ($\approx$ 1000) is consistent with the expected area ratio given the interferometrically observed ratio of mid-IR to near-IR radii. Almost all our sources can then be explained by simply adding an absorbing screen to the hot component. An $A_V$ of 5-15 mag is required for type 1i sources (corresponding to $\approx$ 1 mag in the near-IR) and 15-35 mag is found for type 2 sources; pure type 1 sources cover an $A_V$ range of 0-5 mag. There are only four outliers to this simple scenario. Three of them are Seyfert 2 galaxies with a very red K-N color but hot near-IR color temperature. This can possibly be explained by scattering, non-thermal emission or variations from a screen absorber, but detailed studies are needed to clarify this.
	\item We find that infrared type 1 Seyfert galaxies, where broad lines are only visible in infrared wavebands (such as \brg{}), are more similar to type 1 galaxies also in other infrared indicators (e.g. \fnir{} and K band color temperature) and suggest to treat these Seyfert 1i galaxies as type 1 sources. A mis-classification of these sources may otherwise complicate results from infrared studies.
\end{itemize}

\section*{Acknowledgements}
We thank the anonymous referee for comments that helped to improve the paper. We also thank Thomas Ott and Alex Agudo Berbel for developing and distributing the QFitsView software. This research has made use of the SIMBAD database, operated at CDS, Strasbourg, France. This research has made use of the NASA/IPAC Extragalactic Database (NED) which is operated by the Jet Propulsion Laboratory, California Institute of Technology, under contract with the National Aeronautics and Space Administration. Based on data obtained from the ESO Science Archive Facility. This research has made use of NASA's Astrophysics Data System Bibliographic Services. This research made use of Astropy, a community-developed core Python package for Astronomy \citep{astropy2013}.

\bibliographystyle{aa}
\bibliography{../apj-jour,../papers}

\end{document}